\documentclass[a4paper,10pt]{article}
\pdfoutput=1

\usepackage[usenames,dvipsnames,table]{xcolor}
\usepackage{float}
\usepackage[utf8]{inputenc}
\usepackage{jheppub}
\usepackage{comment}
\usepackage{ytableau}
\usepackage{cancel} 
\usepackage{xcolor}
\usepackage{amsmath,bm}
\usepackage{mathtools}
\usepackage{float}
\usepackage{graphicx}
\usepackage{amsmath, amssymb}
\usepackage{youngtab}
\usepackage{changepage}
\usepackage{etoolbox}
\usepackage{mathdots}

\usepackage{booktabs}

\usepackage{mathtools}

\usepackage{pdflscape}

\pretocmd{\eqref}{Eq.\ }{}{}

\def\cI{{\cal I}}

\def\cM{{\cal M}}

\def\cN{{\cal N}}

\def\CQ{{\cal Q}}

\def\cS{{\cal S}}

\def\cZ{{\cal Z}}

\def\D{\Delta}

\newcommand{\nn}{\nonumber}
\newcommand{\PFS}[1]{\mathcal{Z}_{#1}}

\newcommand{\mf}[1]{\mathfrak{#1}}

\newcommand{\Kah}{K\"ahler }

\newcommand{\MM}[1]{\mathcal{M}\left(#1\right)}
\newcommand{\Nek}{\mathsf{N}^{\beta}}

\title{From pq-webs to magnetic quivers and back I: \\ dictionary and deformations} 
 
\author[a,b]{Sebastiano Garavaglia}
\author[a,b]{Sara Pasquetti}
\author[a,b]{Palash Singh}
\affiliation[a]{Dipartimento di Fisica, Università di Milano--Bicocca, Piazza della Scienza 3, I-20126 Milano, Italy}
\affiliation[b]{INFN, Sezione di Milano--Bicocca, Piazza della Scienza 3, I-20126 Milano, Italy}
\emailAdd{s.garavaglia18@campus.unimib.it}
\emailAdd{sara.pasquetti@gmail.com}
\emailAdd{palash.singh@unimib.it}

\abstract{This is the first of two companion papers in which we study five-dimensional $\mathcal{N}=1$ theories engineered by $(p,q)$ five-brane webs and their associated three-dimensional $\mathcal{N}=4$ magnetic quivers. We focus on the $N+\mathrm{SU}(N)_0^{k-1}+N$ box quiver theories. We propose an exact parameter dictionary relating the five-dimensional flavour masses and instanton couplings to the Fayet--Iliopoulos parameters of the corresponding strong-coupling magnetic quiver.

We test this dictionary through a broad class of deformations, including mass deformations, weak-coupling limits, and vacuum expectation values. On the five-dimensional side, we implement these flows directly at the level of the superconformal index. The dictionary fixes the corresponding FI deformations and monopole VEVs. We analyse them in the magnetic-quiver squashed-sphere partition function using a new method that tracks Higgsing, decoupled sectors, and infrared theories. Our results promote magnetic quivers to tools for the quantitative study of how the five-dimensional superconformal index responds to deformations. Once the parameter dictionary is known, exact three-dimensional observables can be used to reconstruct the corresponding
five-dimensional physics and to navigate the full deformation network in both directions.}

\begin{document} 

\maketitle

\flushbottom

\section{Introduction}
\label{sec:intro}

Magnetic quivers provide a powerful framework for studying the moduli spaces of theories with eight supercharges in various dimensions
\cite{Benini:2009gi,Cremonesi:2015lsa,Ferlito:2016grh,Ferlito:2017xdq,Cabrera:2018ann,Cabrera:2018jxt,Eckhard:2020jyr}. A magnetic quiver encodes the Higgs branch of a higher-dimensional theory as a three-dimensional Coulomb branch. In general, this quiver description need not have a conventional Lagrangian gauge-theory realisation; non-simply-laced magnetic quivers already arise in systematic classifications \cite{BourgetHananyMiketa2020,BourgetGrimmingerMartoneZafrir2021}.
The magnetic quivers  studied here, however, are ordinary unitary $\mathcal N=4$ gauge theories.

In this paper and its companion \cite{two}, we revisit the relation between magnetic quivers and five-dimensional theories beyond the identification of their moduli spaces, using their supersymmetric observables. We ask how deformations of a five-dimensional theory are reflected in the partition function of its magnetic quiver, and how much information about the five-dimensional flow can be recovered from the three-dimensional description.

On the five-dimensional side, we consider $\mathcal{N}=1$ theories engineered by $(p,q)$ five-brane webs in Type IIB string theory \cite{Aharony:1997bh,Aharony:1997ju}, or equivalently by M-theory on non-compact toric Calabi--Yau threefolds \cite{Katz:1996fh,Leung:1997tw}. Our main observable is the superconformal index, equivalently the $S^4\times S^1$ partition function \cite{Kim:2012gu,Terashima:2012ra}. The superconformal index probes the ultraviolet fixed point while retaining dependence on the flavour masses and instanton counting parameters of a chosen infrared gauge-theory description. For the theories considered here, it can be computed from the refined topological string partition function associated with the web \cite{Iqbal:2012xm}. On the magnetic-quiver side, we study the squashed-sphere partition function $\cZ_{S_b^3}$ \cite{Kapustin:2009kz,Hama:2011ea}, as a function of the three-dimensional mass and Fayet--Iliopoulos (FI) parameters.\footnote{A parallel analysis could be formulated using the three-dimensional supersymmetric index \cite{Imamura:2011su, Kapustin:2011jm}.}

\begin{figure}[h!]
  \centering
  \includegraphics[width=\textwidth]{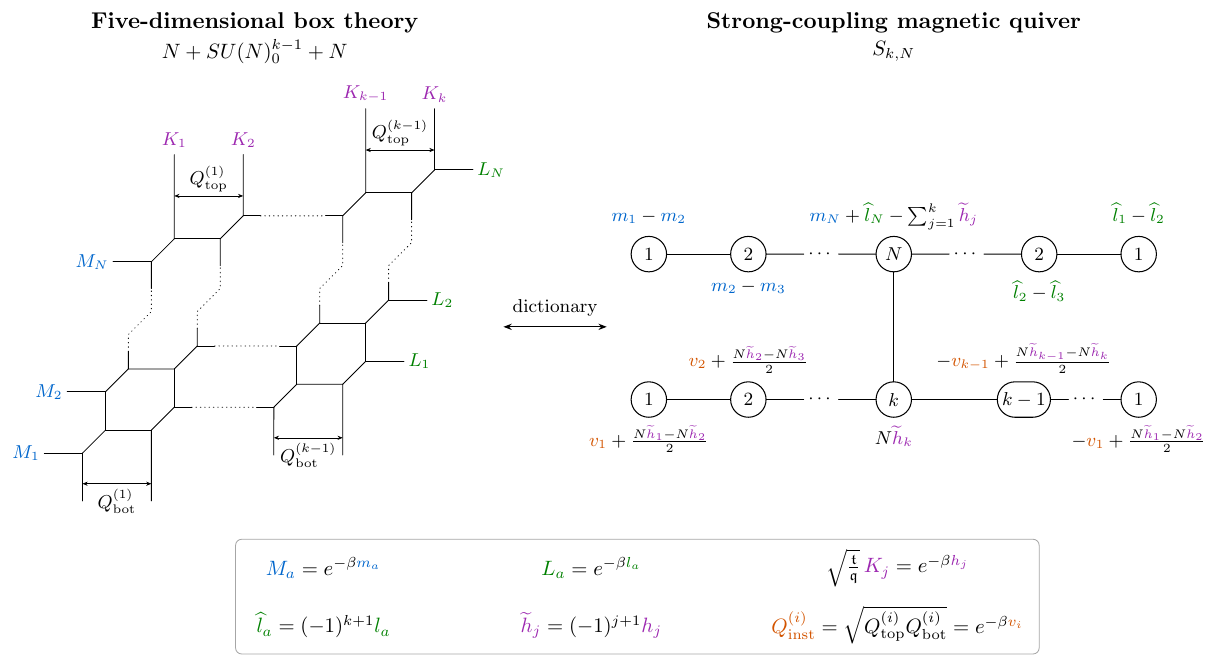}
  \caption{Proposed parameter dictionary between the five-dimensional box theory $N+SU(N)_0^{k-1}+N$ and its strong-coupling magnetic quiver $S_{k,N}$. \textcolor[HTML]{0066CC}{Blue} and \textcolor[HTML]{008000}{green} denote left and right flavour parameters, \textcolor[HTML]{9C27B0}{purple} the abelian strip masses, and \textcolor[HTML]{D35400}{orange} the instanton parameters. The same colours identify each parameter family in additive and multiplicative variables and its contributions to the three-dimensional FI parameters.}
  \label{fig:Intro_dict}
\end{figure}

To compare deformations through these two observables, we need a precise map between their parameters. We focus on the five-dimensional $N+SU(N)_0^{k-1}+N$ gauge theories engineered by rectangular five-brane webs, which we call \emph{box theories}. We propose an exact dictionary relating their independent flavour masses, bifundamental masses, and instanton parameters to the FI parameters of their strong-coupling magnetic quivers $S_{k,N}$, as summarised in Figure \ref{fig:Intro_dict}. In this way, parameters defined using an infrared gauge-theory description acquire definite counterparts in the magnetic quiver associated with the ultraviolet fixed point.

We test our proposed map  by following deformations on both sides. Our direct five-dimensional calculations use an instanton expansion selected by a ruling of the dual toric diagram and the corresponding choice of preferred direction in the refined topological vertex computation. This makes some limits accessible in the perturbative contribution or order by order in instanton number, while other limits require a resummation of the instanton expansion. We therefore focus our direct checks on deformations that can be treated explicitly in both dimensions. 

We also develop an analytic method based on partition functions for studying magnetic-quiver deformations. Earlier work analysed FI flows through vacuum equations and quiver algorithms \cite{vanBeest:2021xyt,Bourget:2023uhe}. Here we implement large FI limits and monopole vacuum expectation values (VEVs) directly in the $\mathcal{N}=2^*$-refined $S_b^3$ localisation integral. Large FI limits produce vacuum-dependent asymptotic factors that must be retained alongside the finite partition functions. Tracking these factors, the surviving background couplings, and the inherited parameters allows us to follow the reduction or splitting of gauge nodes and the appearance of decoupled sectors.

\begin{figure}[b!]
  \centering
  \includegraphics[width=\textwidth]{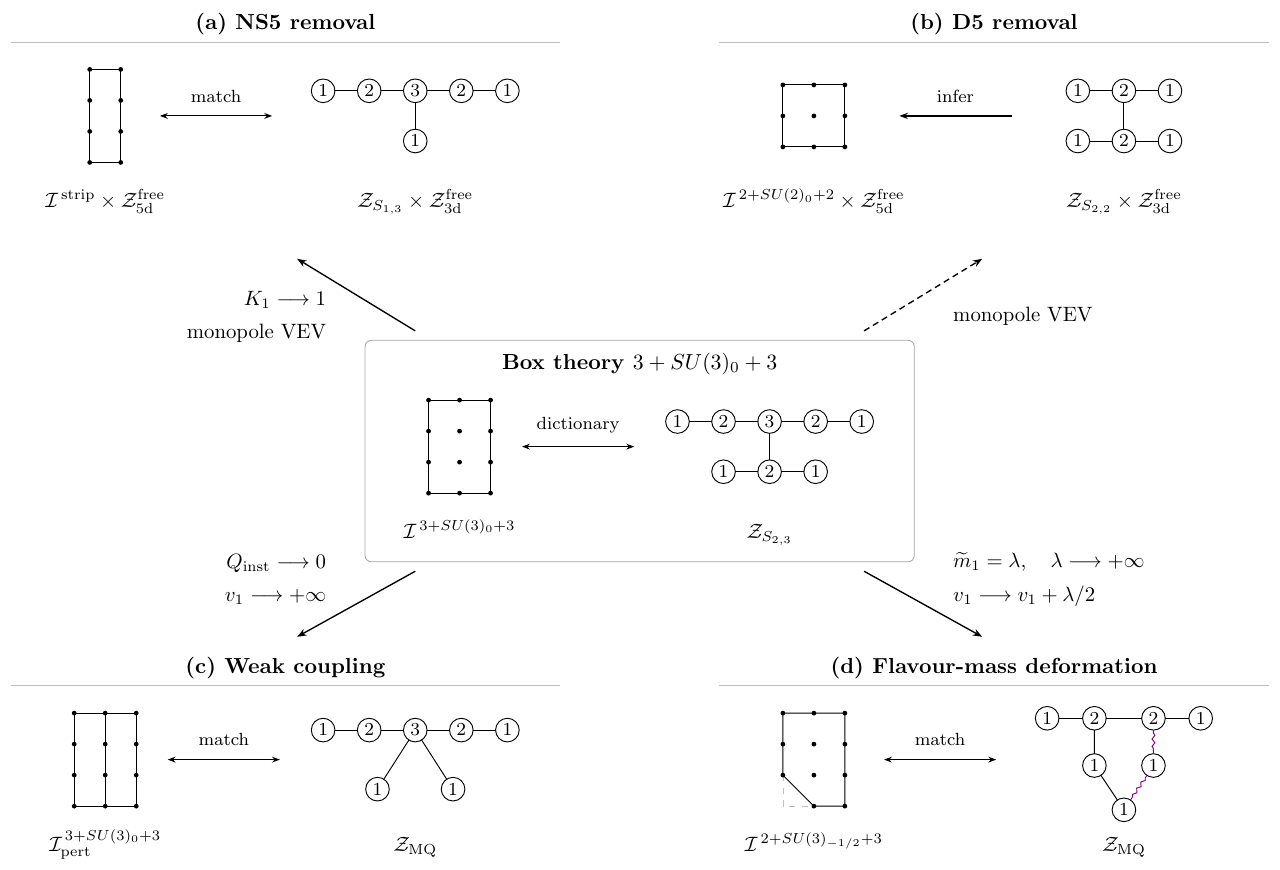}
  \caption{Schematic summary of the classes of deformations studied here, illustrated for the $3+SU(3)_0+3$ box theory  and its magnetic quiver $S_{2,3}$: (a) NS5 removal; (b) D5 removal; (c) a weak-coupling limit; and (d) a flavour-mass deformation. Schematic labels beneath the toric diagrams and magnetic quivers identify the corresponding five-dimensional index and three-dimensional sphere partition function, with free sector contributions displayed explicitly in (a) and (b). Arrows labelled {\it match} compare deformation conditions, inherited parameters, and decoupled sectors. In (b), the arrow labelled {\it infer} indicates the reconstruction of five-dimensional deformation data from a monopole VEV in the magnetic quiver; the dashed branch distinguishes this reverse use of the dictionary. For compactness, the five-dimensional configurations are shown as toric diagrams dual to their $(p,q)$-webs. Wavy links in (d) mark hypermultiplets whose $R$ and axial charge assignments are inherited from the FI flow.}
  \label{fig:Summary}
\end{figure}

We carry out three classes of direct checks. First, removing an external NS5 brane shortens the box. We compute the resulting degeneration of the five-dimensional index and compare it with the monopole-VEV flow of $S_{k,N}$. Second, we take a weak-coupling limit of the box theory and reproduce the corresponding magnetic quiver through the corresponding FI deformation. Third, we integrate out five-dimensional flavours. These limits shift the Chern--Simons level and require an accompanying rescaling of the instanton fugacity. The dictionary maps the combined mass and instanton-parameter flows to correlated FI deformations, whose endpoints we compare with the magnetic quivers of the five-dimensional descendants.

The five-dimensional superconformal index and the three-dimensional sphere partition function are distinct observables, so our tests compare the information carried by their deformations. We find agreement in the deformation conditions and inherited parameters, identify the expected magnetic quivers of the five-dimensional descendants, and match decoupled sectors. The three classes of flows probe different entries of the proposed dictionary and provide complementary evidence that it is inherited under the deformations studied here.

The dictionary also allows us to study five-dimensional flows that are difficult to access directly through the index. For example, the endpoint of removing a D5 brane is evident from the web, but implementing this flow in the index requires at least a partial resummation of the instanton expansion. By analysing the corresponding monopole VEV in the magnetic quiver and applying the dictionary, we determine the five-dimensional VEV condition, reconstruct the relation between the ultraviolet and infrared parameters, and identify the charges of the decoupled sectors. Figure \ref{fig:Summary} summarises the direct checks and this use of the dictionary in the reverse direction.

In the companion paper \cite{two}, we apply the same dictionary to  study the effects of  nilpotent moment-map VEVs. We study deformations for which the five-dimensional index becomes distributional and compare its structure with that of the partition functions of the associated magnetic quivers, which become {\it bad} or {\it broken} under the corresponding VEVs \cite{Giacomelli:2023zkk,Giacomelli:2024laq,Comi:2025zwu}.

A natural next step is a systematic application of the magnetic quiver analyses to five-dimensional deformations that require instanton resummations. Another application would be to investigate flows towards five-dimensional theories without an infrared gauge theory description, such as those discussed in \cite{Eckhard:2020jyr}. Their magnetic quivers could provide a parameterised three-dimensional setting in which to investigate deformations that are difficult to formulate directly in five dimensions. A further open direction is to extend the dictionary beyond box theories and their descendants, even beyond five dimensions.

The paper is organised as follows. In Section \ref{subsec:review_5d}, we review the box five-brane web,  its superconformal index from refined topological strings, describe its strong-coupling magnetic quiver $S_{k,N}$, and propose the 5d/3d parameter dictionary. Section \ref{sec:3dtechnology} develops the sphere-partition-function methods used to implement large FI deformations and monopole VEVs. Readers primarily interested in the applications can proceed to Section \ref{sec:5d3dmap} and return to these technical details as needed. In Section \ref{sec:5d3dmap}, we test the dictionary by removing an NS5 brane, taking weak-coupling limits and integrating out flavours. We also use the magnetic quiver to reconstruct the parameter flow associated with removing a D5 brane.

\section{The $N+SU(N)^{k-1}+N$ box theory and the 5d/3d dictionary}
\label{subsec:review_5d}

In this section, we propose a precise dictionary between the parameters of the five-dimensional box theories and those of their associated three-dimensional magnetic quivers at strong coupling. We focus on five-dimensional $N+SU(N)_0^{k-1}+N$ gauge theories engineered by rectangular $(p,q)$ five-brane webs, whose dual toric diagrams resemble a box.

We begin by reviewing the $S^4\times S^1$ partition function of these theories, constructed from the topological string partition function of the corresponding five-brane web using the refined topological vertex. We then introduce the strong-coupling magnetic quiver $S_{k,N}$ of the box theory and its $S^3$ partition function. Finally, we present the 5d/3d dictionary, which maps the five-dimensional mass and instanton parameters to the Fayet--Iliopoulos parameters of $S_{k,N}$.

\subsection{Box theory:  five-brane web and partition functions}
\label{subsec:5d_partfn}

The $S^4\times S^1$ partition function, or equivalently the 5d superconformal index, of five-dimensional $\cN=1$ SCFTs admitting an infrared Lagrangian description was computed via supersymmetric localisation in \cite{Kim:2012gu,Terashima:2012ra}. The resulting expression is a matrix integral whose integrand contains two copies of the Nekrasov instanton partition function on $\mathbb R^4_{\epsilon_1,\epsilon_2}\times S^1$ \cite{Nekrasov:2002qd,Nekrasov:2003rj}.

The box theories that we consider can be engineered by $(p,q)$ five-brane webs in type IIB string theory \cite{Aharony:1997ju,Aharony:1997bh} or, equivalently, by compactifying M-theory on non-compact Calabi--Yau threefolds described by toric polygons dual to the five-brane webs \cite{Katz:1996fh,Leung:1997tw}. In this geometric engineering perspective, the K-theoretic Nekrasov partition function is identified with the refined topological string partition function of the Calabi--Yau threefold, with the latter being calculable via the refined topological vertex \cite{Aganagic:2003db,Iqbal:2007ii} (also see \cite{Awata:2005fa,Awata:2008ed}).

The associated superconformal index can therefore be schematically written as
\begin{equation}\label{eq:index_gen}
    \cI \equiv \cZ^{S^4\times S^1} = \oint \prod_i \left[ \frac{{\rm d}A_i}{2\pi iA_i} \left| M(\mf t,\mf q) \right|^2 \right] \frac{\left| \cZ_{\rm top} \right|^2}{\left| \cZ_{||} \right|^2} ~.
\end{equation}
Here $A_i$ are the exponentiated Coulomb-branch parameters associated with the infrared gauge theory description, $\cZ_{\rm top}$ denotes the refined topological string partition function, and $M(\mf t,\mf q)$ is the special function defined in \eqref{eq:Mstr_def}. The factor $\cZ_{||}$ collects contributions associated to external parallel legs of the five-brane web. These contributions describe decoupled sectors and must therefore be removed from the physical superconformal index.

Throughout the paper, we denote the exponentiated $\Omega$-deformation parameters of the Nekrasov partition function on $\mathbb R^4_{\epsilon_1,\epsilon_2}\times S^1_{\beta}$ by
\begin{equation}
    \mf q = e^{-\beta\epsilon_1} \;,\;\; \mf t = e^{\beta\epsilon_2} ~,
\end{equation}
where $\beta$ is the radius of the circle. Additionally, the notation $|\cdot|^2$ denotes the standard gluing of the contributions from the two fixed points of $S^4$, rather than the usual complex modulus square, {\it i.e.}, for a generic function $f$, we have
\begin{equation}
    \Big| \, f(U_1,U_2\dots;\mf t,\mf q) \, \Big|^2 = f(U_1,U_2\dots;\mf t,\mf q) \; f(U_1^{-1},U_2^{-1}\dots;\mf t^{-1},\mf q^{-1}) ~.
\end{equation}

\subsubsection{The strip geometry}
\label{subsubsec:strip}

\begin{figure}[]
    \centering
    \includegraphics[width=0.9\linewidth]{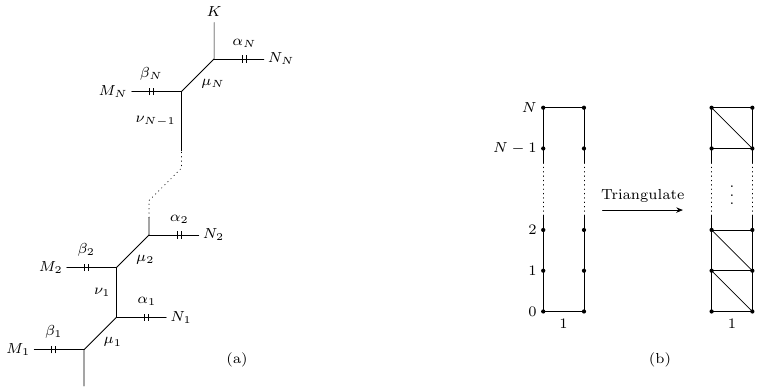}
    \caption{(a) The symmetrical strip of length $N$, engineering free bifundamental hypermultiplets with manifest $SU(N)_{\vec M}\times SU(N)_{\vec N}\times U(1)_K$ global flavour symmetry. The double ticks on the D5 branes mark the horizontal preferred direction  while $\beta_a,\alpha_a$ are  partitions  assigned to the  external legs. The finite internal \Kah parameters are denoted by $\mu_a,\nu_a$. (b) The dual toric diagram and the triangulation corresponding to the resolved five-brane web in panel (a).}
    \label{fig:strip}
\end{figure}

The {\it strip} geometry depicted in Figure \ref{fig:strip}(a) engineers free bifundamental hypermultiplets and serves as the basic building block for the $(p,q)$ five-brane webs of interest. We focus on the symmetrical strip of length $N$ with $N$ D5 branes extending on both sides of an NS5 brane. Its dual toric diagram is a rectangle of height $N$ and unit length, whose triangulation is shown in Figure \ref{fig:strip}(b).

We parameterise the two $SU(N)$ flavour symmetries by the moment map parameters $\vec M, \vec N$, subject to $\prod_{a=1}^N M_a = \prod_{a=1}^N N_a = 1$. Together with $K$ associated to the $U(1)_K$ flavour symmetry, these are related to the internal \Kah parameters through
\begin{equation}\label{eq:strip_par_std}
    \mu_a = \frac{M_aN_a}{K}\sqrt{\frac{\mf q}{\mf t}} \;,\;\; \nu_b = \frac{K}{M_{b+1}N_b}\sqrt{\frac{\mf t}{\mf q}} \;\;,
    \begin{gathered}
        a = 1,\dots,N ~, \\
        \quad\;\; b = 1,\dots,N-1 ~,
    \end{gathered}
\end{equation}
satisfying 
$ \mu_a\nu_a = \frac{M_{a}}{M_{a+1}}$ and $\nu_a\mu_{a+1} = \frac{N_{a+1}}{N_{a}}$. These relations identify the adjacent products of \Kah parameters with the ratios of the non-abelian flavour fugacities. The remaining independent parameter $K$ can be thought of as being an intrinsically internal parameter associated to the strip. This parameterisation is therefore naturally adapted to the interpretation of the strip as engineering free bifundamental hypermultiplets in five dimensions, which we discuss further below as carrying manifest $SU(N)_{\vec M}\times SU(N)_{\vec N}\times U(1)_K$ global symmetry.

Throughout this work, we choose the preferred direction to be horizontal and assign external partitions $\bm\beta=\{\beta_1,\dots,\beta_N\}$ and $\bm\alpha=\{\alpha_1,\dots,\alpha_N\}$ to the horizontal legs. Notably, the sums over internal partitions in the refined topological vertex computation associated to the strip can be performed explicitly, yielding a closed form expression for the strip partition function \cite{Iqbal:2004ne,Taki:2007dh}. We separate this expression into a ``perturbative" and an ``instantonic" part
\begin{equation}
    Z^{\rm strip} (\vec M,\vec N,K;\bm{\beta},\bm{\alpha}) = \cZ_{\rm pert}^{\rm strip} (\vec M,\vec N,K) \; \cZ^{\rm strip}_{\bm{\beta},\bm{\alpha}} (\vec M,\vec N,K) ~,
\end{equation}
where the ``perturbative" part takes the form
\begin{equation}\label{eq:Zstrip_pert}
    \cZ_{\rm pert}^{\rm strip} (\vec M,\vec N,K) = \frac{\prod_{1\leq a<b\leq N} \cM\left( \frac{M_a}{M_b} \right) \cM\left( \frac{N_b}{N_a}\frac{\mf t}{\mf q} \right)}{\prod_{1\leq a<b\leq N }\cM\left( \frac{K}{M_bN_a}\frac{\mf t}{\mf q} \right) \prod_{1\leq a\leq b\leq N }\cM\left( \frac{M_aN_b}{K} \right)} ~,
\end{equation}
while the ``instantonic" part can be expressed as
\begin{equation}\label{eq:Zstrip_inst}
    \begin{split}
        \cZ^{\rm strip}_{\bm{\beta},\bm{\alpha}} (\vec M,\vec N,K) &= \prod_{a=1}^N \Big[ \mf q^{\frac{||\alpha_a||^2}{2}} \mf t^{\frac{||\,{}^t\beta_a||^2}{2}} \widetilde Z_{\alpha_a}(\mf t,\mf q) \widetilde Z_{\,{}^t\beta_a}(\mf q,\mf t) \Big] \\
        &\quad \times \; \frac{\prod_{1\leq a<b\leq N }\cN_{\alpha_a\beta_b}\left( \frac{K}{M_bN_a}\frac{\mf t}{\mf q} \right) \prod_{1\leq a\leq b\leq N }\cN_{\beta_a\alpha_b}\left( \frac{M_aN_b}{K} \right)}{\prod_{1\leq a<b\leq N} \cN_{\alpha_a\alpha_b}\left( \frac{N_b}{N_a}\frac{\mf t}{\mf q} \right) \cN_{\beta_a\beta_b}\left( \frac{M_a}{M_b} \right)} ~.
    \end{split}
\end{equation}
The special functions that appear here are defined in Appendix \ref{app:special_func}; their dependence on $\mf t$, $\mf q$ is suppressed when unambiguous. The terms ``perturbative" and ``instantonic" are used provisionally since the isolated strip engineers a free theory and has no gauge-instanton expansion. However, the terminology still makes sense when these contributions are viewed as the building blocks for configurations that engineer gauge theories where the quotes can truly be dropped.

We now construct the 5d superconformal index by combining two topological string blocks according to the general index formula in \eqref{eq:index_gen}. For an isolated five-brane web, the external partitions must be set to be trivial. When using the web as a building block for a larger configuration, however, we retain arbitrary partitions on the external legs to be glued. The gluing procedure requires a sum over these partitions with appropriate \Kah weights and framing factors as discussed in Appendix \ref{app:topstringPF}. When the resulting configuration engineers a gauge theory, these sums reproduce the Young-diagram sums in the corresponding Nekrasov partition function.

The isolated strip has no compact faces and hence no Coulomb moduli to integrate over. Its decoupled contributions arise from the parallel external D5 legs on either side of the web: the $\beta\beta$ sector on the left and the $\alpha\alpha$ sector on the right. The corresponding contributions are given by
\begin{equation}\label{eq:parallel_std}
    \cZ_{||}^{\rm left} (\vec M) = \prod_{1\leq a<b\leq N} \MM{\frac{M_a}{M_b}} \quad,\quad \cZ_{||}^{\rm right} (\vec N) = \prod_{1\leq a<b\leq N} \MM{\frac{\mf t}{\mf q}\frac{N_b}{N_a}} ~.
\end{equation}
Setting the external partitions to be empty and removing these factors gives
\begin{equation}
     \cI^{\rm strip} (\vec M,\vec N,K) = \left| \frac{ Z^{\rm strip}(\vec M,\vec N,K;\bm\emptyset,\bm\emptyset) }{ \cZ_{||}^{\rm left} (\vec M) \; \cZ_{||}^{\rm right} (\vec N) } \right|^2 = \left| \prod_{a,b=1}^N \cM\left( \frac{M_aN_b}{K} \right) \right|^{-2} ~.
\end{equation}
This is the 5d superconformal index of $N^2$ free hypermultiplets in the $SU(N)\times SU(N)$ bifundamental representation. This has manifest flavour symmetry $SU(N)\times SU(N)\times U(1)$, where the parameter $K$ in \eqref{eq:strip_par_std} can now be realised as the fugacity for this $U(1)$ symmetry.\footnote{The full flavour symmetry of $N^2$ free hypermultiplets is $USp(2N^2)$. This can be seen by redefining $X_{ab}=\frac{M_a N_b}{K}$ and then promoting them to $N^2$ independent parameters corresponding to the Cartan of $USp(2N^2)$.}

Upon gauging the $SU(N)$ flavour symmetry parametrised by $\vec N$, the parameters $\vec M$ and $K$ naturally combine into $U(N)$ flavour fugacities $\widetilde{M}_a=M_a/K$ such that $\prod_{a=1}^N\widetilde{M}_a=K^{-N}$. The hypermultiplet factors then depend on $\widetilde M_aN_b$, as appropriate for $N$ fundamental hypermultiplets of the gauge $SU(N)$.

\subsubsection{The $N+SU(N)_0^{k-1}+N$ box gauge theory}
\label{subsubsec:box}

\begin{figure}[h!]
  \centering
  \includegraphics[width=\textwidth]{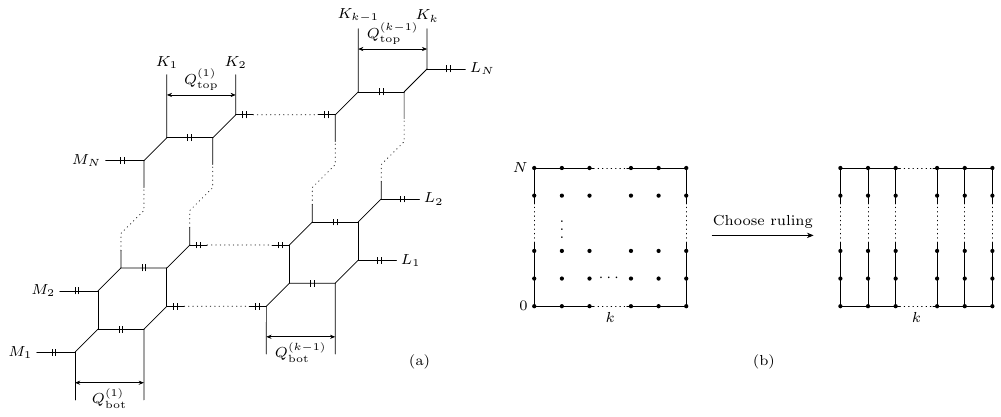}
  \caption{(a) Rectangular five-brane web engineering the $N+SU(N)_0^{k-1}+N$ box theory. The double ticks mark the horizontal preferred direction, and $K_1,\dots,K_k$ label the parameters of the constituent strips. The instanton fugacity of the $i$th gauge node is $Q^{(i)}_{\rm inst}=\sqrt{Q^{(i)}_{\rm bot}Q_{\rm top}^{(i)}}$, where $Q_{\rm bot}^{(i)}$ and $Q_{\rm top}^{(i)}$ are the exponentiated separations of the adjacent NS5 branes at the bottom and the top of the web, respectively. (b) The dual rectangular toric diagram and the vertical ruling defining the strip decomposition.}
  \label{fig:braneweb}
\end{figure}

We can now connect $k$ symmetrical strips of length $N$ to construct the rectangular box five-brane web shown in Figure \ref{fig:braneweb}(a). This configuration engineers the linear quiver gauge theory $N+SU(N)_0^{k-1}+N$, with a bifundamental hypermultiplet between each pair of adjacent gauge nodes and $N$ fundamental hypermultiplets at each end. In this construction, the Chern--Simons level for each gauge group vanishes. The dual toric diagram is a rectangular {\it box} of height $N$ and length $k$; the vertical ruling shown in Figure \ref{fig:braneweb}(b) decomposes it into $k$ constituent strips.

The associated refined topological string partition function is obtained by gluing a sequence of strips according to the prescription reviewed in Appendix \ref{app:topstringPF}. Each glued horizontal leg carries a partition $\alpha_a^{(m)}$ and a base \Kah parameter $Q_{m,a}$, where the gluing entails a sum over these partitions with weights $Q_{m,a}^{|\alpha_a^{(m)}|}$. The resulting expression then takes the form
\begin{equation}\label{eq:Ztop_SUNkm12N}
    \begin{split}
         &Z^{N+SU(N)_0^{k-1}+N}(\vec Q_{\rm inst},\vec M,\vec L,K_1,\dots,K_k; \bm\beta, \bm\gamma) \\
         &\hspace{2cm} = \sum_{\bm{\alpha}^{(1)}}\sum_{\bm{\alpha}^{(2)}}\cdots\sum_{\bm{\alpha}^{(k-1)}} \left[ \prod_{m=1}^{k-1}\prod_{a=1}^N Q_{m,a}^{|\alpha^{(m)}_{a}|} \right] \; Z^{\rm strip} (\vec M,\vec N^{(1)},K_1; \bm{\beta},\bm{\alpha}^{(1)}) \\
         &\hspace{4cm} \times \; \left[ \prod_{m=1}^{k-1} Z^{\rm strip} (\vec N^{(m)},\vec N^{(m+1)},K_{m+1}; \bm{\alpha}^{(m)},\bm{\alpha}^{(m+1)}) \right]' ~,
    \end{split}
\end{equation}
where we define $\bm\alpha^{(k)} \equiv \bm\gamma$ and $\vec N^{(k)} \equiv \vec L$. The prime denotes that the product alternates between the two parameterisations of the strip given in Eqs.\ (\ref{eq:strip_par_std}) and (\ref{eq:strip_par_alt}). The alternate parameterisation of the strip boils down to the following simple definition
\begin{equation}\label{eq:alt_strip}
    \begin{split}
     [Z^{\rm strip}(\vec N, \vec L, K_2; \bm\alpha,\bm\gamma)]' = Z^{\rm strip}(\vec N^{-1}, \vec L^{-1}, \frac{\mf q}{\mf t} K_2^{-1}; \bm\alpha,\bm\gamma) ~.
    \end{split}
\end{equation}
As explained in Appendix \ref{app:topstringPF}, alternating the parameterisation of the strip ensures that the appropriate $SU(N)$ gauge measure is constructed by combining parallel leg contributions.

The refined topological string partition function for the box theory, given in \eqref{eq:Ztop_SUNkm12N}, is equal to the Nekrasov partition function of the $N+SU(N)_0^{k-1}+N$ theory \cite{Nekrasov:2002qd,Nekrasov:2003rj}. We discuss this correspondence in Appendix \ref{app:topstringPF}, to which we relegate most of the technical details. Here, we simply note that the base parameters in each column are directly proportional to the corresponding instanton parameter, $Q_{m,a}\propto Q_{\rm inst}^{(m)}$, where the proportionality factor is expressed in terms of ratios of the remaining parameters and the refinement parameters $\mf t$ and $\mf q$. We therefore use the $k-1$ independent instanton fugacities $\vec Q_{\rm inst}$, rather than all the base parameters, as the argument in \eqref{eq:Ztop_SUNkm12N}. Geometrically, these instanton fugacities are determined by $Q_{\rm inst}^{(i)}=\sqrt{Q_{\rm bot}^{(i)}Q_{\rm top}^{(i)}}$, namely, as the geometric mean of the separations between the NS5-branes at the top and bottom of the five-brane web, see Figure \ref{fig:braneweb}(a). We relegate the presentation of the exact relation between $Q_{i,a}$ and $Q^{(i)}_{\rm inst}$ to Appendix \ref{app:topstringPF}. 

The vertical ruling selects the fibre and base curves appropriate to the $N+SU(N)_0^{k-1}+N$ box gauge theory. We choose the preferred direction of the refined topological vertex to be horizontal, as in the individual strips. The sums over internal partitions within each strip have already been performed, while the remaining sums over the glued horizontal legs give the gauge-instanton expansion of the box theory.

This expression is thus naturally organised around the weakly coupled gauge-theory description.  Nevertheless, information about the enhanced ultraviolet flavour symmetry can also be extracted by organising the superconformal index into symmetry characters order by order \cite{Kim:2012gu, Bergman:2013aca}.

The superconformal index of the $N+SU(N)_0^{k-1}+N$ box theory can then be expressed as
\footnote{ Here $\delta$ denotes the periodic delta function. Its action on a test function is defined via integration over the unit circle:
$$
\oint_{|A|=1}\frac{{\rm d}A}{2\pi i A}\,
\delta\!\left(\frac{A}{A_\star}\right)f(A)
=
f(A_\star)\,.
$$}
\begin{equation}\label{eq:Index_SUNkm12N}
    \begin{split}
        &\cI^{N+SU(N)_0^{k-1}+N} (\vec Q_{\rm inst},\vec M,\vec L, K_1, \dots, K_k) = \oint \prod_{m=1}^{k-1} \left[ \frac{1}{N!} \prod_{i=1}^N\frac{{\rm d}N_i^{(m)}}{2\pi iN_i^{(m)}} \times \delta\left( \prod_{i=1}^N N_i^{(m)} \right) \right] \\
        &\quad \times \left| M(\mf t,\mf q) \right|^{2(k-1)(N-1)} \times \left| \frac{\cZ_{\rm pert}^{\rm strip} (\vec M, \vec N^{(1)}, K_1) \left[ \prod_{m=1}^{k-1} \cZ^{\rm strip}_{\rm pert} (\vec N^{(m)}, \vec N^{(m+1)}, K_{m+1}) \right]' }{ \cZ_{||}^{k,N} (\vec M, \vec L, \vec Q_{\rm bot}, \vec Q_{\rm top}) } \right|^2  \\
        &\quad \times \left| \sum_{\bm{\alpha}^{(1)}}\cdots\sum_{\bm{\alpha}^{(k-1)}} \; \prod_{m=1}^{k-1}\prod_{a=1}^N Q_{m,a}^{|\alpha^{(m)}_{a}|} \cZ^{\rm strip}_{\bm\emptyset,\bm\alpha^{(1)}} (\vec M, \vec N^{(1)}, K_1) \left[ \prod_{m=1}^{k-1} \cZ^{\rm strip}_{\bm\alpha^{(m)},\bm\alpha^{(m+1)}} (\vec N^{(m)}, \vec N^{(m+1)}, K_{m+1}) \right]' \right|^2 ~,
    \end{split}
\end{equation}
where once again the primes denote alternating parameterisations of the strip. Since the superconformal index is computed from the five-brane web in isolation, we need to set the external partitions to be trivial $\bm\alpha^{(k)}  = \bm\emptyset$.
The associated decoupled contribution can then be expressed as
\begin{equation}\label{eq:box_dec}
    \cZ_{||}^{k,N} (\vec M, \vec L, \vec Q_{\rm bot}, \vec Q_{\rm top}) = \cZ_{||}^{\rm NS5}( \vec Q_{\rm bot}, \vec Q_{\rm top} ) \times
    \begin{cases}
        \cZ_{||}^{\rm left}(\vec M) \; \cZ_{||}^{\rm right}(\vec L) \qquad\,, \text{ for } k \text{ odd}, \\
        \cZ_{||}^{\rm left}(\vec M) \; \cZ_{||}^{\rm alt \; right}(\vec L) \quad, \text{ for } k \text{ even}.
    \end{cases}
\end{equation}
where
\begin{equation}\label{eq:parallel_alt}
    \cZ_{||}^{\rm left} (\vec M) = \prod_{1\leq a<b\leq N} \MM{\frac{M_a}{M_b}} \quad,\quad
    \begin{aligned}
        \cZ_{||}^{\rm  right} (\vec L) &= \prod_{1\leq a<b\leq N} \MM{\frac{\mf t}{\mf q}\frac{L_b}{L_a}} ~, \\
        \cZ_{||}^{\rm alt \; right} (\vec L) &= \prod_{1\leq a<b\leq N} \MM{\frac{\mf t}{\mf q}\frac{L_a}{L_b}} ~,
    \end{aligned}
\end{equation}
are the parallel leg contributions from the external D5 branes. The parity dependence of the D5 contribution arises solely from the parameterisation of the final strip: the first strip always uses the standard parameterisation, whereas the final strip uses the alternate parameterisation when $k$ is even. Meanwhile the factor $\cZ_{||}^{\rm NS5}$ captures the contribution from the parallel external NS5 branes and takes the form
\begin{equation}
    \cZ_{||}^{\rm NS5}( \vec Q_{\rm bot}, \vec Q_{\rm top} ) = \prod_{1\leq i<j\leq k} \left[ \MM{\prod_{l=i}^{j-1} Q_{\rm top}^{(l)}} \MM{\frac{\mf t}{\mf q} \prod_{l=i}^{j-1} Q_{\rm bot}^{(l)}} \right] ~,
\end{equation}
where $Q_{\rm bot}^{(i)}$ and $Q_{\rm top}^{(i)}$ denote the exponentiated separations of adjacent NS5 branes on the bottom and top, respectively, as shown in Figure \ref{fig:braneweb}. As we will show in Appendix \ref{app:topstringPF}, these can be expressed in terms of the strip and base \Kah parameters. There we also show that each base parameter $Q_{i,a}$ is directly proportional to the corresponding instanton fugacity $Q_{\rm inst}^{(i)}$ with a proportionality factor determined by the strip parameters.

\subsection{Magnetic quiver of the box}
\label{subsec:boxMQ}

\begin{figure}[h!]
  \centering
  \includegraphics[width=0.5\textwidth]{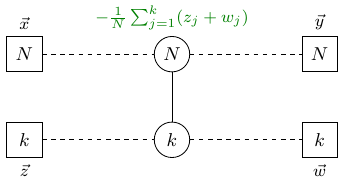}
  \caption{The three-dimensional strong-coupling magnetic quiver $S_{k,N}$ of the five-dimensional $N+SU(N)_0^{k-1}+N$ box theory. We use the dashed-line compact notation for the $T[U(N)]$ and $T[U(k)]$ tails. Accordingly, the FI parameters are specified as the vector labels while the parameter associated to the $U(k)$ node is $z_k+w_k$, explained in Appendix \ref{app:3dstuff}. The overall gauge $U(1)$ is understood to be decoupled.}
  \label{fig:SkN}
\end{figure}

We now introduce the other main character of this work: the strong-coupling magnetic quiver associated to the $N+SU(N)_0^{k-1}+N$ box theory. As explained in \cite{Cabrera:2018jxt}, this quiver can be obtained from the five-brane web consisting of $N$ D5 branes and $k$ NS5 branes. We denote the resulting three-dimensional theory, depicted in Figure \ref{fig:SkN}, by $S_{k,N}$.

The quiver consists of two $T[U(N)]$ tails attached to the common $U(N)$ gauge node and two $T[U(k)]$ tails attached to the common $U(k)$ gauge node. The two central nodes are connected by a bifundamental hypermultiplet. In Figure \ref{fig:SkN}, the tails are represented using the compact dashed-line notation; the definition and further details on their parameterisation are provided in Appendix \ref{app:3dstuff}.

The Coulomb branch global symmetry on $S_{k,N}$ is at least $SU(N)^2\times SU(k)^2\times U(1)$ with further enhancement at $k=2$ or $N=2$ and an even larger enhancement for $k=N=2$. The Cartan subalgebra of this global symmetry has dimension $2(N-1)+2(k-1)+1=2N+2k-3$. Since all matter fields are bifundamentals, the overall diagonal $U(1)$ gauge symmetry acts trivially on the matter and must therefore be decoupled. This sector is associated with a monopole operator that has a vanishing $R$ charge under the ultraviolet assignment.

To manifest this decoupling at the level of the squashed sphere partition function, we can shift all gauge Cartan variables by\footnote{
Constraining $\widetilde u_{a,\alpha}$ to sum to zero is a convenient choice, 
though any fixed value for the sum would suffice. After the redefinition, we 
must preserve the number of independent gauge Cartans; therefore, introducing 
the decoupled $U(1)$ fugacity $U$ requires imposing a single constraint on the 
new variables $\widetilde u_{a,\alpha}$.}
\begin{equation}
    u_{a,\alpha} = \widetilde u_{a,\alpha} + U \;\;\text{ such that } \; \sum_{a,\alpha} \widetilde u_{a,\alpha} = 0 ~,
\end{equation}
where $a$ labels the gauge node and $\alpha$ labels the Cartan component at each node. Under this change of variables, the FI terms in the partition function produce a factor
\begin{equation}
    e^{2\pi i\,UC} \;,\; \text{ where } \; C = \sum_a \zeta_a\,{\rm rk}(G_a) ~,
\end{equation}
where $\zeta_a$ is the FI parameter associated to the gauge group $G_a$. Integrating over the decoupled $U(1)_U$, including the contribution of the adjoint singlet, gives
\begin{equation}
    s_b\left( 2m-\frac{iQ}{2} \right) \int {\rm d}U \,e^{2\pi i\,UC} = \delta(C)\,s_b\left( 2m-\frac{iQ}{2} \right) ~.
\end{equation}
We choose the  FI parameters of the theory satisfying the $C=0$ condition and decouple the divergent factor
\begin{equation}
    \delta(0) \, s_b\left( 2m-\frac{iQ}{2} \right) ~,
\end{equation}
to isolate the interacting part of $S_{k,N}$. Throughout this work, magnetic quivers are understood to have this diagonal sector removed. In the cases where the magnetic quiver reduces to a Wess--Zumino theory, however, we explicitly strip off the divergent and adjoint-singlet factor.

Finally, we choose to parameterise the two $T[U(k)]$ tails off-shell, using unconstrained $U(k)$ parameters $\vec w$ and $\vec z$. The BF couplings included in the compact tail notation induce the FI parameter $w_k+z_k$ on the central $U(k)$ node, as explained in Appendix \ref{app:3dstuff}. For the two $T[U(N)]$ tails, we instead use the on-shell parameterisation in terms of the $SU(N)$ parameters $\vec x$ and $\vec y$ satisfying $\sum_{i=1}^N x_i = \sum_{i=1}^N y_i = 0$. The FI parameter for the central $U(N)$ node is then fixed to be
\begin{equation}
    -\frac{1}{N}\sum_{j=1}^k(z_j+w_j) \nn
\end{equation}
by the condition $C=0$, which sets the FI parameter of the diagonal $U(1)$ to zero. Following Appendix \ref{app:3dstuff} and these conventions, the $S^3_b$ partition function can then be expressed as
\begin{equation}
    \begin{split}
         \PFS{S_{k,N}}&(\vec{x},\vec{y},\vec{z},\vec{w};m) = \int {\rm d}\vec{t}_k \, \D^{\rm HB}_k(\vec{t};m) \int {\rm d}\vec{u}_N \, \D^{\rm HB}_N(\vec{u};m) \, e^{-2\pi i\frac{1}{N} \sum_{j=1}^k(w_j+z_j)\sum_{i=1}^N u_i} \\
         &\qquad \times \; \mathcal{Z}^{T[U(N)]}(\vec{u},\vec{x};m) \, \mathcal{Z}^{T[U(N)]}(\vec{u},\vec{y};m) \, \mathcal{Z}^{T[U(k)]} (\vec{t},\vec{w};m) \, \mathcal{Z}^{T[U(k)]}(\vec{t},\vec{ z};m) \\
         &\qquad\quad \times \; \prod_{i=1}^N\prod_{j=1}^k s_b\left( \frac{i Q}{2}-m\pm (u_i-t_j) \right) ~.
    \end{split}
\end{equation}

\subsection{The 5d/3d dictionary}
\label{subsec:dictionary}

\begin{figure}[h!]
  \centering
  \includegraphics[width=\textwidth]{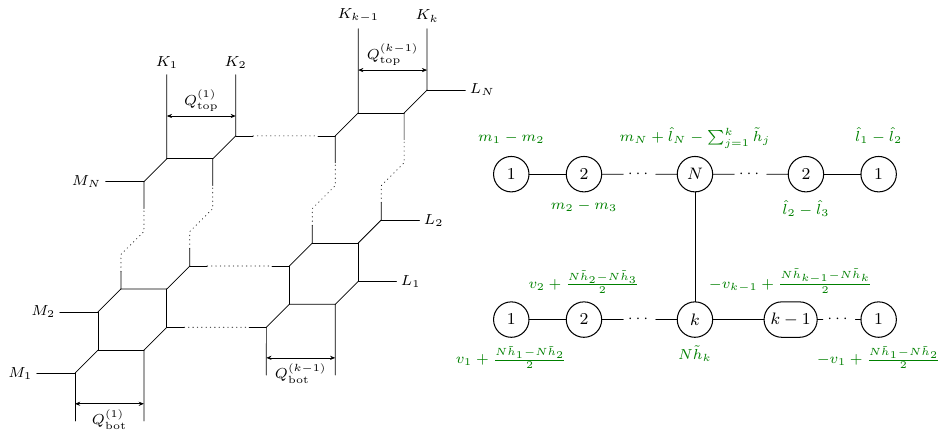}
  \caption{The rectangular five-brane web that engineers the $N+SU(N)_0^{k-1}+N$ box theory and its strong-coupling magnetic quiver $S_{k,N}$, with parameters related by the proposed 5d/3d dictionary. The five-dimensional flavour masses map to the FI parameters of the two $T[U(N)]$ tails, while the abelian strip masses and instanton parameters determine the FI parameters of the remaining nodes. Recall that the instanton fugacities can be expressed as $Q^{(i)}_{\rm inst}=\sqrt{Q_{\rm bot}^{(i)}Q_{\rm top}^{(i)}}$.}
  \label{fig:Dictionary}
\end{figure}

We now propose a 5d/3d dictionary between the mass and instanton parameters of the five-dimensional $N+SU(N)_0^{k-1}+N$ box theory and the FI parameters of its strong-coupling magnetic quiver $S_{k,N}$. We illustrate this in Figure \ref{fig:Dictionary}.

We introduce additive parameters for the two sets of $SU(N)$ masses
\begin{equation}
    M_i = e^{-\beta m_i} \;,\;\; L_i = e^{-\beta l_i} \;\;,\;\;\; i=1,\dots,N ~,
\end{equation}
with $\sum_im_i=\sum_il_i=0$. Similarly, the instanton and $U(1)$ bifundamental masses
\begin{equation}
    Q_j = e^{-\beta v_j} \;,\;\; j=1,\dots,k-1 \;\;,\quad \sqrt{\frac{\mf t}{\mf q}}K_j = \sqrt{\frac{\mf t}{\mf q}} e^{-\beta k_j} = e^{-\beta h_j} \;,\;\; j=1,\dots,k ~.
\end{equation}
For the dictionary below, we identify
\begin{equation}
    \sqrt{\frac{\mf t}{\mf q}} = e^{-\beta\left(\frac{iQ}{4}\right)} \;\Longrightarrow\;\sqrt{\frac{\mf t}{\mf q}}K_j = e^{-\beta\left(\frac{iQ}{4}+k_j\right)} = e^{-\beta h_j} ~.
\end{equation}
The map is then given by
\begin{equation}\label{eq:dict}
    \boxed{
    \begin{gathered}
        \vec x = \vec m \;\;,\;\;\; \vec y = \vec{\widehat l} ~, \\
        \vec w = \left( \frac{N\widetilde h_1}{2}+\sum_{i=1}^{k-1}v_i \,,\; \frac{N\widetilde h_2}{2}+\sum_{i=2}^{k-1}v_i \,,\; \dots \;, \frac{N\widetilde h_{k-1}}{2}+v_{k-1} \,,\; \frac{N\widetilde h_k}{2} \right) ~, \\
        \vec z = \left( \frac{N\widetilde h_1}{2}-\sum_{i=1}^{k-1}v_i \,,\; \frac{N\widetilde h_2}{2}-\sum_{i=2}^{k-1}v_i \,,\; \dots \;, \frac{N\widetilde h_{k-1}}{2}-v_{k-1} \,,\; \frac{N\widetilde h_k}{2} \right) ~,
    \end{gathered}}
\end{equation}
where we have defined
\begin{equation}\label{eq:hl_redef}
    \widetilde h_i \coloneq
    \begin{cases}
        -h_i \;&,\; i \text{ even} \\
        h_i \;&,\; i \text{ odd}
    \end{cases} \quad,\quad \widehat l_i = (-1)^{k+1} l_i ~.
\end{equation}

The signs in the dictionary reflect the alternate strip parameterisation used in the five-dimensional construction of the topological string partition function of $N+SU(N)_0^{k-1}+N$ in Subsection \ref{subsec:5d_partfn}. For a strip in the $i$th position, the replacements for even $i$ specified in \eqref{eq:alt_strip} lead to 
\begin{equation}
    \sqrt{\frac{\mf t}{\mf q}} K_i \longrightarrow \sqrt{\frac{\mf q}{\mf t}} K_i^{-1} \quad,\quad\; L_i \longrightarrow L_i^{-1} ~,
\end{equation}
implying the definitions in \eqref{eq:hl_redef}.

As shown in the first line of the map in \eqref{eq:dict}, the 5d $SU(N)^2$ flavour symmetry parameters map directly onto the parameters of the two $T[U(N)]$ tails. The instanton parameters $v_i$ and the abelian masses $h_i$ instead combine into the two enhanced $SU(k)$ moment map parameters associated with the two $T[U(k)]$ tails $\vec w$ and $\vec z$. More explicitly the sums $w_i+z_i=N\widetilde h_i$ isolate the abelian masses associated with the $i$th strip, whereas the differences $w_i-z_i=2\sum_{j=i}^{k-1}v_j$ contain cumulative sums of the instantonic parameters. Taking consecutive differences gives the FI parameters along the two $T[U(k)]$ tails
\begin{equation}
    w_i-w_{i+1} = v_i+\frac{N}{2}\left(\widetilde h_i-\widetilde h_{i+1}\right) \quad,\quad\; z_i-z_{i+1} = -v_i+\frac{N}{2}\left(\widetilde h_i-\widetilde h_{i+1}\right)
\end{equation}
with instantonic parameters entering the two $T[U(k)]$ tails with opposite signs. The FI parameter of the central $U(k)$ node is $z_k+w_k=N\widetilde h_k$, while that of the central $U(N)$ node is fixed by decoupling the overall $U(1)$ and is equal to
\begin{equation}
    -\frac{1}{N}\sum_{j=1}^k(z_j+w_j) = -\sum_{j=1}^k \widetilde h_j ~. \nn
\end{equation}

As a preliminary consistency check of the proposal, it can be easily checked that the number of parameters in both the 5d theory and the 3d $S_{k,N}$ theory is identical and equal to $2N+2k-3$. Moreover, the map in \eqref{eq:dict} is invertible
\begin{equation}
    \widetilde h_i = \frac{w_i+z_i}{N} \quad,\quad\; v_i = \frac12\left(w_i-w_{i+1}+z_{i+1}-z_i\right) ~,
\end{equation}
so it identifies all independent parameters.

In Section \ref{sec:5d3dmap}, we subject the 5d/3d dictionary in \eqref{eq:dict} to a series of consistency checks. First, we will study the deformation associated with removing the $i$th NS5 brane, thereby shortening the box theory according to
\begin{equation}
    N+SU(N)_0^{k-1}+N \quad\longrightarrow\quad N+SU(N)_0^{k-2}+N ~. \nn
\end{equation}
This deformation involves tuning the parameter $K_i$ and therefore tests the identification of $\widetilde h_i$ on the magnetic quiver side. As we will see, the corresponding tuning condition is mapped to a monopole VEV that shortens both $T[U(k)]$ tails to $T[U(k-1)]$ tails.

Second, we will consider weak-coupling limits, which test the identification and relative signs of the instanton parameters $v_i$. Under the dictionary, these limits become large FI deformations that induce flows to the expected weak-coupling magnetic quivers.

Finally, we will study flavour mass deformations, which probe non-trivial combinations of flavour masses, abelian strip masses, and instanton parameters. Integrating a flavour out in five dimensions requires a simultaneous rescaling of the instanton fugacity. The dictionary translates this rescaling into an accompanying shift of the associated FI parameter. The resulting flow therefore tests the endpoint magnetic quiver as well as the direction of the deformations and the vacuum selected along the flow.

These three classes of deformations probe distinct and independent sectors of the proposed dictionary. As we will discuss in greater detail, we restrict the analysis to deformations that can be implemented exactly on both of the five- and three-dimensional sides, thereby allowing for a direct comparison of the corresponding supersymmetric observables.

Although formulated for the five-dimensional box theory and its strong-coupling magnetic quiver, the resulting 5d/3d parameter map is inherited by all descendant theories obtained through these Higgsings, mass deformations, and weak-coupling limits. A single ultraviolet dictionary therefore controls the full family of deformations generated by the box construction.

Before carrying out these tests, we develop the tools needed to implement large FI deformations and monopole VEVs at the level of the squashed sphere partition function. This is the subject of the next section.

\section{Magnetic quiver toolbox: $S_b^3$ technology}
\label{sec:3dtechnology}

In this section, we review and develop the partition function techniques needed to implement the deformations that will appear in this work. We focus on large FI limits and monopole VEVs in magnetic quivers, describing their implementation at the level of the sphere partition function. Through the 5d/3d dictionary in \eqref{eq:dict}, these correspond to mass and gauge coupling deformations, as well as operator VEVs, in the corresponding five-dimensional theories.

\subsection{FI deformations with $S^3_b$ partition function}
\label{subsec:FIfromPF}

The first class of deformations relevant to this work arises from large real mass deformations in the five-dimensional theory. Through the proposed parameter dictionary in \eqref{eq:dict}, these map to large FI deformations of the magnetic quiver. Such deformations have previously been studied by using vacuum equations and quiver-subtraction techniques to determine the Higgsing pattern and the magnetic quiver describing the endpoint of the flow \cite{vanBeest:2021xyt,Bourget:2023uhe}.

Here we develop a complementary approach that implements the FI deformations directly at the level of the $S_b^3$ partition function. Besides reproducing the endpoint quiver, this method determines the infrared parametrisation and tracks the divergent asymptotic phases and background contact terms generated along the flow. It also captures multiplicities associated with degenerate vacua and, when several such vacua contribute to the limit, expresses the result as a sum of their infrared partition functions. The proposed 5d/3d dictionary specifies the FI deformation associated with a chosen five-dimensional limit, allowing us to follow the resulting flow within the localisation computation.

A key feature of real mass deformations of supersymmetric partition functions is the appearance of divergent asymptotic factors multiplying the finite partition function of the infrared theory reached at the end of the renormalization-group flow \cite{Benini:2011mf,Aharony:2013dha}. These phases are sensitive to both the deformation and the precise way the limit is implemented. For instance, shifting the Coulomb branch integration variables together with the real masses can retain different light degrees of freedom and produce different asymptotic factors. From this perspective, the divergent phases act as labels of the different supersymmetric vacua. Consequently, when deforming a dual pair, the matching of these phases provides a powerful criterion for identifying the corresponding infrared dual descriptions.

In the remainder of this section we illustrate this approach with two representative examples, which provide the basic building blocks for all subsequent applications.

\subsubsection{Warm up: FI deformations in the $\mathcal{N}=4$ SQED mirror pair}
\label{subsubsec:warmup}

We begin by studying a real mass deformation across an abelian mirror dual pair from the point of view of the sphere partition function.

Consider the SQED mirror pair on the left in Figure \ref{fig:FI_mirror}. The electric theory is the $U(1)$ SQED with three flavours, whose sphere partition function in the $\cN=2^*$ conventions of Appendix \ref{app:3dstuff} is given by
\begin{equation}
    \PFS{\rm Electric}(\vec x,\zeta) = \int {\rm d}z\,e^{2\pi iz\zeta} \prod_{j=1}^3 \Big[ s_b(m\pm(z-x_j)) \Big] s_b\left(-2m+\frac{iQ}{2}\right) ~.
\end{equation}
The partition function of the magnetic theory is
\begin{equation}
    \begin{split}
        \PFS{\rm Magnetic}(\vec x,\zeta) &= e^{2\pi i \zeta x_3} \int {\rm d}z_1 {\rm d}z_2 \, e^{2\pi i z_1 (x_1 - x_2)} \, e^{2\pi i z_2 (x_2 - x_3)} \times s_b\left(\frac{iQ}{2}-m \pm z_1\right) \\
    &\quad \times\; s_b\left(\frac{iQ}{2}-m \pm (z_1 - z_2)\right) s_b\left(\frac{iQ}{2}-m \pm (z_2 - \zeta)\right)  s_b\left(2m - \frac{iQ}{2}\right)^2
    \end{split}
\end{equation}
These theories are mirror dual and hence we have $\PFS{\rm Electric}(\vec x,\zeta)=\PFS{\rm Magnetic}(\vec x,\zeta)$. Crucially, the mixed background Chern-Simons term $e^{2\pi i\zeta x_3}$ is part of the proper duality statement (see \cite{Closset:2012vp} for a detailed discussion of Chern-Simons contact terms)  and will be essential for matching the asymptotic prefactors generated by the deformations.

\subsubsection*{Real mass deformation of the electric theory}

The first deformation we study is a real (vector-like) mass deformation for one flavour in the SQED implemented by setting $x_1=\lambda$ and then sending $\lambda\to+\infty$. Using the double sine asymptotics
\begin{equation}\label{eq:sb_asym}
    s_b(x+\xi) \sim e^{\pm\frac{i\pi}{2}(x+\xi)^2} \;,\;\; x\to\pm\infty ~,
\end{equation}
we find
\begin{equation}\label{eq:electricmass}
    \begin{split}
        &\lim_{\lambda\to+\infty} \PFS{\rm Electric}(\lambda,x_2,x_3,\zeta) \\
        &\qquad\quad = \lim_{\lambda\to+\infty} e^{2\pi i\lambda m} \int {\rm d}z\,e^{2\pi iz(\zeta-m)} \prod_{j=2}^3 \Big[ s_b(m\pm(z-x_j)) \Big] \,s_b\left( -2m+\frac{iQ}{2} \right)
    \end{split}
\end{equation}
After extracting the asymptotic prefactor $e^{2\pi i\lambda m}$, the remaining expression is the partition function of $U(1)$ SQED with two flavours. The asymptotic factor will be important when matching the correct deformation on the mirror side. Indeed it is clear on the electric side itself that the flavour mass deformation we are implementing is not unique. For instance, shifting the gauge parameter $z\to z+\lambda$ gives the final theory $U(1)$ SQED with one flavour and a different asymptotic factor.

\begin{figure}[t!]
  \centering
  \includegraphics[width=\textwidth]{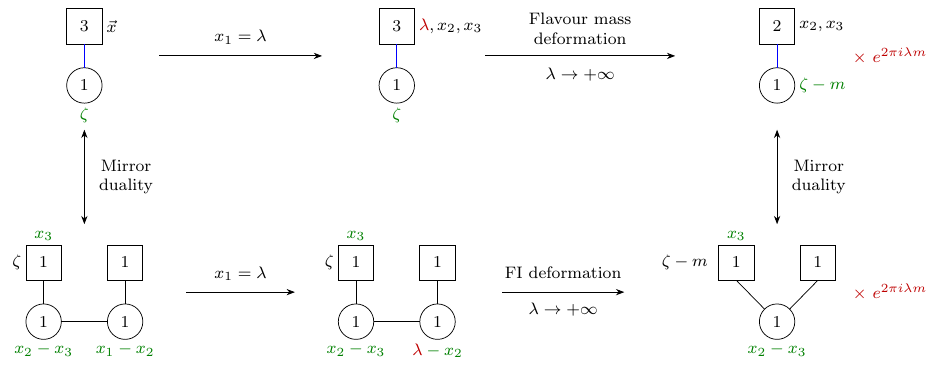}
  \caption{A large real mass deformation of $U(1)$ SQED with three flavours and the corresponding large FI deformation of its mirror dual. Taking $x_1=\lambda\to+\infty$ yields the self-mirror dual pair of $U(1)$ SQED with two flavours. The matching asymptotic prefactors identify the corresponding vacuum on each side.}
  \label{fig:FI_mirror}
\end{figure}

We now turn to the magnetic side, where the parameters $\vec x$ enter through the FI coupling and hence the deformation $x_1=\lambda$ and $\lambda\to+\infty$ corresponds to a large FI deformation. The partition function of the magnetic theory in this limit behaves as
\begin{equation}\label{eq:magnetiFI_pre}
    \begin{split}
        &\lim_{\lambda \to +\infty} \PFS{\rm Magnetic}(x_1=\lambda,x_2,x_3,\zeta) \\
        &\qquad = \lim_{\lambda \to +\infty} \int {\rm d}z_1 {\rm d}z_2 \, e^{2\pi i z_1 (\lambda - x_2)}  \, e^{2\pi i z_2 (x_2 - x_3)} \, e^{2\pi i \zeta x_3} \times s_b\left(\frac{iQ}{2}-m \pm z_1\right) \\
        &\qquad\quad \times \;  s_b\left(\frac{iQ}{2}-m \pm (z_1 - z_2)\right) s_b\left(\frac{iQ}{2}-m \pm (z_2 - \zeta)\right)  s_b\left(2m - \frac{iQ}{2}\right)^2 ~.
    \end{split}
\end{equation}
Here, the $\lambda$ dependent factor $e^{2\pi i\lambda z_1}$ depends on one of the integration variables and hence cannot be factored out of the integral.

To bypass this problem, we derive a useful identity starting from the following partition function identity relating the SQED with one flavour to its $XYZ$ dual:
\begin{equation}
    \int {\rm d}x  e^{-2\pi i x \xi} \, s_b\left(\frac{iQ}{4}+\frac{m}{2}\pm x\right) = s_b(m)s_b\left(\frac{i Q}{4}-\frac{m}{2}\pm\xi\right) ~.
\end{equation}
Shifting $x\to x+s$ and applying the double sine asymptotics as $s\to+\infty$ from \eqref{eq:sb_asym} gives the formal relation
\begin{equation}
    \lim_{s\to\infty} e^{2\pi is\left(\frac{iQ}{4}+\frac{m}{2}-\xi\right)} \int {\rm d}x\,e^{-2\pi ix\left(-\frac{iQ}{4}-\frac{m}{2}+\xi\right)} = \lim_{s\to\infty} s_b(m)s_b\left(\frac{iQ}{4}-\frac{m}{2}\pm\xi\right) ~.
\end{equation}
Recognising the Fourier representation of the delta function, we get the master formula for implementing FI deformations
\begin{equation}\label{eq:masterFIdef}
    \boxed{\lim_{s\to\infty} e^{-2\pi is\left(\frac{iQ}{4}+\frac{m}{2}-\xi\right)} s_b(m)s_b\left(\frac{iQ}{4}-\frac{m}{2}\pm\xi\right) = \delta\left(\xi-\frac{iQ}{4}-\frac{m}{2}\right)} ~.
\end{equation}
Note that \eqref{eq:masterFIdef} should be understood as a distributional identity in $\xi$: when inserted into an integral over $\xi$, the LHS acts as the Dirac delta function on the RHS. In what follows, we use \eqref{eq:masterFIdef} exclusively in such a distributional sense.\footnote{The analysis presented here extends in a straightforward manner to the SQED with one flavour/$XYZ$ duality at the level of the $S^2\times S^1$ index and more generally to supersymmetric partition functions, leading to the corresponding analogue of \eqref{eq:masterFIdef}.}

Coming back to the large FI deformation on the magnetic theory in \eqref{eq:magnetiFI_pre}, we multiply and divide the integrand by a suitable phase factor highlighted in \textcolor{blue}{blue}
\begin{equation}
    \begin{split}
        &\lim_{\lambda \to +\infty} \PFS{\text{Magnetic}}(x_1=\lambda,x_2,x_3,\zeta) \\
        &\qquad = \lim_{\lambda \to +\infty} \int {\rm d}z_1 {\rm d}z_2 \, e^{2\pi i z_1 (\lambda - x_2)} \textcolor{blue}{e^{2\pi i \lambda \left(-m+m\right)}} e^{2\pi i z_2 (x_2 - x_3)} e^{2\pi i \zeta x_3} \, s_b\left(\frac{iQ}{2}-m \pm z_1\right) \\
        &\qquad\quad \times \; s_b\left(\frac{iQ}{2}-m \pm (z_1 - z_2)\right) s_b\left(\frac{iQ}{2}-m \pm (z_2 - \zeta)\right) s_b\left(2m - \frac{iQ}{2}\right)^2 ~.
    \end{split}
\end{equation}
Substituting $m\to2m-\frac{iQ}{2}$ in the master formula gives
\begin{equation}\label{eq:tunedmaster}
    \lim_{\lambda \to +\infty} e^{2\pi i \lambda \left(z_1 -m\right)} s_b\left(2m - \frac{iQ}{2}\right) s_b\left(\frac{iQ}{2}-m \pm z_1\right) = \delta\left(z_1 - m\right) ~,
\end{equation}
which can then be used  to perform the $z_1$ integral giving
\begin{equation}\label{eq:magneticFI}
    \begin{split}
        &\lim_{\lambda \to +\infty} \PFS{\text{Magnetic}}(x_1=\lambda,x_2,x_3,\zeta) = \lim_{\lambda \to +\infty} e^{2\pi i\lambda m} e^{-2\pi ix_2 m} \int {\rm d}z_2 \, e^{2\pi i z_2 (x_2 - x_3)} e^{2\pi i \zeta x_3} \\
        &\hspace{2.5cm} \times \; s_b\left(\frac{iQ}{2}-m \pm (m - z_2)\right) s_b\left(\frac{iQ}{2}-m \pm (z_2 - \zeta)\right) s_b\left(2m - \frac{iQ}{2}\right) ~.
    \end{split}
\end{equation}
The remaining integral is the partition function of $U(1)$ SQED with two flavours. Moreover, the asymptotic factor exactly agrees with that of the electric theory in \eqref{eq:electricmass}. Removing this common factor leaves an equality between the finite partition functions \eqref{eq:electricmass} and \eqref{eq:magneticFI}, expressing the self-duality of the $U(1)$ SQED with two flavours, as illustrated in Figure \ref{fig:FI_mirror}.

\subsubsection*{Large FI deformation of the electric theory}

\begin{figure}[t!]
  \centering
  \includegraphics[width=\textwidth]{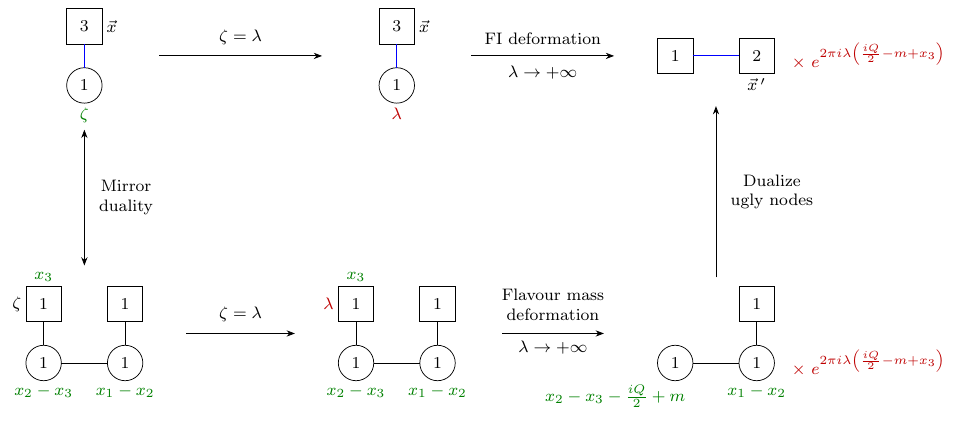}
  \caption{A large FI deformation of $U(1)$ SQED with three flavours and the corresponding large real mass deformation of its mirror dual, implemented by taking $\zeta=\lambda\to+\infty$. In the vacuum shown, the electric theory flows to two free hypermultiplets, while the magnetic theory reaches the same endpoint after successively dualising its ugly nodes. The matching asymptotic prefactors are displayed in red.}
  \label{fig:Mass_mirror}
\end{figure}

We now return to the original mirror pair and consider the large FI limit $\zeta=\lambda\to+\infty$ shown in Figure \ref{fig:Mass_mirror}. In this limit the electric partition function becomes
\begin{equation}
    \begin{split}
        \lim_{\lambda \to +\infty} \PFS{\text{Electric}}(\vec{x}, \zeta=\lambda) &= \lim_{\lambda \to +\infty} \int {\rm d}z \, e^{2\pi iz\lambda} \prod_{j=1}^3 \Big[ s_b\left(m \pm (z - x_j)\right) \Big] s_b\left(-2m + \frac{iQ}{2}\right) ~. 
    \end{split}
\end{equation}
As before the rapidly oscillating exponential depends on the gauge integration variables. To be able to apply the master formula \eqref{eq:masterFIdef} we multiply and divide by the phase indicated in \textcolor{blue}{blue}:
\begin{equation}\label{eq:electric_blue}
    \begin{split}
        \lim_{\lambda \to +\infty} \PFS{\text{Electric}}(\vec{x}, \zeta=\lambda) &= \lim_{\lambda \to +\infty} \int {\rm d}z \, e^{2\pi iz\lambda} \prod_{j=1}^3 \Big[ s_b\left(m \pm (z - x_j)\right) \Big] s_b\left(-2m + \frac{iQ}{2}\right) \\
        &\hspace{3cm} \times \; \textcolor{blue}{e^{2\pi i\lambda \left(\frac{iQ}{2}-m -\frac{iQ}{2}+ m + x_3 - x_3\right)}} ~.
    \end{split}
\end{equation}
Substituting $m\to-2m+\frac{iQ}{2}$ in the master formula \eqref{eq:masterFIdef} then gives
\begin{equation}
    \lim_{\lambda \to +\infty} e^{2\pi i \lambda \left(z - \frac{iQ}{2}+m - x_3\right)} s_b\left(-2m + \frac{iQ}{2}\right) s_b\left(m \pm (z - x_3)\right) = \delta\left(z - x_3 - \frac{iQ}{2}+m\right) ~.
\end{equation}
Using this identity to perform the integral over $z$ then yields the contribution associated with this choice of vacuum
\begin{equation}\label{eq:electricFI}
    \lim_{\lambda \to +\infty} \PFS{\text{Electric}}(\vec{x}, \zeta=\lambda) = \lim_{\lambda \to +\infty} e^{2\pi i \lambda \left(\frac{iQ}{2}-m + x_3\right)} \prod_{j=1}^2 \left[ s_b\left(m \pm (\frac{iQ}{2}-m + x_3 - x_j)\right) \right] ~.
\end{equation}
The finite part is the partition function of two free hypermultiplets while the accompanying asymptotic prefactor must be matched on the magnetic side to identify the corresponding vacuum.

On the magnetic side, $\zeta$ is now a flavour parameter and hence the limit $\zeta=\lambda\to+\infty$ is a large real mass deformation that integrates out one flavour. Using the double-sine asymptotics in \eqref{eq:sb_asym} we obtain
\begin{equation}\label{eq:magneticmasses_pre}
    \begin{split}
        &\lim_{\lambda \to +\infty} \PFS{\text{Magnetic}}(\vec{x}, \zeta=\lambda) = \lim_{\lambda \to +\infty} e^{2\pi i \lambda \left(x_3 +\frac{iQ}{2}- m\right)} \int {\rm d}z_1 {\rm d}z_2 \, e^{2\pi i z_1 (x_1 - x_2)} e^{2\pi i z_2 \left(x_2 - x_3 -\frac{iQ}{2}+ m\right)} \\
        &\hspace{4cm} \times \; s_b\left(\frac{iQ}{2}-m \pm z_1\right) s_b\left(\frac{iQ}{2}-m \pm (z_1 - z_2)\right) s_b\left(2m - \frac{iQ}{2}\right)^2 ~.
    \end{split}
\end{equation}

The asymptotic factor precisely agrees with the electric one in \eqref{eq:electricFI}.\footnote{Choosing the fugacity corresponding to the first or second flavour $x_1,x_2$ instead of $x_3$ in \eqref{eq:electric_blue} would require examining the corresponding shifted asymptotic region of the magnetic integral.} On the other hand, the finite part is the partition function of a $U(1)\times U(1)$ gauge theory with one fundamental hypermultiplet and one bifundamental hypermultiplet, shown in the bottom right corner of Figure \ref{fig:Mass_mirror}. However this theory is ugly since the first $U(1)$ node, associated to the gauge parameter $z_2$, is underbalanced. We can then sequentially confine the two $U(1)$ nodes using the mirror duality between SQED with one flavour and a free (twisted) hypermultiplet to obtain two free multiplets. At the level of the partition function, this follows by applying the following confining duality twice in \eqref{eq:magneticmasses_pre}:
\begin{equation}
    \int {\rm d}z \, s_b\left(\frac{iQ}{2}-m \pm (z - x)\right) s_b\left(2m-\frac{iQ}{2}\right)e^{2\pi i z \eta} = s_b\left(m \pm \eta\right) e^{2\pi i x \eta} ~,
\end{equation}
to yield
\begin{equation}\label{eq:magneticmasses}
    \lim_{\lambda \to +\infty} \PFS{\text{Magnetic}}(\vec{x}, \zeta=\lambda) = \lim_{\lambda \to +\infty} e^{2\pi i\lambda\left(x_3 +\frac{iQ}{2}- m\right)} \prod_{j=1}^2s_b\left(m \pm (x_j - x_3 - \frac{iQ}{2}+m)\right) \,,
\end{equation}
precisely matching the electric counterpart, including the asymptotic prefactor, in \eqref{eq:electricFI}.

\subsubsection{More examples of FI deformations}
\label{sec:anotherfidef}

In this section we consider more intricate examples of FI deformations that illustrate features that we will encounter in later sections. In particular, we show how an FI deformation can cause a magnetic quiver to split into a sum of magnetic quivers. The appearance of multiple magnetic quivers under an FI deformation has already been observed in \cite{vanBeest:2021xyt} through an analysis of the equations of motion. Here, we show that implementing these deformations directly at the level of the partition function can produce several contributions with the same asymptotic prefactors
which therefore survive together in the
large FI limit. Their finite parts need not be identical: they may
describe distinct infrared theories, or equivalent contributions
related by a symmetry. We determine which occurs in each example and
sum the surviving contributions with their appropriate multiplicities.

\subsection*{Example 1 :}

\begin{figure}[t!]
  \centering
  \includegraphics[width=\textwidth]{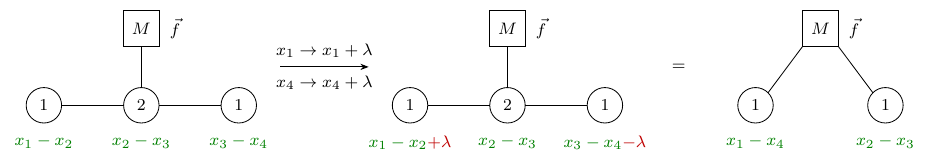}
  \caption{Large FI deformation of the $U(1)\times U(2)\times U(1)$ quiver with $M$ fundamental hypermultiplets at the central node, implemented by $x_1\to x_1+\lambda$ and $x_4\to x_4+\lambda$, with $\lambda\to+\infty$. The resulting infrared theory consists of two decoupled $U(1)$ gauge theories, with $M$ flavours and respective FI parameters $x_1-x_4$ and $x_2-x_3$.}
  \label{fig:FI_ex1}
\end{figure}

We begin with an FI deformation that will be directly relevant to Section \ref{subsubsec:rule_3d}. Consider the quiver on the left of Figure \ref{fig:FI_ex1} whose partition function can be expressed as
\begin{equation}
    \begin{split}
        \cZ &= \frac12\int {\rm d}z_1 {\rm d}z_2^{(1)} {\rm d}z_2^{(2)} {\rm d}z_3 \; \Delta_2^{\rm HB}(\vec{z}_2; m) \; \times e^{2\pi i z_1 (x_1 - x_2)} e^{2\pi i (z_2^{(1)} + z_2^{(2)})(x_2 - x_3)} e^{2\pi i z_3 (x_3 - x_4)} \\
        &\qquad \; \times s_b\left(2m - \frac{iQ}{2}\right)^2 \prod_{j=1}^2 \prod_{k=1}^M \left[ s_b(\frac{iQ}{2}-m \pm (z_2^{(j)} - f_k)) \right] \\
        &\qquad\quad \times \; \prod_{j=1}^2 \left[ s_b(\frac{iQ}{2}-m \pm (z_1 - z_2^{(j)})) s_b(\frac{iQ}{2}-m \pm (z_3 - z_2^{(j)})) \right] 
    \end{split}
\end{equation}
We consider the deformation  implemented by shifting $x_1\to x_1+\lambda$ and $x_4\to x_4+\lambda$ and then taking $\lambda\to+\infty$. The FI parameters on the two outer nodes acquire opposite shifts while the central node remains unchanged. In preparation for applying the master identity \eqref{eq:masterFIdef}, we insert the phase factor, highlighted in \textcolor{blue}{blue}, and write the deformed partition function as
\begin{equation}\label{eq:FIdef_pre}
    \begin{split}
        \lim_{\lambda \to +\infty}\PFS{} &= \lim_{\lambda \to +\infty} \frac12\int {\rm d}z_1 {\rm d}z_2^{(1)} {\rm d}z_2^{(2)} {\rm d}z_3 \; \Delta_2^{\rm HB}(\vec{z}_2; m) \; s_b\left(2m - \frac{iQ}{2}\right)^2 \\
        &\times \; e^{2\pi i z_1 (\lambda+x_1 - x_2)} e^{2\pi i (z_2^{(1)} + z_2^{(2)})(x_2 - x_3)} e^{2\pi i z_3 (x_3-x_4 - \lambda)} \times \textcolor{blue}{e^{2\pi i \lambda \left(z_2^{(\ell)}  -z_2^{(\ell)}+2m-2m\right)} } \\
        &\times \; \prod_{j=1}^2 \left[ s_b(\frac{iQ}{2}-m \pm (z_1 - z_2^{(j)})) s_b(\frac{iQ}{2}-m \pm (z_3 - z_2^{(j)})) \right] \\
        &\times \; \prod_{j=1}^2 \prod_{k=1}^M \left[ s_b(\frac{iQ}{2}-m \pm (z_2^{(j)} - f_k)) \right] ~.
    \end{split}
\end{equation}
Here $\ell=1,2$ specifies which of the two Cartans of the central $U(2)$ node enters this phase factor. In either case, we obtain the following delta functions
\begin{equation}\label{eq:FIex1_delta}
    \begin{split}
        &\lim_{\lambda \to +\infty} e^{2\pi i \lambda \left(z_1 - z_2^{(\ell)} - m \right)} e^{2\pi i \lambda \left(z_2^{(\ell)} - z_3 - m \right)} \times s_b\left(\frac{iQ}{2}-m \pm (z_1 - z_2^{(\ell)})\right) \\
        &\quad \times \; s_b\left(\frac{iQ}{2}-m \pm (z_2^{(\ell)} - z_3)\right)  s_b\left(2m - \frac{iQ}{2}\right)^2 = \delta\left(z_1 - z_2^{(\ell)} - m\right) \delta\left(z_2^{(\ell)} - z_3 - m\right)\,.
    \end{split}
\end{equation}
understood in the same distributional sense as the master identity \eqref{eq:masterFIdef}.

Let us first consider the $\ell=1$ case. Integrating over $z_1$ and $z_3$ imposes the delta conditions leading to the contributions of the $\cN=2$ adjoint chiral and vector multiplets of the $U(2)$ node simplifying to that of a $U(1)\times U(1)$ node:
\begin{equation}
    \begin{split}
        &\left. s_b\left(\frac{iQ}{2}-m \pm (z_1 - z_2^{(2)})\right) s_b\left(\frac{iQ}{2}-m \pm (z_2^{(2)} - z_3)\right) \Delta_2^{\rm HB}(\vec{z}_2; m) \right|_{\substack{z_1=z_2^{(1)}+m\\z_3=z_2^{(1)}-m}} \\
        &\hspace{6cm} = s_b\left(2m-\frac{iQ}{2}\right)^2 ~.
    \end{split}
\end{equation}

Therefore the off-diagonal contributions cancel, leaving the determinants associated with the two $U(1)$ gauge nodes. Consequently, the contribution associated with $\ell=1$ can be expressed as
\begin{equation}\label{eq:FIex1}
    \begin{split}
      & \frac{1}{2}\lim_{\lambda \to +\infty} e^{4\pi i \lambda m} e^{2\pi i m \left( x_1-x_2-x_3+x_4 \right)} \int {\rm d}z_2^{(1)} {\rm d}z_2^{(2)} \, s_b\left(2m - \frac{iQ}{2}\right)^2 \\
        &\quad \times \; e^{2\pi i z_2^{(1)} (x_1 - x_4)}e^{2\pi i z_2^{(2)} (x_2 - x_3)} \prod_{j=1}^2 \prod_{k=1}^M \left[ s_b(\frac{iQ}{2}-m \pm (z_2^{(j)} - f_k)) \right] ~.
    \end{split}
\end{equation}
Repeating the calculation with $\ell=2$ in \eqref{eq:FIex1_delta} gives the same expression with $z_2^{(1)}$ and $z_2^{(2)}$ exchanged. A simple re-labelling of the integration variables produces an identical copy of the $U(1)\times U(1)$ theory in \eqref{eq:FIex1}.

In particular, the second contribution has exactly the same asymptotic prefactor phase as the first one and hence we add the two contributions. This multiplicity then cancels the Weyl factor of the original $U(2)$ gauge group converting it into the product of two $U(1)$ gauge groups. The final expression then takes the form:
\begin{equation}
    \begin{split}
        \lim_{\lambda \to +\infty} \cZ &= \lim_{\lambda \to +\infty} e^{4\pi i \lambda m} e^{2\pi i m \left( x_1-x_2-x_3+x_4 \right)} \int {\rm d}z_2^{(1)} {\rm d}z_2^{(2)} \, s_b\left(2m - \frac{iQ}{2}\right)^2 \\
        &\quad \times \; e^{2\pi i z_2^{(1)} (x_1 - x_4)}e^{2\pi i z_2^{(2)} (x_2 - x_3)} \prod_{j=1}^2 \prod_{k=1}^M \left[ s_b(\frac{iQ}{2}-m \pm (z_2^{(j)} - f_k)) \right] ~.
    \end{split}
\end{equation}
As illustrated in Figure \ref{fig:FI_ex1}, the infrared theory obtained by the RG flow triggered by this FI deformation is the product of two $U(1)$ theories with $M$ flavours.\footnote{The same deformation can also be studied as a large real mass deformation of the mirror theory, the linear quiver $2+U(2)^{M-1}+2$.} 
The FI parameter of one resulting $U(1)$ node is the sum of the FI parameters of the three original gauge nodes, while the other inherits the FI parameter of the original $U(2)$ node.

\subsection*{Example 2 :}

\begin{figure}[h!]
  \centering
  \includegraphics[width=\textwidth]{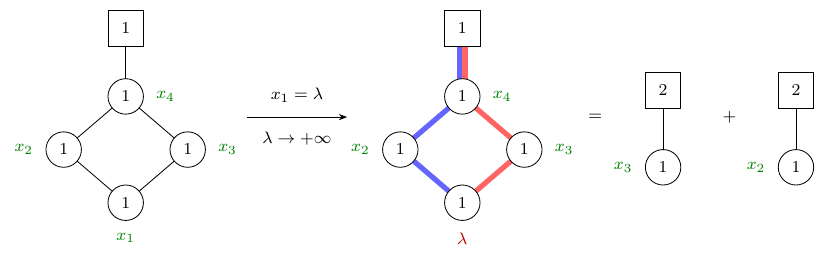}
  \caption{Large FI deformation of an abelian quiver containing a loop, implemented by $x_1=\lambda\to+\infty$. The blue and red paths indicate the sequences of delta function constraints that leave the nodes with FI parameters $x_3$ and $x_2$, respectively. Each contribution describes a $U(1)$ SQCD with two flavours, and their partition functions enter as a sum with an identical asymptotic prefactor.}
  \label{fig:FI_ex2}
\end{figure}

We show another example of the FI deformation implemented at the level of the sphere partition function in full detail. In this case, we consider a quiver gauge theory that contains a loop and demonstrate that after the deformation the resulting partition function takes the form of a sum of two contributions with a common asymptotic prefactor. Concretely, we study the theory shown on the left of Figure \ref{fig:FI_ex2} whose partition function is given by
\begin{equation}
    \begin{split}
        \cZ (x_1, x_2, x_3, x_4) &= \int \prod_{j=1}^4 {\rm d}z_j \times e^{2\pi i\sum_{j=1}^4 x_jz_j} \times s_b\left(2m - \frac{iQ}{2}\right)^4 s_b\left(\frac{iQ}{2} - m \pm z_4\right) \\
        &\qquad \times \; \prod_{j=2}^3 \left[ s_b\left(\frac{iQ}{2} - m \pm (z_1 - z_j)\right) s_b\left(\frac{iQ}{2} - m \pm (z_4 - z_j)\right) \right] ~,
    \end{split}
\end{equation}
where $z_1,z_2,z_3,z_4$ are the Cartan variables of the four $U(1)$ gauge nodes, with respective FI parameters $x_1,x_2,x_3,x_4$.

We implement the large FI deformation by setting $x_1=\lambda$ and then sending $\lambda\to+\infty$. In preparation for applying the master identity \eqref{eq:masterFIdef}, we multiply and divide by suitable phase factors highlighted in \textcolor{blue}{blue} and \textcolor{red}{red}:\footnote{
The two coloured factors in \eqref{eq:FIdef2_pre} are both identically one and display two alternative ways of applying the master identity. We follow one path at a time; the constraints associated with the blue and red paths are not imposed simultaneously.}
\begin{equation}\label{eq:FIdef2_pre}
    \begin{split}
        \cZ (x_1, x_2, x_3, x_4) &= \int \prod_{j=1}^4 {\rm d}z_j \times e^{2\pi i\lambda z_1} \, e^{2\pi i\sum_{j=2}^4 x_jz_j} \times s_b\left(2m - \frac{iQ}{2}\right)^4 s_b\left(\frac{iQ}{2} - m \pm z_4\right) \\
        &\qquad \times \; \prod_{j=2}^3 \left[ s_b\left(\frac{iQ}{2} - m \pm (z_1 - z_j)\right) s_b\left(\frac{iQ}{2} - m \pm (z_4 - z_j)\right) \right] \\
        &\qquad\quad \times \; \textcolor{blue}{e^{2\pi i \lambda (z_2 - z_2 + z_4 - z_4 + 3m - 3m)}} \textcolor{red}{e^{2\pi i \lambda (z_3 - z_3 + z_4 - z_4 + 3m - 3m)}} ~.
    \end{split}
\end{equation}
The \textcolor{blue}{blue} choice uses the bifundamentals connecting nodes 1,2, and 4, followed by the fundamental at node 4, while the \textcolor{red}{red} choice instead passes through node 3 instead of node 2. These two paths are illustrated in Figure \ref{fig:FI_ex2} and we apply the master identity \eqref{eq:masterFIdef} along each path separately.

With the \textcolor{blue}{blue} path, using the master identity \eqref{eq:masterFIdef} we obtain
\begin{equation}
    \begin{split}
        \lim_{\lambda \to +\infty} &e^{2\pi i \lambda [(z_1 - z_2) + (z_2 - z_4) + z_4 - 3m]} \times s_b\left(\frac{iQ}{2} - m \pm (z_1 - z_2)\right) s_b\left(\frac{iQ}{2} - m \pm (z_2 - z_4)\right) \\
        &\times \; s_b\left(\frac{iQ}{2} - m \pm z_4\right) s_b\left(2m - \frac{iQ}{2}\right)^3 = \delta(z_1 - z_2 - m) \delta(z_2 - z_4 - m) \delta(z_4 - m) ~,
    \end{split}
\end{equation}
where the constraints do not fix  $z_3$. We can then integrate over $z_1,z_2,z_4$ in \eqref{eq:FIdef2_pre} implementing the constraints obtained from the above delta functions to find
\begin{equation}\label{eq:FIdef2_first}
    \begin{split}
        & \lim_{\lambda \to +\infty} e^{6\pi i \lambda m}\, e^{2\pi i m (2x_2 + x_4)} \int {\rm d}z_3\, e^{2\pi i x_3 z_3} s_b\left(2m - \frac{iQ}{2}\right) \\
        &\quad \times \; s_b\left(\frac{iQ}{2} - m \pm (3m - z_3)\right) s_b\left(\frac{iQ}{2} - m \pm (m - z_3)\right)\ ~.
    \end{split}
\end{equation}
The integral corresponds to the partition function of $U(1)$ SQED with two flavours and FI parameter $x_3$.

Repeating the same procedure for the \textcolor{red}{red} path in Figure \ref{fig:FI_ex2} gives the following delta functions
\begin{equation}
    \delta(z_1-z_3-m) \delta(z_3-z_4-m) \delta(z_4-m) ~,
\end{equation}
leaving $z_2$ unconstrained. Performing the $z_1,z_3,z_4$ integrations yields
\begin{equation}\label{eq:FIdef2_second}
    \begin{split}
        &\lim_{\lambda \to +\infty} e^{6\pi i \lambda m}\, e^{2\pi i m (2x_3 + x_4)} \int {\rm d}z_2\, e^{2\pi i x_2 z_2} s_b\left(2m - \frac{iQ}{2}\right) \\
        &\quad \times \; s_b\left(\frac{iQ}{2} - m \pm (3m - z_2)\right) s_b\left(\frac{iQ}{2} - m \pm (m - z_2)\right)\ ~,
    \end{split}
\end{equation}
which describes another copy of $U(1)$ SQED with two flavours, now with FI parameter $x_2$.

Note that both the contributions in Eqs.\ (\ref{eq:FIdef2_first}) and (\ref{eq:FIdef2_second}) carry the same $\lambda$-dependent asymptotic prefactor $e^{6\pi i\lambda m}$. Therefore these correspond to degenerate vacua and are hence summed to obtain the resulting infrared partition function as the following sum
\begin{equation}
    \begin{split}
        \lim_{\lambda \to +\infty} &Z(x_1 = \lambda, x_2, x_3, x_4) = \lim_{\lambda \to +\infty} e^{6\pi i \lambda m} \left[ e^{2\pi i m (2x_3 + x_4)} \int {\rm d}z_2\, e^{2\pi i x_2 z_2} \times s_b\left(2m - \frac{iQ}{2}\right) \right. \\
        &\hspace{3.5cm} \times \;  s_b\left(\frac{iQ}{2} - m \pm (3m - z_2)\right) s_b\left(\frac{iQ}{2} - m \pm (m - z_2)\right) \\
        &\hspace{2.5cm} + \; e^{2\pi i m (2x_2 + x_4)} \int {\rm d}z_3\, e^{2\pi i x_3 z_3} s_b\left(2m - \frac{iQ}{2}\right) \\
        &\hspace{3cm} \left. \times \; s_b\left(\frac{iQ}{2} - m \pm (3m - z_3)\right) s_b\left(\frac{iQ}{2} - m \pm (m - z_3)\right) \right]\ ~.
    \end{split}
\end{equation}
Thus, the effect of this FI deformation is to decompose the original theory into the sum of the theories associated to the \textcolor{blue}{blue} and \textcolor{red}{red} paths illustrated in Figure \ref{fig:FI_ex2}.

The contributions in Eqs.~(\ref{eq:FIdef2_first}) and
(\ref{eq:FIdef2_second}) are not the only formal outcomes of repeated
applications of \eqref{eq:masterFIdef}. Longer sequences traversing all
the gauge nodes can produce terms with the same leading asymptotic factor
$e^{6\pi i\lambda m}$. Their constraints include those of the \textcolor{blue}{blue} or \textcolor{red}{red} path, together with an additional condition that localises the
remaining $z_3$ or $z_2$ integral.
For example, consider the constraints
\begin{equation}
\delta(z_1-z_2-m)\,
\delta(z_2-z_4-m)\,
\delta(z_4-z_3-m)\,
\delta(z_3).
\end{equation}
The last two set $z_4=m$ and $z_3=0$. Thus the blue constraints are
recovered, with an additional condition on its residual SQED integral.
We do not count such terms as separate frames here.

The procedure illustrated in these examples is reminiscent of the Jeffrey--Kirwan residue prescription. In both cases, the FI chamber selects a set of admissible localisation contributions such that the final result is obtained by summing over such contributions. In our approach, these contributions are identified through successive applications of the distributional identity \eqref{eq:masterFIdef} which imposes delta function constraints along admissible paths in the quiver. We do not, however, evaluate the original integral via residues. Instead the large FI limit directly localises the relevant gauge integrations while retaining the associated asymptotic prefactor, contact terms, and multiplicities.

\subsection{Monopole VEVs}
\label{sec:monopolevevs}

The second class of deformations that we consider consists of turning on VEVs for  operators in the five-dimensional theory. Through the proposed parameter dictionary in \eqref{eq:dict}, these deformations are mapped to monopole operator VEVs in the corresponding magnetic quivers. Here, we focus on the class of VEVs associated with removing NS5 or D5 branes from the five-brane web of the box theory. More general nilpotent VEVs for the moment map operators will instead be studied in detail in the companion paper \cite{two}. As in the case of the FI deformations discussed in the previous subsection, our goal is to implement these monopole VEVs directly at the level of the $S^3_b$ partition function. We describe the relevant procedure below.

The direct analysis of a flow triggered by a monopole VEV is generally a difficult problem. However, several complementary approaches provide useful alternatives. One useful strategy is to employ three-dimensional mirror symmetry under which monopole operators are mapped to mesonic operators, whose VEVs can be studied using standard Higgsing techniques. Crucially, three-dimensional mirror symmetry can also be applied locally within a three-dimensional theory \cite{Hwang:2021ulb,Comi:2022aqo}. This then allows us to study monopole VEVs even when the complete theory does not admit a Lagrangian mirror description, as will be the case for most of the strong-coupling magnetic quivers considered here. More precisely, one starts by temporarily ungauging the appropriate node(s) to isolate a subquiver containing the entire support of the monopole operator whose VEV is to be turned on. One then considers the mirror dual of this subquiver, studies the corresponding mesonic VEV, and finally glues the deformed quiver back into the original theory.\footnote{An example of this procedure is the flow illustrated in Figure \ref{fig:boxNS5rem}.}

In the mirror description, the mesonic VEV can be analysed using standard techniques, such as the equations of motion, or more directly at the level of supersymmetric partition functions; we follow the latter approach. The problem then reduces to studying the effect of tuning the mass parameters to the values dictated by the VEV condition, namely those for which the corresponding operator has vanishing total charge. This tuning modifies the analytic structure of the partition function: poles move in the complex plane until they pinch the integration contour. The deformed partition function is then obtained by evaluating the residues associated with these pinching poles \cite{Gaiotto:2012xa}. We illustrate this mechanism below with an explicit example that will serve as a building block for the monopole VEV triggered flows in Section \ref{3dremNs}.

\subsubsection*{A monopole VEV flow}

\begin{figure}[h!]
  \centering
  \includegraphics[width=0.9\textwidth]{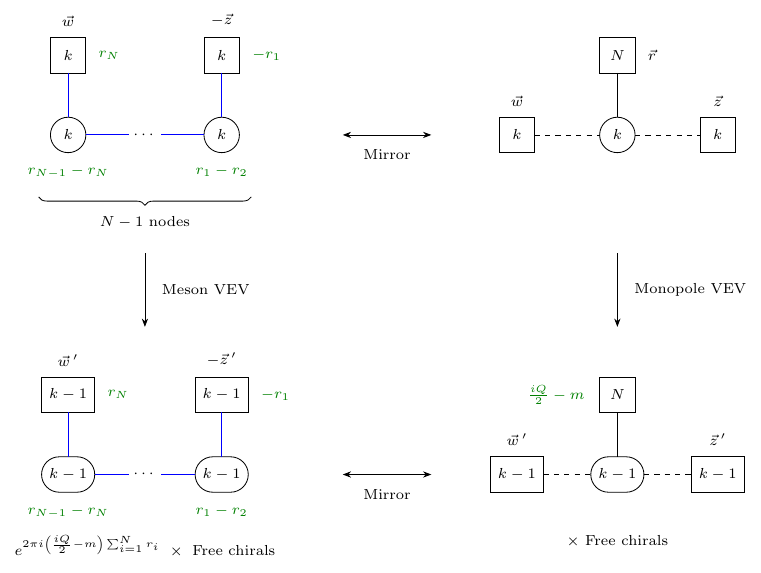}
  \caption{Mirror description of a monopole VEV flow. A VEV for a bifundamental monopole component in the magnetic theory corresponds to a long meson VEV in the electric quiver, lowering each electric gauge rank from $k$ to $k-1$. On the magnetic side, the central node becomes $U(k-1)$ and both $T[U(k)]$ tails shorten to $T[U(k-1)]$. The vectors $\vec w'$ and $\vec z'$ are obtained by deleting a single entry from $\vec w$ and $\vec z$, respectively, with the exact entry to be deleted selected by the exact VEV. The accompanying free chirals and background contact terms are retained when applying mirror symmetry to the interacting sector.}
  \label{fig:MonopoleVEV_ex}
\end{figure}

Our starting point is the mirror pair shown in the first row of Figure \ref{fig:MonopoleVEV_ex}. The electric theory is the linear quiver $k+U(k)^{N-1}+k$, while its magnetic dual consists of a central $U(k)$ gauge node with $N$ fundamental hypermultiplets and two $T[U(k)]$ tails. We use the charge conventions of Appendix \ref{app:3dstuff}. \footnote{In our conventions, the chiral multiplets in the blue electric hypermultiplets have $R_0=1$ and axial charge $-1$, while those in the magnetic hypermultiplets have $R_0=0$ and axial charge $+1$. The tails on the mirror side are represented using the compact dashed line notation. For more details refer to Appendix \ref{app:3dstuff}.}

In the magnetic theory, consider monopole operators carrying flux $(+1,\vec 0)$ or $(-1,\vec 0)$ along strings of consecutive gauge nodes extending to the left and to the right from the central node into the two tails. For each sign, these operators can be assembled into a $k\times k$ matrix; the two matrices transform in conjugate bifundamental representations of $U(k)_{\vec w}\times U(k)_{\vec z}$. All of these monopoles have total charge $\mathcal Q^{\pm}=\pm(w_i+z_j)+N(iQ/2-m)$, as defined in \eqref{eq:monopolecharge}. Under mirror symmetry, these operators map to long mesons connecting the two flavour nodes of the electric quiver, with the same total charges and in the bifundamental representation of $U(k)_{\vec w}\times U(k)_{\vec z}$. In particular, the mesons whose constituent arrows all point to the left map to the positive flux monopoles, while those with the opposite orientation map to the negative flux monopoles.\footnote{Conversely the mesons in the magnetic theory transforming in the adjoint of $SU(N)_{\vec r}$ map to the Coulomb branch moment map operators of the electric theory.}

We are interested in turning on VEVs for individual components of these bifundamental monopole matrices. For a negative flux monopole labelled by $i,j$, whose support extends from the $U(i)$ node on the left to the $U(j)$ node on the right, the VEV condition is given by $-w_i-z_j=N(m-iQ/2)$. The effect of this VEV can be studied in the electric description. For simplicity, we take $N=3$ so that the electric theory contains two $U(k)$ gauge nodes, and choose $i=j=1$. The VEV condition then becomes $-w_1-z_1=3(m-iQ/2)$. It corresponds to a VEV for a long meson connecting the first flavour component at one end of the electric quiver to the first flavour component at the other.

Let $\vec u, \vec v$ denote the Cartan variables of the two gauge nodes and $\vec w, \vec z$ parameterise the global flavour symmetries. Then the contribution of the three hypermultiplets is
\begin{equation}
    \prod_{i,j=1}^k s_b(m\pm(w_i-u_j)) s_b(m\pm(u_i-v_j)) s_b(m\pm(v_i+z_j)) ~.
\end{equation}
To study the VEV, we focus on the double sine functions involving $w_1$ and $z_1$, since these parameters appear in the VEV condition. For simplicity, we also restrict to double sine functions with $u_1$ and $v_1$; an identical argument holds for generic $u_i$ and $v_j$. The relevant double sine functions contain the following set of poles (see Appendix \ref{app:3dstuff}):
\begin{equation}
    \left\{
    \begin{aligned}
        m+w_1-u_1&= \frac{i Q}{2}+ i m_1 b + i \frac{n_1}{b} ~, \\
        m+u_1-v_1 &= \frac{i Q}{2}+ i m_2 b + i \frac{n_2}{b} ~, \\
        m+v_1+z_1&= \frac{i Q}{2}+ i m_3 b + i \frac{n_3}{b} ~,
    \end{aligned}
    \right. \quad\;\; m_a,n_a \in \mathbb Z_{\geq0} ~. 
\end{equation}
The first condition can be solved to give the pole positions
\begin{equation}
    u_1 = m+w_1-\frac{iQ}{2}-i\left(m_1b+\frac{n_1}{b}\right) ~,
\end{equation}
whereas combining the second and third gives
\begin{equation}
    u_1 = iQ-2m-z_1+i\left( (m_2+m_3) b+\frac{n_2+n_3}{b} \right) ~.
\end{equation}
Therefore for the VEV condition $-w_1-z_1=3(m-iQ/2)$, the integration over $u_1$ is pinched, or trapped between two sets of colliding poles from inside and outside, to fix
\begin{equation}
    u_1 = m-\frac{iQ}{2}+w_1 ~,
\end{equation}
for $m_a=n_b=0$ for $a,b=1,2,3$.

Similarly, the third condition gives
\begin{equation}
 v_1=-m-z_1+\frac{iQ}{2}+i\left(m_3 b+\frac{n_3}{b}\right) ~,
\end{equation}
while the first two give
\begin{equation}
    v_1=2m-iQ+w_1-i\left((m_1+m_2)b+\frac{(n_1+n_2)}{b} \right) ~.
\end{equation}
The leading poles coincide at the same VEV condition, pinching the integration contour to fix
\begin{equation}
    v_1 = -z_1-m+\frac{iQ}{2} = w_1+2m-iQ ~,
\end{equation}
for $m_a=n_b=0$ for $a,b=1,2,3$.

Evaluating the contour in the presence of this pinching freezes $u_1$ and $v_1$ at these values. The contributions involving the broken gauge generators combine with those of the corresponding adjoint and hypermultiplet components and cancel. The remaining gauge-dependent factors reproduce the vector, adjoint, and matter contributions of the reduced quiver $(k-1)+U(k-1)\times U(k-1)+(k-1)$, where the flavour symmetries are now parameterised by $\vec w'=(w_2,\dots,w_k)$ and $\vec z'=(z_2,\dots,z_k)$. The flow also produces gauge-neutral chiral multiplets whose contribution can be expressed as
\begin{equation}
    s_b\left(\frac{iQ}{2}\right) s_b\left(2m-\frac{iQ}{2}\right) \prod_{i=2}^k s_b\left(m\pm\left(w_i-w_1-m+\frac{iQ}{2}\right)\right) s_b\left(m\pm\left(z_i-z_1-m+\frac{iQ}{2}\right)\right) ~,
\end{equation}
and a background Chern--Simons term $e^{2\pi i(iQ/2-m)\sum_{i=1}^3r_i}$.

Clearly, the choice of $u_1$ and $v_1$ was arbitrary and we could have chosen any of the $k$ Cartan components at each gauge node leading to their freezing. The resulting $k^2$ equivalent contributions consistently convert the original Weyl factor of $U(k)^2$ to that of $U(k-1)^2$ as $k^2\times1/(k!)^2=1/(k-1)!^2$. Therefore, tuning $-w_1-z_1=3(m-iQ/2)$ implements a long meson VEV that Higgses
\begin{equation}
    U(k)_{\vec u}\times U(k)_{\vec v} \;\longrightarrow\; U(k-1)_{\vec u'}\times U(k-1)_{\vec v'} ~,
\end{equation}
together with the free sector and background factor described above. The same analysis extends to general $N$ which implies that turning on a VEV for one long meson in the electric theory in the upper left corner of Figure \ref{fig:MonopoleVEV_ex} triggers a flow to the bottom left corner of the same figure. Applying mirror symmetry to the interacting sector then identifies the corresponding monopole VEV flow in the magnetic theory. The central node becomes $U(k-1)$ and the two tails shorten from $T[U(k)]$ to $T[U(k-1)]$, together with the free chirals and background CS term, as shown on the RHS of Figure \ref{fig:MonopoleVEV_ex}.

\section{Consistency checks of the 5d/3d dictionary}
\label{sec:5d3dmap}

In this section, we test the proposed dictionary in \eqref{eq:dict} between the parameters of the five dimensional $N+SU(N)_0^{k-1}+N$ box theory and the FI parameters of its strong-coupling magnetic quiver $S_{k,N}$. Through this dictionary, five-dimensional mass deformations map to FI deformations of the magnetic quiver, as argued in \cite{vanBeest:2021xyt}, while Higgs branch VEVs map to monopole VEVs. We analyse the corresponding three-dimensional flows at the level of the $S^3_b$ partition function using the techniques developed in Section \ref{sec:3dtechnology}.

We primarily consider deformations that can be implemented explicitly in both five and three dimensions, allowing a direct comparison of their effects on the respective supersymmetric observables. On the five-dimensional side, the calculation is organised according to a chosen ruling of the toric diagram and preferred direction in the refined topological string computation. Deformations compatible with this description can be implemented directly, whereas others require a resummation or analytic continuation of the instanton expansion.

We therefore focus our direct consistency checks on the removal of an external NS5 brane, specific mass deformations, and weak coupling limits accessible in this description. In each case, we first implement the deformation in five dimensions and then use the 5d/3d dictionary to identify and analyse its three-dimensional magnetic quiver counterpart. In every case, we find that the resulting magnetic quiver precisely reproduces the expected five-dimensional physics, providing a strong consistency check of our dictionary. We leave the analysis of nilpotent VEVs for moment map operators to the companion paper \cite{two}.

The magnetic quiver framework also allows us to study deformations whose direct five-dimensional derivation is more involved. In Subsection \ref{3dremD5} we use the dictionary in the opposite direction: the three-dimensional analysis of the monopole VEV corresponding to the removal of a D5 brane determines the associated five-dimensional parameter flow.

\subsection{Removing an NS5 brane from the $(p,q)$ web: shortening the box}
\label{subsec:shorten_box}

We first consider the removal of an NS5 brane from the box five-brane web. This deformation is realised by a baryonic Higgs branch VEV associated with the $U(1)$ symmetry of the corresponding strip. Via the dictionary in \eqref{eq:dict}, this operation maps into a monopole VEV in the magnetic quiver.

We study the removal of an external NS5 brane explicitly in five and three dimensions and compare the results. We then discuss the removal of internal NS5 branes and the deformation obtained by removing a D5 brane from the magnetic quiver perspective.

\subsubsection{5d computation}
\label{subsubsec:shorten_5d}

Removing the $i$-th NS5 brane corresponds to a baryonic Higgs branch VEV associated with the hypermultiplets of the $i$-th strip. In our parameterisation, this deformation is implemented by the tuning
\begin{equation}\label{eq:NS5removal_cond}
    K_i \longrightarrow 1 ~.
\end{equation}
The $k$ parameters $K_i$ encode the $k$ baryonic $U(1)$ symmetries, rotating the $i$-th bifundamental hypermultiplets, associated with the strips forming the box.\footnote{Upon circle reduction, this operation corresponds to closing a minimal puncture of the associated theory of class $\cS$. A symmetrical strip of length $N$ can be viewed as the uplift of a trinion with two maximal and one minimal puncture. }

We focus on the removal of the leftmost NS5 brane, corresponding to $K_1\rightarrow1$. The rightmost NS5 brane can be treated analogously. Removing an internal NS5 brane introduces an additional complication in the partition sums, which we discuss at the end of this subsection.

\subsubsection*{Distribution from strip degeneration}

We first isolate the first strip and study its degeneration before gluing it to the remainder of the box. Using \eqref{eq:Zstrip_pert}, the modulus square of its perturbative contribution to the topological string partition function of the $N+SU(N)_0^{k-1}+N$ box theory can be written as\footnote{The parameter $K_1$ also enters the relation between the base parameters and the instanton fugacities. Once \eqref{eq:identitywall_5d_inst} sets $\bm\alpha^{(1)}=\bm\emptyset$, the base parameter weights associated with the first gauge node become trivial.}
\begin{equation}\label{eq:NS5_strippert}
    \left| \cZ^{\rm strip}_{\rm pert}(\vec M,\vec N^{(1)},K_1) \right|^2  = \left| \frac{\prod_{1\leq a<b\leq N} \MM{\frac{M_a}{M_b}} \MM{\frac{N^{(1)}_a}{N^{(1)}_b}} }{\prod_{a,b=1}^N \cM\left( \frac{M_aN^{(1)}_b}{K_1} \right) } \right|^2 ~,
\end{equation}
where we have used \eqref{eq:Minv_identity} to simplify the expression. For the isolated strip, both sets of external parallel leg contributions are removed. When the strip is glued into the box, however, only the left external contribution is removed: the factors associated with $\vec N^{(1)}$ must be retained to reconstruct the first $SU(N)$ gauge measure together with the adjacent strip.

As reviewed in Appendix \ref{app:special_func}, the ``modulus square" of the $\cM$ functions has two families of zeroes: one inside and another one outside the unit circle. Therefore, the denominator of \eqref{eq:NS5_strippert} contains poles located at
\begin{equation}
    (\bar P) \;:\; N_b^{(1)} = \frac{K_1}{M_a} \mf t^{-n_2}\mf q^{n_1} \quad,\quad (P) \;:\;  N_b^{(1)} = \frac{K_1}{M_a} \mf t^{n_1+1}\mf q^{-n_2-1} \;\;\text{ with } n_1,n_2\in\mathbb Z_{\geq0} ~.
\end{equation}
To describe the pinching mechanism, consider $N=3$. 
As we take the limit $K_1\to1$, there is a collision between different sets of poles in \eqref{eq:NS5_strippert}.

It is enough to examine the leading poles, with $n_1=n_2=0$.
Two poles of the $(\bar P)$ family occur at
\begin{equation}
    N_1^{(1)}=\frac{K_1}{M_1} \;\;,\quad
    N_2^{(1)}=\frac{K_1}{M_2} ~.
\end{equation}
A third $(\bar P)$ pole occurs at
$N_3^{(1)}=K_1/M_3$. Although it belongs to the same family
as a pole in $N_3^{(1)}$, the $SU(3)$ constraint
$N_1^{(1)}N_2^{(1)}N_3^{(1)}=1$ expresses it as a pole in
the independent variable $N_2^{(1)}$:
\begin{equation}
    N_2^{(1)}
    =\frac{M_3}{K_1N_1^{(1)}}
    \quad\xrightarrow{\;N_1^{(1)}=K_1/M_1\;}\quad
    \frac{1}{K_1^2M_2} ~,
\end{equation}
where we also used $M_1M_2M_3=1$. With the flavour
fugacities on the unit circle and $K_1$ approaching one from
inside it, the pole at $K_1/M_2$ lies inside the
$N_2^{(1)}$ contour, whereas the pole at
$1/(K_1^2M_2)$ lies outside. They collide as $K_1\to1$,
pinching the contour at $N_2^{(1)}=M_2^{-1}$.
Together with $N_1^{(1)}=M_1^{-1}$, the $SU(3)$ constraint
then fixes $N_3^{(1)}=M_3^{-1}$.

For general $N$, the analogous final $(\bar P)$ condition,
expressed through the $SU(N)$ constraint, gives an outside
pole at $N_{N-1}^{(1)}=1/(K_1^{N-1}M_{N-1})$ after the
first $N-2$ leading poles have been selected. It collides
with the inside pole at $N_{N-1}^{(1)}=1/M_{N-1}$
as $K_1\to1$.

For generic flavour fugacities, there are $N!$ such pinchings and the perturbative contribution takes the form:\begin{equation}\label{eq:identitywall_5d_pert}
    \left| \frac{ \cZ^{\rm strip}_{\rm pert}(\vec M, \vec N, K_1\rightarrow1) }{ \cZ_{||}^{\rm left}(\vec M) } \right|^2 = \sum_{\sigma\in S_N} \left|\frac{ M(\mf t,\mf q)^{-N} }{ \prod_{1\leq a<b\leq N}\MM{\frac{M_a}{M_b}} }\right|^2 \times \delta(0) \prod_{a=1}^{N-1} \delta\left( N_{\sigma(a)}^{(1)} M_a \right) ~.
\end{equation}
The factors of $M(\mf t,\mf q)$ arise from the regular part of the vanishing $\cM$ factors, as defined in \eqref{eq:Mstr_def}. Upon gluing the degenerate strip to the remainder of the box, the factor $|M(\mf t,\mf q)|^{-2N}$ combines with the Cartan contribution $|M(\mf t,\mf q)|^{2(N-1)}$ of the first gauge node leaving $\delta(0)/|M(\mf t,\mf q)|^2$. Since no zero cancels this divergence, it must be retained in the computation. This factor can be interpreted as a contribution from free fields generated by the VEV; an analogous contribution appears in the three-dimensional calculation discussed in the next subsection.

The remaining mass-dependent factor precisely cancels the parallel leg contribution involving $\vec N^{(1)}$ from the second strip. Before the degeneration, this contribution supplies one half of the $SU(N)_{\vec N^{(1)}}$ measure with the other half being supplied by the first strip. After the degeneration, the delta functions freeze $\vec N^{(1)}$ and the corresponding legs of the second strip become external. The same cancellation can therefore be understood as removing the external parallel leg contribution at the newly exposed boundary of the shortened box. Finally, the terms in the sum over $S_N$ are related by the Weyl group of $SU(N)$. Upon integrating over $\vec N^{(1)}$, they give identical contributions against the remaining gauge invariant integrand. Their sum therefore produces a multiplicity $N!$, equal to the order of the Weyl group, which cancels the $1/N!$ in the integration measure of the first gauge node.

We now turn to the effect of the degeneration on the ``instantonic" contribution of the first strip. The contribution is evaluated on the support of the Dirac delta functions obtained above and can be written as
\begin{equation}
    \begin{split}
        &\cZ^{\rm strip}_{\bm\emptyset,\bm\alpha^{(1)}} (\vec M, \vec N^{(1)}, K_1) \\
        &\quad = \prod_{a=1}^N \Big[ \mf q^{\frac{||\alpha^{(1)}_a||^2}{2}} \widetilde Z_{\alpha^{(1)}_a}(\mf t,\mf q) \Big] \; \frac{ \prod_{1\leq a<b\leq N} \cN_{\alpha^{(1)}_a\emptyset} \left(\frac{\mf t}{\mf q}\frac{K_1}{M_b N^{(1)}_a}\right) \prod_{1\leq a\leq b\leq N} \cN_{\emptyset\alpha^{(1)}_b}\left(\frac{M_a N^{(1)}_b}{K_1}\right) }{ \prod_{1\leq a<b\leq N} \mathcal{N}_{\alpha^{(1)}_a\alpha^{(1)}_b}\left( \frac{\mf t}{\mf q}\frac{N^{(1)}_b}{N^{(1)}_a} \right) } ~.
    \end{split}
\end{equation}
On the support of each term in the perturbative distribution in \eqref{eq:identitywall_5d_pert}, the identities \eqref{eq:Ntrivial_identity} force all partitions $\alpha_a^{(1)}$ to be empty. Thus
\begin{equation}\label{eq:identitywall_5d_inst}
    \cZ^{\rm strip}_{\bm\emptyset,\bm\alpha^{(1)}} (\vec M, \vec N^{(1)}, K_1\rightarrow1) \Big\vert_{\rm \delta \; from \; (\ref{eq:identitywall_5d_pert})} = \prod_{a=1}^N \,\delta_{\alpha_a^{(1)},\emptyset} ~.
\end{equation}
The instantonic contribution therefore collapses to a product of Kronecker deltas, restricting the sum over $\bm\alpha^{(1)}$ to the empty partitions. Combining this result with the perturbative distribution in \eqref{eq:identitywall_5d_pert} gives the modulus square of the degenerate strip contribution, with only the left external parallel leg factor removed:
\begin{equation}\label{eq:identitywall_5d}
    \begin{split}
        &\left| \frac{ Z^{\rm strip}(\vec M, \vec N, K_1\rightarrow1; \bm\emptyset, \bm\alpha^{(1)}) }{ \cZ_{||}^{\rm left} (\vec M) } \right|^2 \\
        &\quad = \sum_{\sigma\in S_N} \left| \; \frac{ M(\mf t,\mf q)^{-N} }{ \prod_{1\leq a<b\leq N}\MM{\frac{M_a}{M_b}} } \times \prod_{a=1}^N \,\delta_{\alpha_a^{(1)},\emptyset} \; \right|^2 \times \delta(0) \prod_{a=1}^{N-1} \delta\left( M_aN_{\sigma(a)}^{(1)}  \right) ~.
    \end{split}
\end{equation}
Consequently, the superconformal index associated with this configuration is itself distributional and takes the form of a sum of delta-function contributions. This provides our first example of a \emph{broken} five-dimensional theory. In the companion paper \cite{two}, we will investigate systematically how five-dimensional partition functions become distributional when an excessive VEV is turned on.

This behaviour of the degenerate strip is not unexpected. The five-brane web obtained by removing an NS5 brane from the strip geometry consists of $N$ parallel D5 branes and therefore represents a tube. It may be regarded as the five-dimensional uplift of the four-dimensional $\cN=2$ class-$\cS$ theory associated with a sphere with two maximal punctures. Through the AGT correspondence \cite{Alday:2009aq,Wyllard:2009hg}, the latter is related to the two-point function of generic primary fields in the $A_{N-1}$ Toda CFT. Such a two-point function is distributional and identifies the momenta of the two primary fields up to the action of the Weyl group. This accounts for the sum of delta functions appearing in \eqref{eq:identitywall_5d}.
Equivalently, through the five-dimensional version of the AGT correspondence, the topological-string partition function of the strip geometry is related to a $q$-Toda three-point function involving two generic primary fields and one semi-degenerate primary, whose momentum is proportional to the first or last fundamental weight \cite{Nieri:2013yra,Nieri:2013vba,Mitev:2014isa,Isachenkov:2014eya}. The limit $K_1\rightarrow1$ further degenerates this three-point function to the two-point function of the two generic primaries. The latter is again a sum of delta functions imposing the identification of their momenta up to Weyl transformations.

In Subsection \ref{subsubsec:shorten_3d}, we will compare this distributional five-dimensional result directly with the three-dimensional identity wall. The latter is obtained by implementing the three-dimensional analogue of this degeneration in the strong-coupling magnetic quiver associated with the strip.

\subsubsection*{Distributional action of the degenerate strip}

We now return to our goal of studying the effect of removing the leftmost NS5 brane from the box configuration. Armed with the topological string partition function of the degenerated strip \eqref{eq:identitywall_5d}, we can compute the effect of this degeneration on the remainder of the box. The Dirac delta functions fix $\vec N^{(1)}$ to $\vec M^{-1}$, up to Weyl permutations, while the Kronecker deltas set $\bm\alpha^{(1)}=\bm\emptyset$. The first gauge integration and its partition sum therefore disappear. Keeping the inherited alternating parameterisation of the surviving strips, the resulting superconformal index becomes 
\begin{equation}\label{eq:index5d_shortened}
    \begin{split}
         &\cI^{N+SU(N)_0^{k-1}+N} (\vec Q_{\rm inst},\vec M,\vec L, K_1\rightarrow1, K_2, \dots, K_k) \\[0.3mm]
         &\;= \frac{\delta(0)}{\left| M(\mf t,\mf q) \right|^2} \times \cZ^{\rm 5d}_{\rm free}(\vec Q_{\rm bot}, \vec Q_{\rm top}) \times \oint \prod_{m=2}^{k-1} \left[ \frac{1}{N!} \prod_{i=1}^N\frac{{\rm d}A_i^{(m)}}{2\pi iA_i^{(m)}} \times \delta\left( \prod_{i=1}^N A_i^{(m)} \right) \times \left| M(\mf t,\mf q) \right|^{2(N-1)} \right] \\
        &\hspace{3cm} \times \left| \frac{\left[\cZ_{\rm pert}^{\rm strip} (\vec M^{-1}, \vec A^{(2)}, K_2) \prod_{m=2}^{k-1} \cZ^{\rm strip}_{\rm pert} (\vec A^{(m)}, \vec A^{(m+1)}, K_{m+1}) \right]' }{ \cZ^{k-1,N}_{||} (\vec M, \vec L^{-1}, \vec Q_{\rm bot}^\prime,\vec Q_{\rm top}^\prime) } \right|^2  \\
        &\quad \times \left| \sum_{\bm{\alpha}^{(2)}}\cdots\sum_{\bm{\alpha}^{(k-1)}} \; \prod_{m=2}^{k-1} \prod_{a=1}^N Q_{m,a}^{|\alpha^{(m)}_{a}|} \left[ \cZ^{\rm strip}_{\bm\emptyset,\bm\alpha^{(2)}} (\vec M^{-1}, \vec A^{(2)}, K_2) \prod_{m=2}^{k-1} \cZ^{\rm strip}_{\bm\alpha^{(m)},\bm\alpha^{(m+1)}} (\vec A^{(m)}, \vec A^{(m+1)}, K_{m+1}) \right]' \right|^2 ~,
    \end{split}
\end{equation}
with $\vec A^{(k)}=\vec L$, $\bm\alpha^{(k)}=\bm\emptyset$. The primes in \eqref{eq:index5d_shortened} indicate the alternating parameterisations inherited from the original box: the first surviving strip retains the alternate parameterisation of the original second strip, and the subsequent strips continue to alternate accordingly. Removing the leftmost NS5 brane also deletes the first separation at the top and bottom of the five-brane web. We denote the remaining geometric parameters by $\vec Q_{\rm bot}^\prime=(Q^{(2)}_{\rm bot},\dots,Q^{(k-1)}_{\rm bot})$ and $\vec Q_{\rm top}^\prime=(Q^{(2)}_{\rm top},\dots,Q^{(k-1)}_{\rm top})$. Correspondingly, the instanton fugacities for the shortened box are $\vec Q_{\rm inst}^\prime=(Q^{(2)}_{\rm inst},\dots,Q^{(k-1)}_{\rm inst})$.

The external parallel leg contribution associated with pairs of surviving NS5 branes reproduces the corresponding contribution to $\cZ_{||}^{k-1,N}$ for the shortened box. Meanwhile we denote those involving the removed brane by $\cZ^{\rm 5d}_{\rm free}$ in \eqref{eq:index5d_shortened}. We will see that in three dimensions this corresponds to a free sector associated with the monopole VEV in the magnetic quiver. This contribution can be expressed as
\begin{equation}
    \cZ^{\rm 5d}_{\rm free}(\vec Q_{\rm bot}, \vec Q_{\rm top}) = \prod_{j=2}^k \left| \MM{\prod_{l=1}^{j-1}Q_{\rm top}^{(l)}} \MM{\frac{\mf t}{\mf q}\prod_{l=1}^{j-1}Q_{\rm bot}^{(l)}} \right|^{-2}_{K_1=1}
\end{equation}
This can be expressed in terms of the instanton and strip parameters by using the relations between these geometric separations, instanton fugacities, and the strip parameters described in Appendix \ref{app:topstringPF} to obtain
\begin{equation}\label{eq:freeNS5_5d}
    \begin{split}
        \cZ^{\rm 5d}_{\rm free}(\vec Q_{\rm bot}, \vec Q_{\rm top}) &= \prod_{j=2}^k \left| \MM{ \left(\frac{\mf t}{\mf q}\right)^{\frac{N}{4}} \left(\sqrt{\frac{\mf t}{\mf q}}K_j\right)^{\frac{N}{2}(-1)^j} \prod_{l=1}^{j-1} Q_{\rm inst}^{(l)} } \right. \\
        &\qquad \times \; \left. \MM{ \left(\frac{\mf t}{\mf q}\right)^{1-\frac{N}{4}} \left(\sqrt{\frac{\mf t}{\mf q}}K_j\right)^{\frac{N}{2}(-1)^{j+1}} \prod_{l=1}^{j-1} Q_{\rm inst}^{(l)} } \right|^{-2}
    \end{split}
\end{equation}

The remaining integral in \eqref{eq:index5d_shortened} can be identified as the index of the shortened box and therefore we obtain
\begin{equation}\label{eq:indexmatch_NS5removal}
    \begin{split}
        &\cI^{N+SU(N)_0^{k-1}+N} (\vec Q_{\rm inst},\vec M,\vec L, K_1\rightarrow1, K_2, \dots, K_k) \\
        &\quad = \frac{\delta(0)}{|M(\mf t,\mf q)|^2} \times \cZ_{\rm free}^{\rm 5d}(\vec Q_{\rm bot},\vec Q_{\rm top}) \times \cI^{N+SU(N)_0^{k-2}+N} (\vec Q'_{\rm inst},\vec M,\vec L^{-1}, \frac{\mf q}{\mf t}K_2^{-1}, \dots, \frac{\mf q}{\mf t}K_k^{-1}) ~.
    \end{split}
\end{equation}
Thus the first $SU(N)_{\vec N^{(1)}}$ gauge node is completely Higgsed, leaving the interacting theory associated with a box shortened by one strip, as expected from the five-brane web. The fugacity $Q_{\rm inst}^{(1)}$ no longer enters the interacting index, although it remains in the accompanying free contribution.

\subsubsection*{Comments on removing NS/D5 branes from the web}

This discussion so far concerns the removal of the leftmost NS5 brane. The rightmost brane can be treated analogously by taking $K_k\rightarrow1$, with the appropriate strip parameterisation. Successive removals of external NS5 branes can therefore be described by iterating this calculation.

Removing an internal NS5 brane is technically more involved because the corresponding strip carries non-trivial partitions on both sides. Its perturbative correction has the same pinching mechanism as above, with the parallel leg factors retained according to the internal gluing. The instantonic contribution, however, does not immediately collapse using \eqref{eq:Ntrivial_identity}. Instead one encounters factors of the form
\begin{equation}
    \cN_{\alpha^{(a-1)}\alpha^{(a)}}\left(\frac{\mf t}{\mf q}\right) \;\text{ and }\; \cN_{\alpha^{(a)}\alpha^{(a-1)}}\left(1\right) \nn
\end{equation}
which impose containment conditions of the form $\bm\alpha^{(a-1)}\subseteq\bm\alpha^{(a)}$ rather than forcing the partitions to be empty \cite{Hayashi:2014hfa}. The remaining partition sum therefore does not trivialise in the same way as for the external strip. We expect that a suitable resummation reproduces the effective identification of the adjacent partition data $\bm\alpha^{(a-1)}=\bm\alpha^{(a)}$ and yields the shortened box partition function.

Since our primary aim is to study deformations that can be treated exactly in both five and three dimensions, we do not pursue a perturbative analysis of this resummation. In the unrefined topological string partition function, by contrast, an identity that trivializes the sum over partitions is available. In this limit, one can show analytically that the contribution of the $a$-th column, or equivalently of the corresponding $SU(N)$ gauge node, drops out, leaving the partition function of the shortened box theory $N+SU(N)_0^{k-2}+N$.

It is also natural to consider the deformation obtained by removing a D5 brane from the box five-brane web. This lowers the height of the box by one unit and produces the Higgsing
\begin{equation}
    N+SU(N)_0^{k-1}+N \quad\longrightarrow\quad (N-1)+SU(N-1)_0^{k-1}+(N-1) ~.
\end{equation}
Although the endpoint is apparent from the five-brane web, the implementation of this Higgsing in the superconformal index is less direct in the gauge theory frame selected by our ruling. For an external NS5 brane, a perturbative pole collision completely Higgses one gauge group. Removing a D5 brane instead reduces the rank of every gauge node by one and involves the instanton sector in this description. Extracting this Higgsing directly therefore requires a non-trivial resummation of the instanton expansion.\footnote{Fibre-base duality exchanges D5 and NS5 branes and can make the removal of an individual D5 brane accessible in the dual gauge theory frame. The fiber-base swap however, does not help in case of a simultaneous D5 and NS5 brane removal.} Consequently, although the endpoint theory is geometrically evident, determining the Higgsing mechanism and the precise relation between ultraviolet and infrared parameters from this representation of the index is difficult. As we will see in Subsection \ref{3dremD5}, these deformations can be efficiently studied in the associated magnetic quiver theory which also allows us to reconstruct the 5d parameter flow via our proposed 5d/3d dictionary.

\subsubsection{3d computation}
\label{subsubsec:shorten_3d}

In this subsection, we study the three-dimensional magnetic quiver counterpart of the deformation corresponding to the removal of a single NS5 brane.

\subsubsection*{3d counterpart of the strip degeneration}

\begin{figure}[h!]
  \centering
  \includegraphics[width=\textwidth]{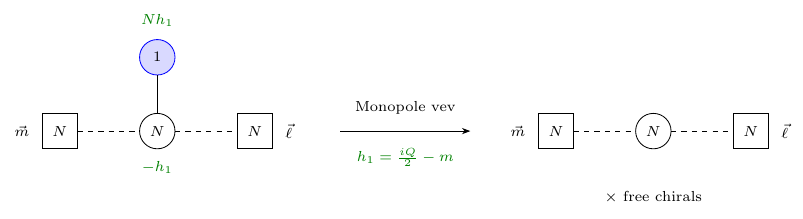}
  \caption{Monopole VEV corresponding to the removal of the NS5 brane from a single strip. At $h_1=iQ/2-m$, the monopole with flux $-1$ under the highlighted $U(1)$ node acquires a VEV. The resulting flow removes this node, leaving the three-dimensional identity wall together with free chirals.}
  \label{fig:monoVEV}
\end{figure}

We begin with the case $k=1$, corresponding to the removal of the NS5 brane from a single symmetrical strip. Its magnetic quiver $S_{1,N}$ is depicted on the left of Figure \ref{fig:monoVEV}. It is a star-shaped quiver with two $T[U(N)]$ tails and a single $U(1)$ tail with FI parameter $Nh_1$.

Removing the first NS5 brane amounts to the limit $K_1\to 1$ on the five-dimensional side. At the canonical $\mathcal{N}=4$ superconformal $R$-charge assignment this corresponds to the condition $h_1=\frac{iQ}{4}$. We now explain how this condition is realised in
the magnetic quiver.

Using the monopole charge formula in \eqref{eq:monopolecharge}, the total charge of the monopole operator with flux $\mathfrak{m}=-1$ under the $U(1)$ node (carrying FI parameter $\xi=Nh_1$) is:
\begin{equation}
    \mathcal{Q}(-1) = \big(iQ-2m\big)\frac{N}{2}-Nh_1 = N\left(\frac{iQ}{2}-m-h_1\right) ~.
\end{equation}
Setting $\mathcal{Q}(-1)=0$, as required for this operator to acquire a VEV, gives
\begin{equation}
    h_1=\frac{iQ}{2}-m ~.
\end{equation}
This is the $\mathcal{N}=2^*$ refined version of the condition. At the canonical $R$ charge value $m=\frac{iQ}{4}$, it reduces to $ h_1=\frac{iQ}{4}$, in agreement with the 5d/3d parameter dictionary.

The effect of this monopole VEV is to Higgs the $U(1)$ node. The resulting three-dimensional quiver, shown on the right in Figure \ref{fig:monoVEV}, consists of the two fused $T[U(N)]$ tails. The central $U(N)$ node is underbalanced, and the resulting theory is therefore a broken theory, known as the three-dimensional identity wall \cite{Bottini:2021vms}. Including the free chirals generated by this VEV, its squashed sphere partition function is
\begin{equation}\label{eq:identitywall_3d}
    \left. \PFS{S_{1,N}} \right|_{h_1=\frac{iQ}{2}-m} = s_b\left(\frac{iQ}{2}\right) s_b\left(2m-\frac{iQ}{2}\right) \sum_{\sigma\in S_N} \frac{ \prod_{j=1}^{N-1} \delta\left( m_j+l_{\sigma(j)} \right) }{ \Delta^{\rm CB}_{SU(N)}(\vec{m};m) } ~.
\end{equation}
The first two factors in \eqref{eq:identitywall_3d}, $s_b\left(iQ/2\right) s_b\left(2m-iQ/2\right)$, are generated by the monopole VEV, as explained in Section \ref{sec:monopolevevs}, while the remaining distributional expression is the partition function of the identity wall.

The fusion of two $T[U(N)]$ tails naturally produces an identity wall between two $U(N)$ symmetries, with a product of $N$ delta functions and the $U(N)$ measure in the denominator. Here, however, we
work with the $SU(N)$ variables $\vec{m}$ and $\vec{l}$. Consequently, once $N-1$ delta-function constraints have been imposed, the remaining one becomes $\delta(0)$. This divergent factor, together with one diagonal component of the adjoint contribution that converts the $U(N)$ measure into the $SU(N)$ measure, must be removed by the overall $U(1)$ decoupling procedure described in Subsection \ref{subsec:boxMQ}. Since the electric dualisation algorithm reduces the identity wall to a collection of free hypers and hence no gauge groups remain in the final theory, this decoupling must be implemented explicitly.

It can be immediately seen that \eqref{eq:identitywall_3d} has precisely the same structure as its five-dimensional counterpart \eqref{eq:identitywall_5d}. In particular, the Weyl sum over $S_N$ of products of delta functions matches directly under the dictionary \eqref{eq:dict}. At the canonical $R$ charge assignment, the three-dimensional measure takes the form:
\begin{equation}
    \left. \Delta_{SU(N)}^{\rm CB}(\vec{m};m) \right|_{m\rightarrow\frac{iQ}{4}} = \prod_{1\leq a<b\leq N} s_b\left( \frac{iQ}{2}\pm(m_a-m_b) \right)^{-1} ~.
\end{equation}
The corresponding five-dimensional denominator factor appearing in \eqref{eq:identitywall_5d} is
\begin{equation}
    \prod_{1\leq a<b\leq N} \left| \MM{\frac{M_a}{M_b}} \right|^2 ~. \nn
\end{equation}
The two expressions match factor by factor under the dictionary in \eqref{eq:dict}: each pair labelled by $a<b$ contributes an inverse double sine pair in three dimensions and a modulus squared $\cM$ factor in five dimensions, with their respective arguments related by the parameter map.

The 5d result \eqref{eq:identitywall_5d} contains the divergent factor $\delta(0)/|M(\mf{t},\mf{q})|^2$, while its three-dimensional counterpart contains $s_b\left(iQ/2\right)s_b(2m-iQ/2)$, which can similarly be interpreted as a $\delta(0)$ divergence via the distributional identification $s_b(iQ/2)\sim\delta(0)$ \cite{Giacomelli:2023zkk}. There is no immediate three-dimensional counterpart of the factor $M(\mf{t},\mf{q})$ produced by the five-dimensional residue. The remaining three-dimensional factor $s_b\left(2m-iQ/2\right)$ becomes trivial upon imposing the $\mathcal{N}=4$ condition $m=iQ/4$, since $s_b(0)=1$.

We thus find a direct correspondence between the squashed-sphere partition function of the three-dimensional identity wall in \eqref{eq:identitywall_3d} and the perturbative part of the topological string partition function of the degenerate strip geometry obtained after removing the NS5 brane.

\subsubsection*{3d counterpart of removing one NS5 from the box}
\label{3dremNs}

We now consider the three-dimensional counterpart of removing the $j$-th NS5 brane from the full box. The associated five-dimensional limit $K_j\to1$, through the refined map discussed above corresponds to
\begin{equation}
    h_j = \frac{iQ}{2}-m ~.
\end{equation}
This condition allows a component of the monopole operator in the bifundamental representation of $U(k)_{\vec w}\times U(k)_{\vec z}$ to acquire a VEV, as discussed in Subsection \ref{sec:monopolevevs}.

More precisely, consider the monopole operators with unit flux (up to sign) for the string of consecutive gauge nodes along the chain extending from the left $U(j)$ node, through the central $U(k)$ node, to the right $U(j)$ node, and zero flux elsewhere. These monopoles carry total charge
\begin{equation}
    \CQ = \pm(w_j+z_j)+N\left(\frac{iQ}{2}-m\right) = \pm N\widetilde h_j + N\left(\frac{iQ}{2}-m\right) ~,
\end{equation}
where the second equality follows from the 5d/3d dictionary \ref{eq:dict}. Since $\widetilde h_j=(-1)^{j+1}h_j$ the condition $h_j=iQ/2-m$ corresponds to the negative flux monopole acquiring a VEV for odd $j$, and the positive flux monopole acquiring a VEV for even $j$.

\begin{figure}[h!]
  \centering
  \includegraphics[width=\textwidth]{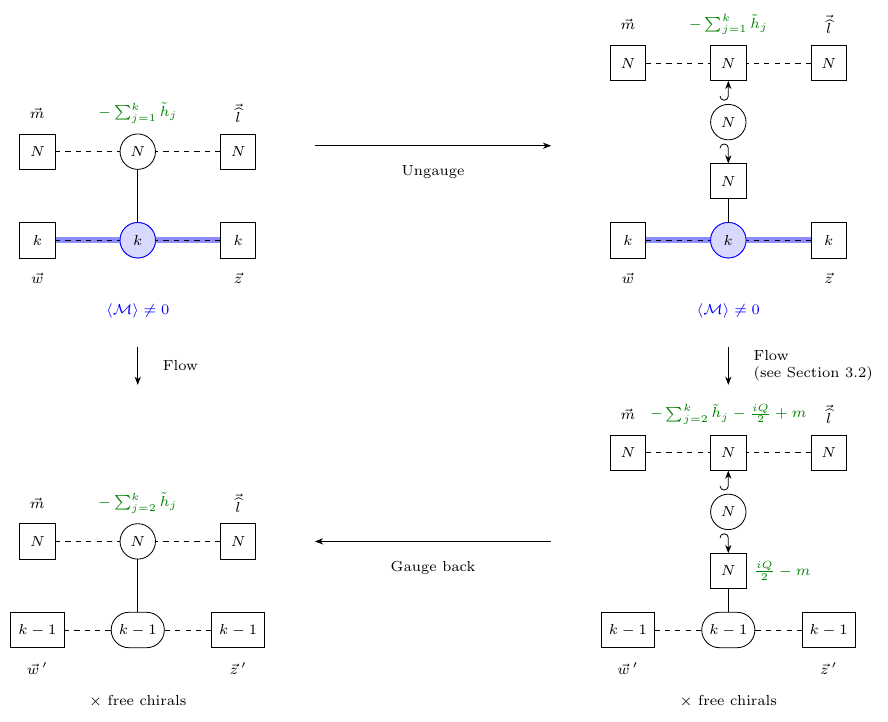}
  \caption{The monopole VEV associated with the removal of the leftmost NS5 brane triggers a flow $S_{k,N}\longrightarrow S_{k-1,N}$. The blue highlight marks the support of the monopole which carries flux $(-1,\vec 0)$ at every gauge node in the lower block, including the central $U(k)$ node. Ungauging the $U(N)$ node isolates this block enabling the application of the analysis of Section \ref{sec:monopolevevs}. The VEV lowers its gauge ranks by one and produces free chirals and a background contact term. Gauging back the $U(N)$ symmetry gives $S_{k-1,N}$ and reduced vectors $\vec w'$ and $\vec z'$. The overall $U(1)$ is understood to be decoupled.}
  \label{fig:boxNS5rem}
\end{figure}

We first remove the leftmost NS5 brane by setting $h_1=iQ/2-m$. The corresponding monopole has flux $(-1,\vec 0)$ at every gauge node in the lower block, including the central $U(k)$ node and both $T[U(k)]$ tails. Its effect can be studied using the local computation of Section \ref{sec:monopolevevs}, as illustrated in Figure \ref{fig:boxNS5rem}. The idea is to first ungauge the $U(N)$ and isolate the subquiver that contains the entire support of the monopole acquiring a VEV, allowing a direct application of the computation in Subsection \ref{sec:monopolevevs}. This subquiver consists of the two $T[U(k)]$ tails joined by the central $U(k)$ node, now coupled to $N$ fundamental hypermultiplets. The VEV lowers the ranks of all gauge nodes in this isolated subquiver by one unit, leaving the $N$ fundamental hypermultiplets untouched as depicted on the bottom right in Figure \ref{fig:boxNS5rem}. The local computation also generates a background contact term, which upon gauging the $U(N)$ symmetry back, shifts its FI parameter by $iQ/2-m$. The quiver obtained upon gauging back this resulting isolated subquiver is simply $S_{k-1,N}$, with the correct FI parameter for the $U(N)$ gauge node, as depicted on the bottom left in Figure \ref{fig:boxNS5rem}. This is exactly as expected for the strong-coupling magnetic quiver of the shortened box theory $N+SU(N)_0^{k-2}+N$.

The parameters of the surviving $U(k-1)\times U(k-1)$ symmetries are
\begin{equation}
    \vec{w}' = \left(w_2,\cdots w_k\right) \;\;,\quad \vec{z}' = \left(z_2,\cdots z_k\right) ~,
\end{equation}
obtained by deleting the first entries of $\vec w$ and $\vec z$ in \eqref{eq:dict}. Through the 5d/3d dictionary, this removes the dependence of the interacting quiver on $v_1$, associated with the instanton fugacity $Q_{\rm inst}^{(1)}$, in agreement with the five-dimensional result in \eqref{eq:indexmatch_NS5removal}.

As noted in Subsection \ref{sec:monopolevevs}, this VEV also generates a free sector whose contribution to the sphere partition function can be expressed as
\begin{equation}
    \begin{split}
        \cZ_{\rm free}^{\rm 3d} &= s_b\left(2m-\frac{iQ}{2}\right) s_b\left(\frac{iQ}{2}\right) \\
        &\quad \times \; \prod_{j=2}^{k} \left\{ s_b\left(m\pm\left[\frac{N}{2}(-1)^{j+1}h_j-\left(\frac N2-1\right)\left(\frac{iQ}{2}-m\right)-\sum_{i=1}^{j-1}v_i\right]\right) \right. \\
        &\qquad\qquad \times \; \left. s_b\left(m\pm\left[\frac{N}{2}(-1)^{j+1}h_j-\left(\frac N2-1\right)\left(\frac{iQ}{2}-m\right)+\sum_{i=1}^{j-1}v_i\right]\right) \right\} ~.
\end{split}
\end{equation}
Comparing it with \eqref{eq:freeNS5_5d}, we find the same dependence on the strip and instanton parameters: for each $j$, the two double-sine pairs correspond to the two modulus-squared $\cM$ factors in the five-dimensional free sector. In particular, both retain the dependence on $v_1$, which disappears from the interacting theory.
Using the distributional identification
$s_b(iQ/2)\sim\delta(0)$ (see \cite{Giacomelli:2023zkk}), the divergent factor in
$\cZ_{\rm free}^{\rm 3d}$ matches the $\delta(0)$ that survives in the 5d result.

As discussed at the end of Subsection \ref{subsubsec:shorten_5d}, removing an internal NS5 brane is subtler in the refined topological string computation. Its three-dimensional counterpart, however, follows from the same local analysis. For an internal brane labelled by $i$, the condition $h_i=iQ/2-m$ activates a monopole operator supported on the chain joining the two $U(i)$ nodes through the central $U(k)$ node, with negative flux for odd $i$ and positive flux for even $i$.

The three-dimensional computation is identical to the case with $h_1=iQ/2-m$, and applying the analysis of Subsection \ref{sec:monopolevevs} again gives the magnetic quiver $S_{k-1,N}$ now with 
\begin{equation}
        \vec{w}' = \left(w_1,\cdots, w_{i-1},w_{i+1},\cdots w_k\right) \;\;,\quad \vec{z}' = \left(z_1,\cdots, z_{i-1},z_{i+1},\cdots z_k\right) ~.
\end{equation}
From the dictionary in \eqref{eq:dict}, it is clear that deleting the $i$-th entries means that $v_{i-1}$ and $v_i$ enter the interacting quiver only through the combination $v_{i-1}+v_i$. Equivalently, the corresponding instanton fugacities combine into $Q_{\rm inst}^{(i-1)}Q_{\rm inst}^{(i)}$. This is natural from the five-brane web: removing an internal NS5 brane combines the separations on either side with the new separation parameters being $Q_{\rm top}^{(i-1)}Q_{\rm top}^{(i)}$ and $Q_{\rm bot}^{(i-1)}Q_{\rm bot}^{(i)}$.

\begin{figure}[h!]
  \centering
  \includegraphics[width=\textwidth]{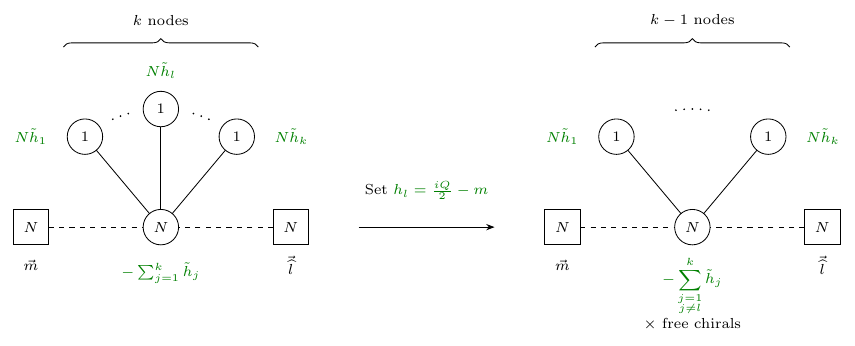}
  \caption{Removal of an NS5 brane in the weak-coupling magnetic quiver. The initial quiver has $k$ abelian nodes with FI parameters $N\widetilde h_j$, coupled to the central $U(N)$ node. Setting $h_i=iQ/2-m$ activates a monopole of the $i$-th abelian node, with flux $-1$ for odd $i$ and $+1$ for even $i$. The VEV removes this node, shifts the central parameter to $-\sum_{j\neq i}\widetilde h_j$, and generates the displayed free chirals.}
  \label{fig:weakNS5rem}
\end{figure}

We conclude this section by discussing the same deformation in the weak-coupling phase of the box theory. Its magnetic quiver is shown on the left in Figure \ref{fig:weakNS5rem}. As we will discuss in the next section, the parameterisation of the weak-coupling magnetic quiver is inherited from that of the strong-coupling magnetic quiver through the large FI deformations corresponding to $v_j\to+\infty$, or equivalently $Q_{\rm inst}^{(j)}\to0$.

In this phase, the $i$-th $U(1)$ node carries the FI parameter $N\widetilde h_i$ and the condition $\tilde h_i=iQ/2-m$ allows the negative (positive) flux monopole to acquire a VEV for odd (even) $i$. Applying the same local analysis using the technique introduced before simply amounts to Higgsing the corresponding $U(1)$ gauge node. The accompanying free sector contributes $s_b(iQ/2)s_b(2m-iQ/2)$ and differs from the free sector generated at strong coupling. 
Thus in this case, removing an NS5 brane and taking the weak coupling limit associated with the chosen ruling commute at the level of the interacting magnetic quiver: either order yields the weak-coupling magnetic quiver of the shortened box.

\subsubsection{Removing one D5 from the web, Higgsing and parameter flow from 3d}
\label{3dremD5}

As discussed at the end of Subsection \ref{subsubsec:shorten_5d}, removing a D5 brane directly in the five-dimensional superconformal index requires resumming the instanton expansion in our chosen ruling. Alternatively, fibre--base duality makes the deformation accessible in a different ruling. This is not helpful, however, when NS5 and D5 branes are removed simultaneously.

The corresponding deformation can nevertheless be studied explicitly in the magnetic quiver, and the resulting parameter flow translated back to five dimensions using the dictionary. This provides an application of the dictionary beyond the deformations for which we can perform a direct comparison of partition functions. Therefore this section is intended to be a display of the potential of this approach rather than another test of the dictionary.

As discussed above, the deformation associated to removing an NS5 brane from the five-brane web, which produces the flow $S_{k,N}\longrightarrow S_{k-1,N}$, is associated to a VEV for a diagonal component for the $U(k)\times U(k)$ bifundamental monopole operator supported on the lower block. Similarly, removing a D5 brane produces the flow $S_{k,N}\longrightarrow S_{k,N-1}$ through a VEV for a diagonal component of a $U(N)\times U(N)$ bifundamental monopole operator supported on the upper block.

For the $a$-th diagonal component, consider the monopole with flux $(-1,\vec 0)$ along the chain extending symmetrically from the central $U(N)$ node to the two $U(a)$ nodes in the $T[U(N)]$ tails, and zero flux elsewhere. The condition associated to turning on a VEV for this monopole in terms of the moment map parameters $\vec m,\vec{\widehat l}$ associated to the two $T[U(N)]$ tails can be expressed as
\begin{equation}\label{eq:D5_vev_33}
    -m_a-\widehat l_a+\sum_{j=1}^k \tilde h_j = k\left(m-\frac{iQ}{2}\right).
\end{equation}
We can study the effect of this VEV by applying the analysis of Subsection \ref{sec:monopolevevs} after exchanging $k\leftrightarrow N$ and by replacing $(\vec w,\vec z)$ with $(\vec m,\vec{\widehat l})$.\footnote{Notice that the central $U(N)$ node of $S_{k,N}$ also carries the FI parameter $-\sum_j\widetilde h_j$, which is absent in Subsection \ref{sec:monopolevevs}. We account for it in the local calculation by shifting both sets of tail variables $\vec m$ and $\vec{\widehat l}$ by $\frac12\sum_j\widetilde h_j$. Together, these shifts reproduce the $\sum_j\widetilde h_j$ term in the VEV condition \eqref{eq:D5_vev_33}.}

\begin{figure}[h!]
  \centering
  \includegraphics[width=\textwidth]{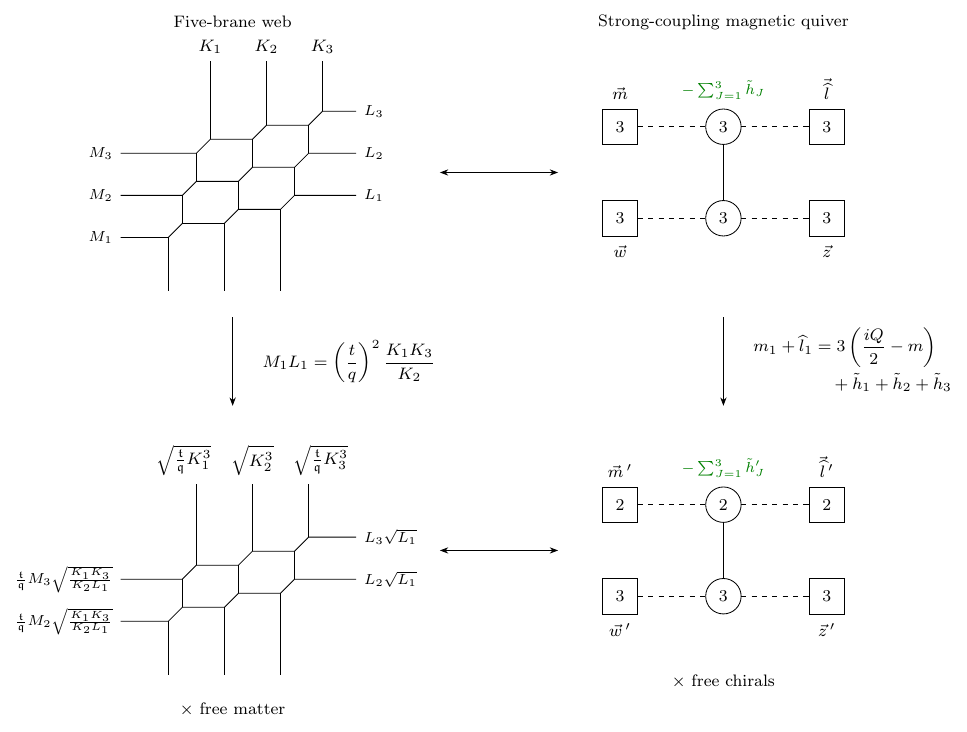}
  \caption{Removal of a D5 brane and the corresponding monopole VEV flow. On the right, a monopole VEV triggers the flow $S_{3,3}\longrightarrow S_{3,2}$. The 5d/3d dictionary determines the corresponding mesonic VEV and the parameter flow in five dimensions, shown on the left, where removing a D5 brane produces the flow $3+SU(3)_0^2+3 \longrightarrow 2+SU(2)_0^2+2$.}
  \label{fig:D5rem}
\end{figure}

We first discuss the effect of this VEV and the procedure to reconstruct the five-dimensional parameter flow for $N=k=3$, choosing $a=1$ (VEV for the first component of the bifundamental monopole). The corresponding deformation is summarised in Figure \ref{fig:D5rem}.
Using the results of Subsection \ref{sec:monopolevevs}, the VEV lowers the rank of all gauge nodes supporting the monopole by one unit, so that the theory flows to $S_{3,2}$, up to free chirals. This flow is depicted on the RHS of Figure \ref{fig:D5rem}. The resulting parameterisation of the $S_{3,2}$ quiver on the bottom right corner of Figure \ref{fig:D5rem} can be expressed as\footnote{The flow deletes the first entries of $\vec m$ and $\vec{\widehat l}$, leaving two vectors that are no longer traceless. We therefore shift the remaining entries by $m_1/2$ and $\widehat l_1/2$, respectively, to restore the $SU(2)$ constraints. The flow also generates an additional FI contribution $iQ/2-m$ for the lower $U(3)$ node. To recover the standard parameterisation, we shift each entry of both the tail parameters $\vec w$ and $\vec z$ by $iQ/4-m/2$. These redefinitions give the primed parameters in \eqref{eq:D5_shift_33} and put the FI parameter of the upper $U(2)$ node in the standard form $-\sum_i \widetilde h_i'$.}
\begin{equation}\label{eq:D5_shift_33}
    \begin{split}
        \vec m\,' &= \Big(m_2+\frac{m_1}{2},\,m_3+\frac{m_1}{2}\Big) \;\;,\quad \vec{\widehat l}\,' = \Big(\widehat l_2+\frac{\widehat l_1}{2},\, \widehat l_3+\frac{\widehat l_1}{2}\Big) \;\;,\quad \widetilde h'_J = \frac{3\widetilde h_J+\frac{iQ}{2}-m}{2} \;\;, \\
        &\vec w\,'= \big(\widetilde h'_1+v_1+v_2,\ \widetilde h'_2+v_2,\ \widetilde h'_3\big) \;\;,\quad \vec z\,'=\big(\widetilde h'_1-v_1-v_2,\ \widetilde h'_2-v_2,\ \widetilde h'_3\big)\ ~.    
    \end{split}
\end{equation}
Note that the instanton parameters $v_i$ remain unchanged. The free chirals generated along this flow have the following contribution to the sphere partition function
\begin{equation}\label{eq:goldstones_3d_D5}
    \cZ_{\rm Free}^{\rm 3d} = s_b\left(\frac{iQ}{2}\right) s_b\left(2m-\frac{iQ}{2}\right) \prod_{i=2}^{3} s_b\!\left(m\pm\left[m_i-m_1+\frac{iQ}{2}-m\right]\right) s_b\!\left(m\pm\left[\widehat l_i-\widehat l_1+\frac{iQ}{2}-m\right]\right) ~.
\end{equation}

In exponentiated variables, the VEV condition in \eqref{eq:D5_vev_33} can be expressed as
\begin{equation}
    M_1L_1 = \left(\frac{\mf t}{\mf q}\right)^2 \frac{K_1K_3}{K_2} ~.
\end{equation}
In the five-dimensional $3+SU(3)_0^2+3$ box theory, this corresponds to a VEV for one component of the long meson connecting the two $SU(3)$ flavour nodes. The primed parameters in \eqref{eq:D5_shift_33} then determine those of the resulting $2+SU(2)_0^2+2$ box theory via the 5d/3d dictionary as depicted on the LHS of Figure \ref{fig:D5rem}
\begin{equation}
    \begin{split}
        M_a' = M_a\sqrt{M_1} \;\;,&\quad L_a' = L_a\sqrt{L_1} \;\;\text{ where }\; a=2,3 \;,\quad Q_{\rm inst}^{(b)}{}' = Q_{\rm inst}^{(b)} \;\;\text{ where }\; b=1,2 ~, \\
       K_1'=&\sqrt{\frac{\mf t}{\mf q}}K_1^{3/2},
\qquad
K_2'=K_2^{3/2},
\qquad
K_3'=\sqrt{\frac{\mf t}{\mf q}}K_3^{3/2}.
    \end{split}
\end{equation}

Thus, the surviving flavour masses are recentered, the instanton fugacities remain unchanged, and the strip parameters undergo a non-trivial transformation.

Following the NS5 removal analysis, we can also infer the five-dimensional free matter contribution from the external parallel D5 leg factors in \eqref{eq:box_dec} as the following:
\begin{equation}\label{eq:goldstone_5d_D5}
    \cZ_{\rm free}^{\rm 5d} = \prod_{i=2}^3 \left| \MM{\frac{M_1}{M_i}} \MM{\frac{\mf t}{\mf q}\frac{L_i}{L_1}} \right|^{-2} ~,
\end{equation}
which originate from the external parallel D5 leg contributions $\cZ_{||}^{\rm left}(\vec M) \cZ_{||}^{\rm right}(\vec L)$ involving the decoupled leg $M_1$. These factors match the parameter-dependent free sector in \eqref{eq:goldstones_3d_D5} factor by factor: each double sine pair corresponds  to an inverse modulus square of $\cM$ with their arguments related by the 5d/3d dictionary.

The analysis extends to general $N$ and $k$. A VEV for the $a$-th component of the bifundamental monopole triggers a flow to the interacting quiver $S_{k,N-1}$ with parameters
\begin{equation}
    \begin{split}
        &\qquad m_b' = m_b+\frac{m_a}{N-1} \;\;,\quad \widehat l_b' = \widehat l_b+\frac{\widehat l_a}{N-1} \;\;\text{ where }\; b\neq a ~, \\
        \widetilde h_j' = &\frac{N\widetilde h_j+\frac{iQ}{2}-m}{N-1} \;\;,\quad w_j' = \frac{N-1}{2}\,\widetilde h_j' + \sum_{l=j}^{k-1} v_l \;\;,\quad z_j' = \frac{N-1}{2}\,\widetilde h_j' - \sum_{l=j}^{k-1} v_l ~,
    \end{split}
\end{equation}
for $j=1,\dots,k$. The free chirals generated along the flow in this case can be expressed as
\begin{equation}\label{eq:goldstones_3d_D5_general}
    \begin{split}
        \cZ_{\rm Free}^{\rm 3d} &= s_b\left(\frac{iQ}{2}\right) s_b\left(2m-\frac{iQ}{2}\right) \\
        &\quad \times \; \prod_{\substack{i=1\\i\neq a}}^{N} s_b\left(m\pm\left(m_i-m_a+\frac{iQ}{2}-m\right)\right) s_b\left(m\pm\left(\widehat l_i-\widehat l_a+\frac{iQ}{2}-m\right)\right) ~.
    \end{split}
\end{equation}

Using the 5d/3d dictionary, the exponentiated VEV condition in this case takes the form
\begin{equation}\label{eq:D5_vev_5d}
    M_a\,L_a^{(-1)^{k+1}} = \left(\frac {\mf t}{\mf q}\right)^{\frac k2} \prod_{j=1}^{k}\left(\sqrt{\frac {\mf t}{\mf q}}\,K_j\right)^{(-1)^{j+1}} ~,
\end{equation}
where the alternating exponents follow from the alternating conventions for the strip parameterisations discussed in Section \ref{subsec:5d_partfn}. Once again, this imposes a VEV for a component of the long meson stretching between the two $SU(N)$ flavour symmetries in the $N+SU(N)_0^{k-1}+N$ box theory. The resulting 5d theory obtained under this flow is the $(N-1)+SU(N-1)_0^{k-1}+(N-1)$ box theory with parameters
\begin{equation}\label{eq:D5_5d_params}
    \boxed{
    \begin{gathered}
        M'_b=M_b\,M_a^{\frac{1}{N-1}} \;\;,\quad L'_b=L_b\,L_a^{\frac{1}{N-1}} \;\;\text{ where }\; b\neq a \;\;,\quad Q_{\rm inst}^{(i)}{}' = Q_{\rm inst}^{(i)} ~, \\[5pt]
        K'_j=\left[ K_j^{\,N}\left(\frac{\mf t}{\mf q}\right)^{\frac{1+(-1)^{j+1}}{2}} \right]^{\frac{1}{N-1}}
        =\begin{cases}
            \left(\frac{\mf t}{\mf q}\right)^{\frac{1}{N-1}} K_j^{\frac{N}{N-1}} &,\;\; j \text{ odd} ~, \\[4pt]
            K_j^{\frac{N}{N-1}} &,\;\; j \text{ even}\ ~.
        \end{cases}
    \end{gathered}
    }
\end{equation}
Thus, removing a D5 brane recentres the surviving flavour masses and transforms the abelian strip parameters, while leaving the instanton fugacities unchanged. The VEV condition \eqref{eq:D5_vev_5d} may additionally be used to eliminate $M_a$ in favour of $L_a$ and $K_j$. 
Up to Weyl permutations that place the removed D5 leg first, the
external parallel D5 leg factors give
\begin{equation}\label{eq:goldstones_5d_D5_general}
    \cZ_{\rm Free}^{\rm 5d} = \prod_{\substack{b=1\\b\neq a}}^{N} \left| \MM{\frac{M_a}{M_b}} \MM{\frac{\mf t}{\mf q}\left( \frac{L_b}{L_a} \right)^{(-1)^{k+1}}} \right|^{-2} ~.
\end{equation}
These factors match the parameter-dependent free sector in \eqref{eq:goldstones_3d_D5_general} factor by factor under the dictionary, as before. The magnetic quiver analysis, therefore, determines both the parameter flow of the interacting five-dimensional theory and the global symmetry charges of the accompanying free sector.

\subsection{Flow between SC and WC: ruling the box}
\label{sec:rule_box}

In this subsection, we study weak-coupling deformations obtained by independently sending the gauge couplings of the five-dimensional $N+SU(N)_0^{k-1}+N$ box theory to zero. We compare these deformations with the corresponding large FI deformations of the associated magnetic quiver $S_{k,N}$, determined by the 5d/3d dictionary.

As discussed earlier, we primarily aim to study deformations that can be studied exactly in both five and three dimensions. On the magnetic quiver side, the FI deformations associated with different rulings of the toric diagram, equivalently different infrared descriptions of the ultraviolet fixed point of the box gauge theory, can be studied efficiently through the techniques developed in Section \ref{sec:3dtechnology}. On the five-dimensional side, however, our calculation is organised around the infrared description as the $N+SU(N)_0^{k-1}+N$ gauge theory, with the corresponding vertical ruling and the fixed choice of preferred direction in the refined topological vertex expansion.

We therefore begin with the weak-coupling deformations associated with this ruling, for which the relevant gauge couplings appear directly through the instantonic fugacities in the refined topological string partition function. We then discuss a deformation associated with a different ruling, illustrating how the magnetic quiver and the 5d/3d parameter dictionary identify a five-dimensional flow that is less transparent in the original gauge theory frame.

\subsubsection{5d computation}
\label{subsubsec:rule_5d}

For simplicity, we begin with the $N+SU(N)_0+N$ theory before extending the discussion to general $k$. Its superconformal index can be obtained from \eqref{eq:Index_SUNkm12N}, by setting $k=2$. The gauge coupling enters the corresponding instanton expansion in the topological string partition function through the instanton counting parameter, $$Q_{\rm inst}\sim e^{-\frac{8\pi^2\beta}{g^2}} ~.$$ Thus the weak-coupling limit $g^2\rightarrow0$ corresponds to $Q_{\rm inst}\rightarrow0$, while the remaining parameters are held fixed.

As explained in Appendix \ref{app:topstringPF}, every base parameter $Q_a$ is proportional to the instanton parameter $Q_{\rm inst}$. Consequently, sending the instanton counting parameter to zero sends every base parameter to 0. This implies that the instanton sum in the associated topological string partition function trivialises since in the limit $Q_a\to0$ only the trivial partitions contribute $\bm\alpha=\bm\emptyset$. Retaining only the empty partition contribution in both copies of the topological string partition function gives the superconformal index
\begin{equation}
    \begin{split}
        \cI^{N+SU(N)+N} (Q_{\rm inst}\to0,\vec M,\vec L&, K_1, K_2) = \frac{1}{N!} \oint \prod_{i=1}^N\frac{{\rm d}A_i}{2\pi iA_i} \times \delta\left( \prod_{i=1}^N A_i \right) \\
        &\times \; \left| M(\mf t,\mf q) \right|^{2(N-1)}
        \; \left| \frac{\cZ_{\rm pert}^{\rm strip} (\vec M, \vec A, K_1) \; \cZ_{\rm pert}^{\rm alt \; strip} (\vec A,\vec L, K_2)}{\prod_{1\leq a<b\leq N} \MM{\frac{M_a}{M_b}} \MM{\frac{\mf t}{\mf q}\frac{L_a}{L_b}} } \right|^2 ~.
    \end{split}
\end{equation}
Therefore the resulting expression retains the perturbative vector and hypermultiplet contributions and their gauge projection, but excludes the instanton and anti-instanton sectors. Physically, this can be understood as the instantons becoming infinitely massive and hence decoupling from the low energy theory.

The same analysis applies to the general $N+SU(N)_0^{k-1}+N$ box theory. There are $k-1$ independent gauge couplings, and the weak coupling regime can be taken for any subset of them. For each selected node $i$, the corresponding perturbative truncation sets $\alpha^{(i)}_a=\bm\emptyset$ for $a=1,\cdots N$ while retaining the instanton sums associated with the other gauge nodes. Applying this to all $k-1$ gauge nodes simultaneously leaves the fully perturbative contribution.

\subsubsection{3d computation}
\label{subsubsec:rule_3d}

\begin{figure}[h!]
  \centering
  \includegraphics[width=0.8\textwidth]{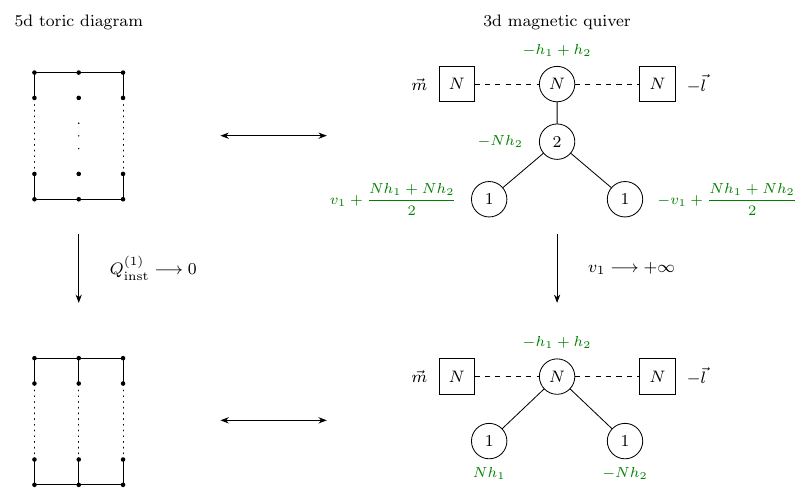}
  \caption{Weak coupling deformation of the $N+SU(N)_0+N$ theory. The left column shows the toric diagram and its vertical ruling associated to the compatible weak coupling limit $Q_{\rm inst}^{(1)}\to0$. The right column shows the corresponding FI deformation $v_1\to+\infty$ of $S_{2,N}$.}
  \label{fig:WCk2}
\end{figure}

We now study the corresponding deformation of the magnetic quiver. For the $N+SU(N)_0+N$ theory, the 5d/3d dictionary maps the weak coupling deformation $Q_{\rm inst}^{(1)}\to0$ to the FI deformation $v_1\to+\infty$ in $S_{2,N}$. Figure \ref{fig:WCk2} depicts the deformation together with the associated vertical ruling of the dual toric diagram.

The divergent FI parameter $v_1$ appears on the two lower $U(1)$ tails with opposite signs, exactly as in the case studied in Figure \ref{fig:FI_ex1} of Subsection \ref{sec:anotherfidef}. Following the discussion there, the resulting quiver is depicted at the bottom right of Figure \ref{fig:WCk2}. Note that the FI parameter of one of the resulting $U(1)$ nodes is the sum of the FI parameters along the original $U(1)-U(2)-U(1)$ sequence of nodes, whereas the second inherits the original $U(2)$ FI parameter. The central $U(N)$ node and the two $T[U(N)]$ tails remain unchanged and the resulting quiver is the corresponding weak-coupling magnetic quiver of the $N+SU(N)_0+N$ theory.

\begin{figure}[h!]
  \centering
  \includegraphics[width=\textwidth]{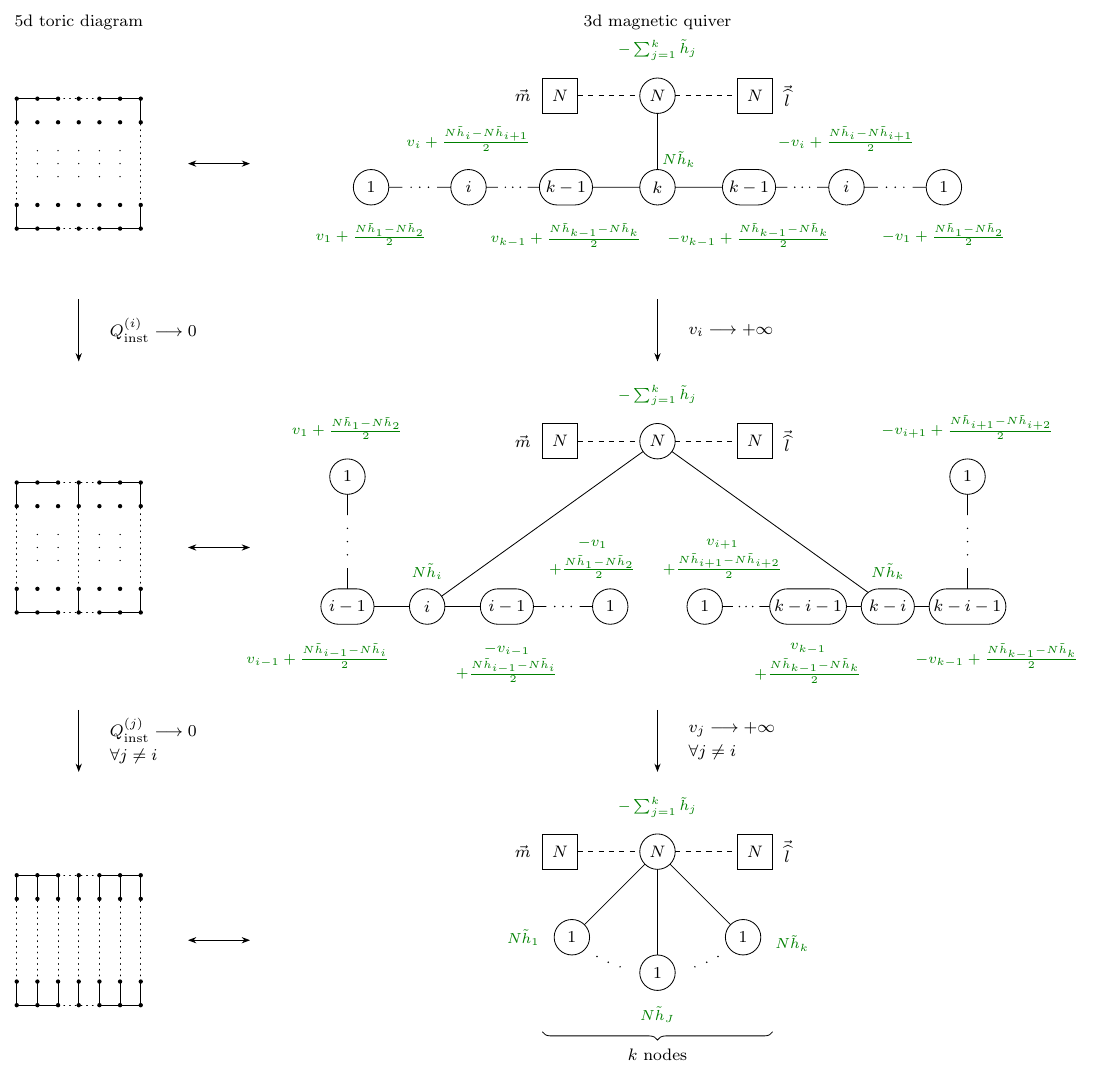}
  \caption{Partial and complete weak-coupling deformations of the box theory, shown through its toric diagram (left) and magnetic quiver (right). The top row corresponds to the unruled toric diagram and hence the strong-coupling magnetic quiver. In the middle row $Q_{\rm inst}^{(i)}\to 0$, equivalently $v_i\to+\infty$, sends the $i$-th gauge coupling to zero. The box divides into groups of $i$ and $k-i$ strips, while the $U(k)$ node splits into $U(i)$ and $U(k-i)$ nodes, each with two corresponding tails. Sending all remaining instanton parameters to zero gives the bottom row: a complete vertical ruling into $k$ strips and a magnetic quiver with $k$ $U(1)$ nodes attached to the central $U(N)$.}
  \label{fig:WCflow_full}
\end{figure}

For the general box theory, the weak-coupling limit can be taken independently at each gauge node. Figure \ref{fig:WCflow_full} displays the successive subdivisions alongside the corresponding FI deformations of $S_{k,N}$. Taking the $i$-th gauge coupling to zero corresponds to $Q_{\rm inst}^{(i)}\to0$ and thus the FI deformation $v_i\to+\infty$ in $S_{k,N}$. At the level of the toric diagram, it introduces a subdivision separating groups of $i$ and $k-i$ strips. On the magnetic quiver side using the technology of section \ref{subsec:FIfromPF} we find that the central $U(k)$ node and its two $T[U(k)]$ tails split into a $U(i)$ node with two $T[U(i)]$ and a $U(k-i)$ node with two $T[U(k-i)]$ tails. Both nodes remain attached to the original central $U(N)$ node, whose two $T[U(N)]$ tails are unaffected, as displayed in the middle in Figure \ref{fig:WCflow_full}.

The enhanced global symmetry in this case is $SU(N)^2\times SU(i)^2\times SU(k-i)^2\times U(1)^2$ with additional enhancement at special ranks. This means that this deformation breaks the original $SU(k)^2\to SU(i)^2\times SU(k-i)^2\times U(1)$. Relative to the original Coulomb symmetry $SU(N)^2\times SU(k)^2\times U(1)$, the rank of the resulting symmetry is reduced by one. The lost Cartan direction is  $v_i$ and the limit $v_i\to+\infty$ corresponds to sending the
instanton fugacity $Q_{\rm inst}^{(i)}$ to zero.

Sending all remaining $Q_{\rm inst}^{(j)}\to0$ describes the five-dimensional theory in the complete weak-coupling frame, in which all $k-1$ gauge couplings vanish. Geometrically, this corresponds to the complete vertical ruling of the toric diagram as shown on the bottom left of Figure \ref{fig:WCflow_full}. The corresponding FI deformations $v_j\to+\infty$ replace the original $U(k)$ node and the two $T[U(k)]$ tails with $k$ $U(1)$ nodes connected to the original $U(N)$ gauge node, as depicted on the bottom right of Figure \ref{fig:WCflow_full}.

\begin{figure}[h!]
  \centering
  \includegraphics[width=\textwidth]{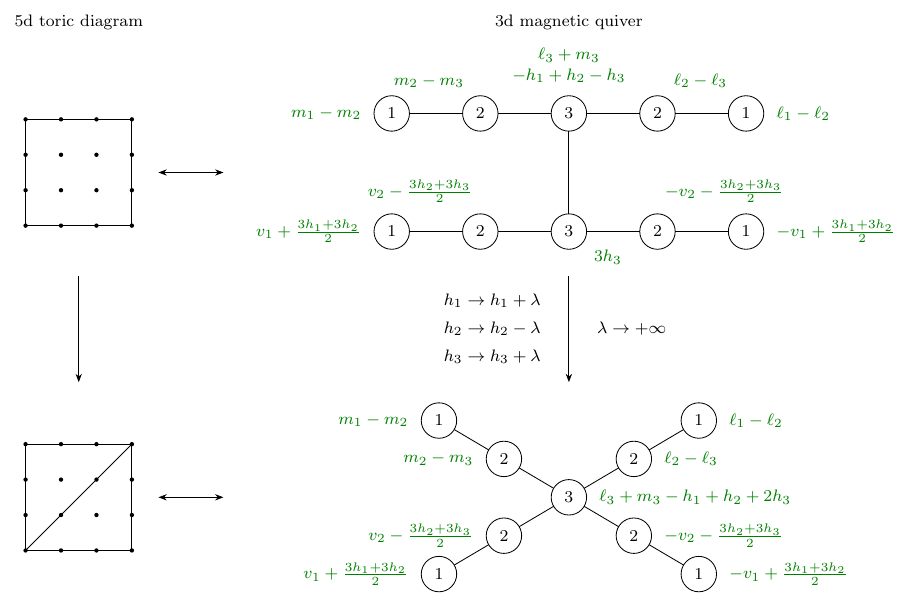}
  \caption{Diagonal ruling of the square box and the corresponding FI deformation of $S_{3,3}$. The left panel shows the diagonal subdivision of the toric diagram. On the right panel, opposite divergent FI parameters combine the two central $U(3)$ nodes into a diagonal $U(3)$ coupled to four $T[U(3)]$ tails.}
  \label{fig:WCclassS}
\end{figure}

We conclude this discussion by considering a deformation associated with a different ruling of the box, in particular one not compatible with the choice of preferred direction in the refined topological vertex computations. Figure \ref{fig:WCclassS} displays the corresponding diagonal subdivision of the toric diagram for $N=k=3$ together with the corresponding weak-coupling magnetic quiver: a star-shaped quiver with four $T[U(3)]$ tails.

We determine the FI deformation leading to this diagonal ruling by applying the  methods of Subsection \ref{subsec:FIfromPF}. The resulting correlated limit, expressed in the untilded variables of Figure \ref{fig:WCclassS}, is
\begin{equation}
    h_1 \longrightarrow h_1+\lambda \;\;,\quad h_2 \longrightarrow h_2-\lambda \;\;,\quad h_3 \longrightarrow h_3+\lambda \;\;,\quad \lambda \longrightarrow +\infty ~.
\end{equation}
Since $\widetilde h_j=(-1)^{j+1}h_j$, this corresponds to $\widetilde h_j\longrightarrow\widetilde h_j+\lambda$. In the vertical ruling corresponding to the infrared gauge theory description $3+SU(3)_0^2+3$, the $h_j$ parameters are associated with the $U(1)_{K_j}$ symmetries that rotate the bifundamental hypermultiplets. Although the diagonal alternative ruling is visible geometrically, its realisation as a correlated mass limit within the parameter space of the vertically ruled gauge theory frame is less immediate. The magnetic quiver description identifies the corresponding FI deformation directly, and  the dictionary identifies the corresponding correlated limit of the vertical-frame parameters.

\subsection{Flavour-mass deformations: deforming the box}
\label{subsec:flavourmass}

Flavour-mass flows and their action on five-brane webs, generalised toric polygons, and magnetic quivers have been studied in \cite{vanBeestBourgetEckhardSchaferNameki2020}. Here we follow the corresponding FI limits directly in the magnetic quiver partition function and compare the inherited parameters with the five-dimensional superconformal index.

Integrating out a five-dimensional flavour modifies the toric polygon, shifts the effective Chern--Simons level, and requires a compensating rescaling of the instanton counting parameter. Through our proposed 5d/3d parameter dictionary \eqref{eq:dict}, the mass limit and instanton rescaling determine a correlated large FI deformation of the magnetic quiver. We follow these operations by comparing the descendant toric geometries with the associated quiver obtained directly from the three-dimensional partition function. The agreement between the resulting three-dimensional theories and the expected strong-coupling magnetic quiver provides a non-trivial test of both the flavour-mass and instanton sectors of the dictionary.

\subsubsection{5d computation}
\label{subsubsec:5dflavourmass}

We begin with the $N+SU(N)_0+N$ box theory. The same analysis applies locally to the general box theory $N+SU(N)_0^{k-1}+N$ since integrating out an external flavour affects only the gauge node adjacent to the corresponding flavour symmetry.

It is convenient to perform the calculation using the instantonic contributions in the Nekrasov partition function. In this representation, both the induced Chern--Simons coupling and the required rescaling of the instanton counting parameter are manifest. The Nekrasov and refined topological string partition functions are equivalent once the base \Kah parameters are expressed in terms of the instanton fugacities, as discussed in Appendix \ref{app:topstringPF}. The instanton partition function of this gauge theory can then be expressed as
\begin{equation}
    Z^{\rm Nek}_{\rm inst}(Q_{\rm inst},\vec M,\vec N,\vec L,K_1,K_2) = \sum_{\bm\alpha} \left[\prod_{a=1}^N Q_{\rm inst}^{|\alpha_a|}\right] \frac{ \prod_{i=1}^N\prod_{a=1}^N \Nek_{\emptyset\alpha_a}\left(\frac{M_iN_a}{K_1}\right) \Nek_{\emptyset\alpha_a}\left(\frac{N_aL_i}{K_2}\right) }{ \prod_{a,b=1}^N \Nek_{\alpha_a\alpha_b} \left(\frac{N_b}{N_a}\right) } ~.
\end{equation}
Here $\vec M$ and $\vec L$ parameterise the $SU(N)$ flavour symmetries, while $\vec N$ parameterises the $SU(N)$ gauge symmetry and hence satisfy $\prod_{i=1}^N M_i = \prod_{i=1}^N N_i = \prod_{i=1}^N L_i = 1$.

Since the variables $M_i$ and $L_i$ are subject to those $SU(N)$ constraints, a single physical flavour mass cannot be sent to infinity while keeping the remaining masses fixed. We therefore combine them with the $U(1)$ parameters $K_1,K_2$ and define
\begin{equation}
    \widetilde M_i = \frac{M_i}{\sqrt{\frac{\mf t}{\mf q}}K_1} = e^{-\beta\,\widetilde m_i} \;\;,\quad \widetilde L_i = \frac{L_i}{\sqrt{\frac{\mf t}{\mf q}}K_2} = e^{-\beta\,\widetilde l_i} ~,
\end{equation}
which are the natural $U(N)$ flavour fugacities and correspond to\footnote{The $SU(N)$ tracelessness condition on the un-tilded variables implies $Nh_1=-\sum_{i=1}^N \widetilde m_i$ and $Nh_2=-\sum_{i=1}^N \widetilde l_i$.}
\begin{equation}
    \widetilde m_i = m_i - h_1 \;\;,\quad \widetilde l_i = l_i - h_2 ~.
\end{equation}
Integrating out a fundamental hypermultiplet corresponds to either limit
\begin{equation}
    \widetilde m_i \longrightarrow +\infty \;\Longleftrightarrow\; \widetilde M_i \longrightarrow 0 \;\;,\quad \widetilde m_i \longrightarrow -\infty \;\Longleftrightarrow\; \widetilde M_i \longrightarrow \infty ~.
\end{equation}
The two choices induce Chern--Simons coupling of opposite signs. For fixed partition $\alpha$, the relevant asymptotics of the constituent special functions in these limits are
\begin{equation}\label{eq:NekAsym}
    \begin{split}
        \Nek_{\emptyset\alpha}(Q) &\sim Q^{-\frac{|\alpha|}{2}} \left(\frac{\mf t}{\mf q}\right)^{\frac{|\alpha|}{4}} \mf t^{-\frac{||\alpha^t||^2}{4}} \mf q^{\frac{||\alpha||^2}{4}} \;\;\quad\text{ as }\;\; Q\longrightarrow0 ~, \\
        \Nek_{\emptyset\alpha}(Q) &\sim (-1)^{|\alpha|} Q^{\frac{|\alpha|}{2}} \left(\frac{\mf t}{\mf q}\right)^{-\frac{|\alpha|}{4}} \mf t^{\frac{||\alpha^t||^2}{4}} \mf q^{-\frac{||\alpha||^2}{4}} \;\;\quad\text{ as }\;\; Q\longrightarrow\infty ~,
    \end{split}
\end{equation}
here $\alpha^t$ denotes the transposed partition and $||\alpha||^2=\sum_i\alpha_i^2$.

We first consider the deformation
\begin{equation}
    \widetilde M_1 = \Lambda \;\;,\quad \Lambda \longrightarrow +\infty ~.
\end{equation}
Using the asymptotic formula for the Nekrasov special function in \eqref{eq:NekAsym}, we obtain
\begin{equation}
    \begin{split}
        \lim_{\Lambda\to+\infty} Z^{\rm Nek}_{\rm inst} \Big |_{\widetilde M_1=\Lambda} &\sim \lim_{\Lambda\to+\infty} \sum_{\bm\alpha} \left[ \prod_{a=1}^N \left(-Q_{\rm inst}\sqrt\Lambda\right)^{|\alpha_a|} \right] \times \textcolor{blue}{\prod_{a=1}^N N_a^{\frac{|\alpha_a|}{2}} \mf t^{\frac{||\alpha_a^t||^2}{4}} \mf q^{-\frac{||\alpha_a||^2}{4}}} \\
        &\qquad \times\; \frac{ \prod_{i=2}^N\prod_{a=1}^N \Nek_{\emptyset\alpha_a} \left(\sqrt{\frac{\mf t}{\mf q}}\widetilde M_iN_a\right) \prod_{i=1}^N\prod_{a=1}^N \Nek_{\emptyset\alpha_a}\left(\sqrt{\frac{\mf t}{\mf q}}N_a \widetilde L_i\right) }{ \prod_{a,b=1}^N \Nek_{\alpha_a\alpha_b}\left(\frac{N_b}{N_a}\right) } ~.
    \end{split}
\end{equation}

This is precisely the Nekrasov instanton partition function of the $(N-1)+SU(N)_{+\frac12}+N$. In particular, the factor highlighted in \textcolor{blue}{blue} above indicates the contribution coming from the Chern--Simons level $+1/2$. To obtain a finite instanton expansion, it is clear that the infinite mass limit must be accompanied with a simultaneous redefinition
\begin{equation}
    Q_{\rm inst} \longrightarrow \frac{Q_{\rm inst}}{\sqrt{\Lambda}} ~.
\end{equation}

The opposite limit works very similarly. To do so, we first set $\widetilde M_1=\Lambda^{-1}$ and then take the limit $\Lambda\to\infty$ and obtain
\begin{equation}
    \begin{split}
        \lim_{\Lambda\to+\infty} Z^{\rm Nek}_{\rm inst} \Big |_{\widetilde M_1=\Lambda^{-1}} &\sim \lim_{\Lambda\to+\infty} \sum_{\bm\alpha} \left[ \prod_{a=1}^N \left(Q_{\rm inst}\sqrt\Lambda\right)^{|\alpha_a|} \right] \times \textcolor{red}{\prod_{a=1}^N N_a^{-\frac{|\alpha_a|}{2}} \mf t^{-\frac{||\alpha_a^t||^2}{4}} \mf q^{\frac{||\alpha_a||^2}{4}}} \\
        &\qquad \times\; \frac{ \prod_{i=2}^N\prod_{a=1}^N \Nek_{\emptyset\alpha_a} \left(\sqrt{\frac{\mf t}{\mf q}}\widetilde M_iN_a\right) \prod_{i=1}^N\prod_{a=1}^N \Nek_{\emptyset\alpha_a}\left(\sqrt{\frac{\mf t}{\mf q}}N_a \widetilde L_i\right) }{ \prod_{a,b=1}^N \Nek_{\alpha_a\alpha_b}\left(\frac{N_b}{N_a}\right) } ~.
    \end{split}
\end{equation}
This is indeed the Nekrasov instanton partition function of $(N-1)+SU(N)_{-\frac12}+N$, where now the factor in \textcolor{red}{red} corresponds to the contribution with Chern--Simons level $-1/2$. The same rescaling of the instanton counting parameter is required to produce a finite instanton expansion.

Summarising, a flavour with mass tending to $+\infty$ shifts the Chern--Simons level by $-1/2$, while a flavour with mass tending to $-\infty$ shifts it by $+1/2$. In both cases, each integrated out flavour requires a simultaneous rescaling of the instanton counting parameter $Q_{\rm inst}\longrightarrow Q_{\rm inst}\Lambda^{-1/2}$. 
Table \ref{tab:mass_limits} summarises the representative one- and two-flavour mass limits, up to exchanging the two flavours, together with the resulting five-dimensional theories and the instanton-parameter rescalings required for a finite Nekrasov instanton expansion.
\begin{table}[htbp]
    \centering
    \renewcommand{\arraystretch}{1.5}
    \setlength{\tabcolsep}{12pt}
    \begin{tabular}{@{}cccc@{}}
        \toprule
        $\widetilde M_1$
        & $\widetilde M_2$
        & Resulting theory
        & Instanton rescaling \\
        \midrule
        $\Lambda$ & $1$
        & $(N-1)+SU(N)_{+\frac12}+N$
        & $Q_{\mathrm{inst}}\longrightarrow Q_{\mathrm{inst}}/\sqrt{\Lambda}$ \\

        $\Lambda^{-1}$ & $1$
        & $(N-1)+SU(N)_{-\frac12}+N$
        & $Q_{\mathrm{inst}}\longrightarrow Q_{\mathrm{inst}}/\sqrt{\Lambda}$ \\

        \addlinespace
        $\Lambda$ & $\Lambda$
        & $(N-2)+SU(N)_{+1}+N$
        & $Q_{\mathrm{inst}}\longrightarrow Q_{\mathrm{inst}}/\Lambda$ \\

        $\Lambda^{-1}$ & $\Lambda^{-1}$
        & $(N-2)+SU(N)_{-1}+N$
        & $Q_{\mathrm{inst}}\longrightarrow Q_{\mathrm{inst}}/\Lambda$ \\

        $\Lambda$ & $\Lambda^{-1}$
        & $(N-2)+SU(N)_0+N$
        & $Q_{\mathrm{inst}}\longrightarrow Q_{\mathrm{inst}}/\Lambda$ \\
        \bottomrule
    \end{tabular}
    \caption{Flavour mass limits of $N+SU(N)_0+N$ gauge theory as $\Lambda\to+\infty$, showing the resulting theory and the instanton rescaling required for a finite expansion.}
    \label{tab:mass_limits}
\end{table}

\subsubsection{3d computation}
\label{subsubsec:deform_3d}

We now implement the corresponding FI deformations of the strong-coupling magnetic quiver directly in its $S^3_b$ partition function, using the large FI deformation technology developed in Section \ref{sec:3dtechnology}. The five-dimensional mass limits and instanton rescaling fix the deformation through the dictionary, while the three-dimensional calculation identifies the contributing infrared vacua and their leading asymptotics. To relate the additive and multiplicative parameters, we write $\Lambda=e^{\beta\lambda}$ with $\lambda\to+\infty$. The associated instanton fugacity rescaling then corresponds to $v\to v+\lambda/2$.

\begin{figure}[h!]
  \centering
  \includegraphics[width=\textwidth]{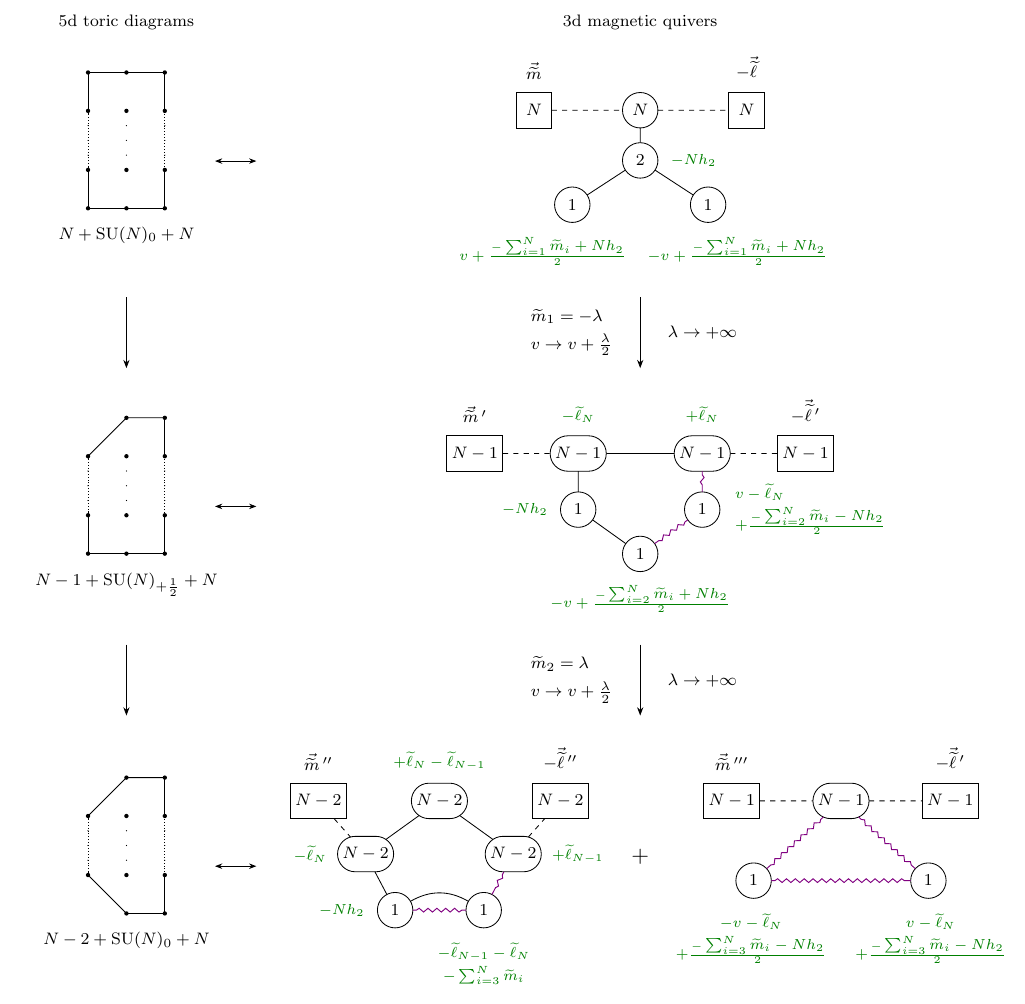}
  \caption{Flavour mass deformations of $N+SU(N)_0+N$, shown through toric diagrams (left) and magnetic quivers (right). Successively decoupling flavours with negative and positive masses gives $(N-1)+SU(N)_{+1/2}+N$ and $(N-2)+SU(N)_0+N$. The final two quivers describe the two Higgs branch cones at strong coupling; they appear with the same leading asymptotic and hence their contributions are summed. The purple links denote hypermultiplets with noncanonical charge assignments.}
  \label{fig:MassOpp}
\end{figure}

\begin{figure}[h!]
  \centering
  \includegraphics[width=\textwidth]{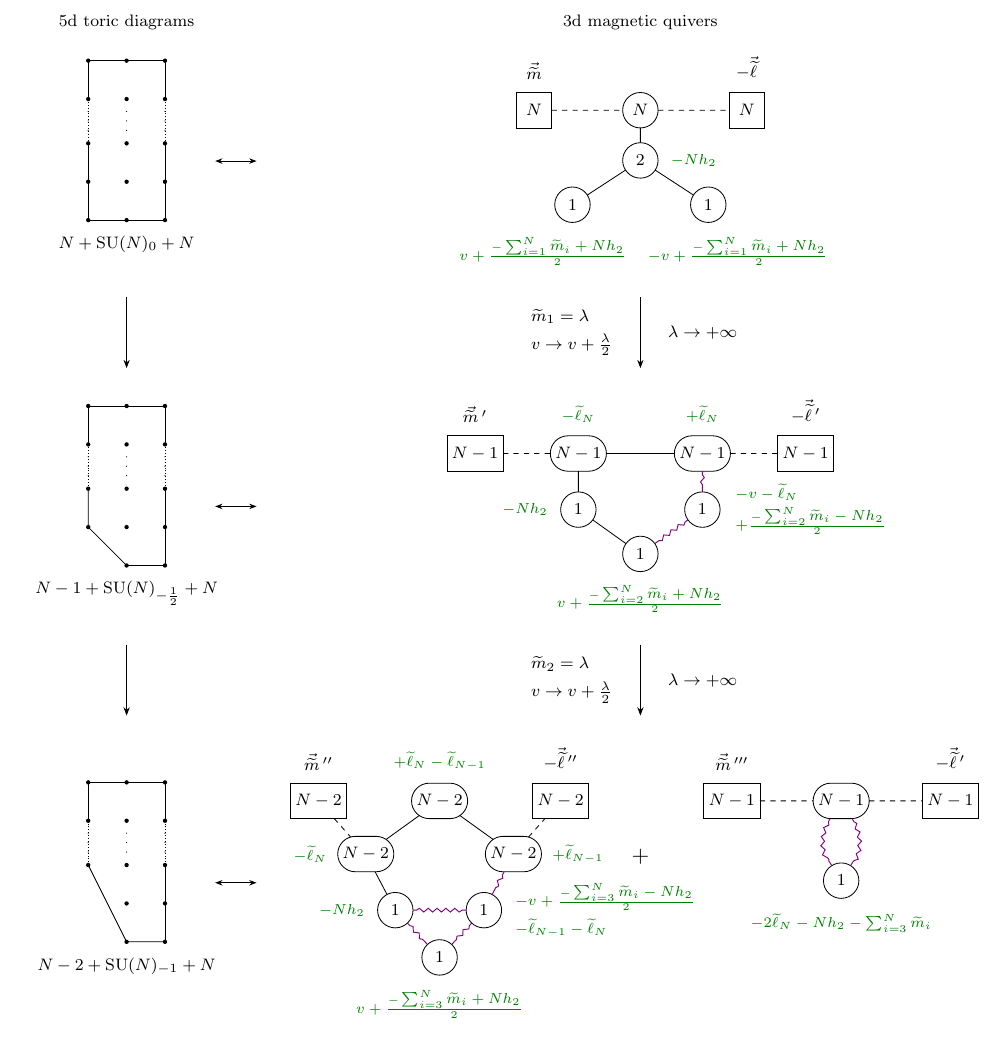}
  \caption{Flavour mass deformations of $N+SU(N)_0+N$, shown through toric diagrams (left) and magnetic quivers (right). Successively decoupling two positive mass flavours gives $(N-1)+SU(N)_{-1/2}+N$ and $(N-2)+SU(N)_{-1}+N$. The final two quivers describe the two Higgs branch cones; they appear with the same leading asymptotic and hence their contributions are summed.  The purple links denote hypermultiplets with noncanonical charge assignments.}
  \label{fig:MassSam}
\end{figure}

We organise the deformations listed in Table \ref{tab:mass_limits} into two sequences displayed in Figures \ref{fig:MassOpp} and \ref{fig:MassSam}. The changes in the toric polygons record the five-dimensional flavour decouplings, while the quiver flows implement the correlated FI deformations fixed by the dictionary. When the strong-coupling Higgs branch of the resulting five-dimensional theory has multiple cones, each cone is described by a magnetic quiver \cite{Cabrera:2018jxt}.

In both figures, we use
\begin{equation}
    \begin{split}
        \vec{\widetilde m}\,{}' = (\widetilde m_2,\dots,\widetilde m_N) \;\;&,\quad \vec{\widetilde m}\,{}'' = (\widetilde m_3,\dots,\widetilde m_N) \;\;,\quad \vec{\widetilde m}\,{}''' = (\widetilde m_3,\dots,\widetilde m_N,\widetilde l_N) ~, \\
        \vec{\widetilde l}' &= (\widetilde l_1,\dots,\widetilde l_{N-1}) \;\;,\quad \vec{\widetilde l}'' = (\widetilde l_1,\dots,\widetilde l_{N-2}) ~.
    \end{split}
\end{equation}
The wavy links do not denote a different type of edge in the underlying magnetic quiver. They record the non-canonical $R$ and axial charge assignments inherited by the $\mathcal N=2^*$-refined partition function after the large FI limit. These inherited charge assignments are essential when we apply further deformations or gauge global symmetries.\footnote{A similar feature appears in the mirror duals of bad special-unitary theories studied in \cite{Giacomelli:2024laq}: their partition functions decompose into frames with constrained parameters, even though the frames contain no explicit delta functions.}

Starting from the $S_{2,N}$ magnetic quiver in the first row of Figure \ref{fig:MassOpp}, we take
\begin{equation}
    \widetilde m_1 = -\lambda \;\;,\quad v \longrightarrow v+\frac{\lambda}{2} \;\;\quad\text{ with }\;\; \lambda \longrightarrow +\infty ~.
\end{equation}
This is the additive counterpart of $\widetilde M_1=\Lambda\to+\infty$, together with the required instanton rescaling $Q_{\rm inst}\to Q_{\rm inst}/\sqrt{\Lambda}$. Applying the technology of section \ref{subsec:FIfromPF} to implement large FI deformations we find the  quiver shown on the second line of Figure \ref{fig:MassOpp}, which is the strong-coupling magnetic quiver of the $$(N-1)+SU(N)_{+1/2}+N \quad \text{theory}$$ in agreement with the first line of Table \ref{tab:mass_limits}. Notice that the end point of this flow gives 
hypermultiplets with non-canonical charge assignments which we indicate with purple wavy links.

We may then integrate out a second flavour with the opposite mass sign
\begin{equation}
    \widetilde m_2 = \lambda \;\;,\quad v \longrightarrow v+\frac{\lambda}{2} \;\;\quad\text{ with }\;\; \lambda \longrightarrow +\infty ~.
\end{equation}
In this limit, the sphere partition function decomposes into the sum of the two magnetic quiver contributions displayed in the final row of Figure \ref{fig:MassOpp}. 
The two infrared vacua share the same leading asymptotics and therefore must be summed, as in the second example of subsection \ref{sec:anotherfidef}.

The two contributions describe the two hyperk\"ahler cones describing the strong-coupling Higgs branch of $$(N-2)+SU(N)_0+N \quad \text{theory ,}$$ whose toric polygon is displayed at the bottom left of Figure \ref{fig:MassOpp}. Note that only one of these contributions depends on the instantonic parameter $v$, and thus the other magnetic quiver describes the Higgs branch cone insensitive to the corresponding weak-coupling deformations.

The second sequence, shown in Figure \ref{fig:MassSam}, integrates out two flavours both with positive masses. This is obtained by taking
\begin{equation}
    \widetilde m_1 = \lambda \;\;,\quad v \longrightarrow v+\frac{\lambda}{2} \;\;\quad\text{ with }\;\; \lambda \longrightarrow +\infty ~.
\end{equation}
This gives the strong-coupling magnetic quiver of the 5d $(N-1)+SU(N)_{-1/2}+N$ theory on the second line. Its parametrisation differs from that of the intermediate quiver in the opposite sign sequence in Figure \ref{fig:MassOpp} by $v\to-v$. We then take
\begin{equation}
    \widetilde m_2 = \lambda \;\;,\quad v \longrightarrow v+\frac{\lambda}{2} \;\;\quad\text{ with }\;\; \lambda \longrightarrow +\infty ~,
\end{equation}
obtaining the two magnetic quivers in the last row of Figure \ref{fig:MassSam}. These describe the two hyperk\"ahler cones in the strong-coupling Higgs branch of $$(N-2)+SU(N)_{-1}+N \quad \text{theory ,}$$ once again in agreement with Table \ref{tab:mass_limits}.

In summary, for each of the flavour mass limits considered here, we implement the precise large FI deformation of the strong-coupling magnetic quiver dictated by the five-dimensional flow via the 5d/3d dictionary in \eqref{eq:dict}. In each case, the resulting three-dimensional theory is exactly the strong-coupling magnetic quiver of the corresponding five-dimensional theory listed in Table \ref{tab:mass_limits}.

\subsubsection*{FI deformations from MQ of $3+SU(3)_0+3$ to pure $SU(3)_0$}

\begin{figure}[htbp]
  \centering
  \includegraphics[width=0.93\textwidth]{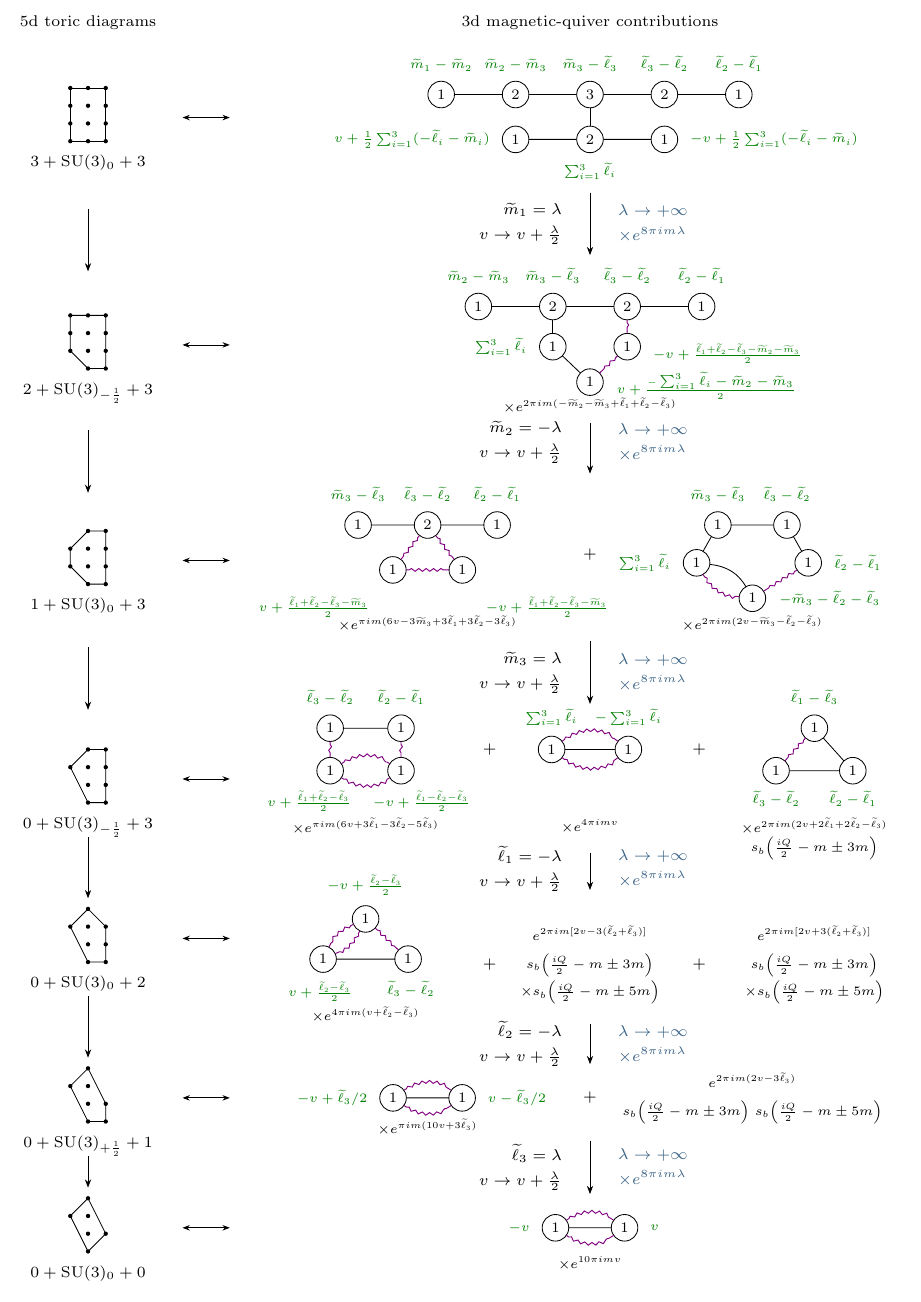}
  \caption{Successive flavour decouplings from $3+SU(3)_0+3$ to pure $SU(3)_0$, shown through toric diagrams (left) and 
  frames produced by the implementation of the corresponding FI deformations. Each mass limit includes the shift $v\to v+\lambda/2$.  Purple links denote hypermultiplets with noncanonical charge assignments.}
  \label{fig:MassCascade}
\end{figure}

We next apply the same procedure to the $3+SU(3)_0+3$ theory, giving mass to its six flavours one at a time until we obtain pure $SU(3)_0$ SYM. Figure \ref{fig:MassCascade} shows the toric geometry at each step alongside the contributions obtained by implementing the corresponding FI deformations in the magnetic-quiver partition function.

At each step, the dictionary fixes both the flavour-mass limit and the accompanying shift of the instanton parameter. We take the six limits in the following order:
\begin{equation}
    \widetilde m_1=\lambda,\qquad
    \widetilde m_2=-\lambda,\qquad
    \widetilde m_3=\lambda,\qquad
    \widetilde l_1=-\lambda,\qquad
    \widetilde l_2=-\lambda,\qquad
    \widetilde l_3=\lambda,
    \qquad \lambda\to+\infty,
\end{equation}
with the appropriate instanton-parameter shift at each step. The first two deformations are the $N=3$ specialisations of the flows discussed in connection with Figure \ref{fig:MassOpp}. We now follow the remaining steps and list the frames produced by the partition-function method of Subsection \ref{subsec:FIfromPF}.

The third deformation produces three frames, displayed in the fourth row of Figure \ref{fig:MassCascade}: two interacting quivers and a third contribution represented by the triangular quiver. The fourth deformation again produces three frames, now comprising one interacting quiver and two Wess--Zumino models. After the fifth deformation, one interacting quiver and one Wess--Zumino model remain. The other Wess--Zumino contribution has a different leading asymptotic phase at this step and therefore drops out of the displayed result. The final deformation removes the remaining Wess--Zumino contribution, leaving a single interacting frame.

The frames shown in Figure \ref{fig:MassCascade} are the distinct contributions we obtain by applying the distributional identity \eqref{eq:masterFIdef}. In particular, we distinguish the additional frames by the constraints generated in the calculation; their physical interpretation requires a separate comparison with the brane-web analysis.

The first two deformations agree with that analysis \cite{Bourget:2019rtl}. After two flavours have been removed, the two interacting frames reproduce the magnetic quivers of the two Higgs-branch cones of the $1+SU(3)_0+3$ theory. After the third deformation, the first two frames agree with the magnetic quivers associated with $SU(3)_{-1/2}+3$, while our calculation also produces a third frame. This extra frame can be obtained by further Higgsing the first one. In the partition function, the corresponding operation pinches an integration contour and produces an accompanying free sector. The extra frame thus describes a vacuum on the moduli space of the first frame, which may explain why it is not listed as an independent magnetic quiver in the brane-web analysis. Likewise, at the fourth and fifth steps, the interacting frame agrees with the expected magnetic quiver, while the additional Wess--Zumino contributions arise through further Higgsings. At the final step, the single interacting frame agrees with the expected magnetic quiver of pure $SU(3)_0$.

Our computation therefore reproduces the magnetic quivers found in \cite{Bourget:2019rtl}, while also yielding additional frames related to them by Higgsing. A further reason to retain these frames comes from the weak-coupling limit. At each stage of the cascade, the dictionary maps this limit to a large-FI deformation of the magnetic quivers. We denote its parameter by $\omega$ to distinguish it from the flavour-mass limits above, and take $v=\omega\to+\infty$. The resulting contributions are expected to reproduce the mirror description of three-dimensional $\mathcal N=4$ $SU(3)$ SQCD with the corresponding number $F$ of flavours. For $F\leq4$, this theory is bad, and its partition function has several frames \cite{Giacomelli:2024laq}.\footnote{This comparison assumes that the weak-coupling limit and the flavour-mass deformations commute.}

Applying the large-$v$ limit to the contributions in each row of Figure \ref{fig:MassCascade}, we find that each flows to a frame of the corresponding bad $SU(3)$ SQCD partition function. The additional strong-coupling frames are needed to recover all the frames in this weak-coupling limit.

\clearpage

\acknowledgments
We thank Antoine Bourget, Noppadol Mekareeya, and  Matteo Sacchi for discussions and comments on the draft, and in particular Riccardo Comi and Andrea Sangiovanni for discussions and collaborations on related projects.

\appendix

\section{Special functions}
\label{app:special_func}

In this appendix, we collect the special functions and identities used in the topological string and Nekrasov partition functions. Throughout, we use the exponentiated $\Omega$ deformation parameters
\begin{equation}\label{eq:Omegadef_Region}
    \mf q=e^{-\beta\epsilon_1} \;\;,\quad \mf t=e^{\beta\epsilon_2} \quad\text{such that}\quad |\mf t|>1 \;,\; |\mf q|<1 ~.
\end{equation}
The five-dimensional superconformal index features a matrix integral of two copies of the topological string partition function. We denote this operation by a modulus square where throughout we follow the following definition:
\begin{equation}\label{eq:mod_square}
    \Big|\, f(U_1,U_2,\dots;\mf t,\mf q) \,\Big|^2 = f(U_1,U_2,\dots;\mf t,\mf q) \, f(U_1^{-1},U_2^{-1},\dots;\mf t^{-1},\mf q^{-1}) ~.
\end{equation}
The notation denotes inversion of the fugacities, with the appropriate analytic continuation of the second factor.

In the region of interest \eqref{eq:Omegadef_Region}, we define the following generalised refined MacMahon function appearing in the perturbative contributions of the topological string partition function
\begin{equation}\label{eq:Mfun_def}
    \MM{U;\mf t, \mf q} \coloneq \prod_{i,j=1}^\infty (1-U\,\mf t^{-i}\mf q^j) ~.
\end{equation}
For generic deformation parameters in the chosen region, this function has zeroes located at $U=\mf t^{m_1} \mf q^{-m_2}$ with $m_1,m_2\in\mathbb Z_{\geq 1}$. We write $\MM{U}\equiv\MM{U;\mf t,\mf q}$ when the deformation parameters are understood. The combination entering the five-dimensional superconformal index is:
\begin{equation}\label{eq:Msquare_def}
    |\MM{U}|^2 \equiv | \MM{U;\mf t,\mf q} |^2 = \MM{U;\mf t,\mf q} \, \MM{U^{-1};\mf t^{-1},\mf q^{-1}} ~.
\end{equation}
After analytic continuation, the second factor admits the following product representation:
\begin{equation}\label{eq:Minvfun_def}
    \MM{U^{-1};\mf t^{-1}, \mf q^{-1}} = \prod_{i,j=1}^\infty (1-U^{-1} \mf t^{1-i}\mf q^{j-1}) ~.
\end{equation}
Its zeroes are located at $U=\mf t^{1-m_1}\mf q^{m_2-1}$ with $m_1,m_2\in\mathbb Z_{\geq 1}$, or equivalently at $U=\mf t^{-n_1}\mf q^{n_2}$ with $n_1,n_2\in\mathbb Z_{\geq0}$. Therefore the zeroes of the squared or glued function $|\MM{U}|^2$ form the two families
\begin{equation}\label{eq:Msq_zeroes}
    (\bar P) :\; U = \mf t^{-n_1}\mf q^{n_2} \quad,\quad (P) :\; U = \mf t^{n_1+1} \mf q^{-n_2-1} \quad\text{ where } n_1,n_2\in\mathbb Z_{\geq0} ~. 
\end{equation}
The product representations also imply the reflection identity
\begin{equation}\label{eq:Minv_identity}
    \left| \MM{U} \right|^2 = \left| \MM{\frac{\mf t}{\mf q}U^{-1}} \right|^2 ~.
\end{equation}
To remove the zero at $U=1$, we define
\begin{equation}\label{eq:regpart_def}
    R(U;\mf t,\mf q) = \frac{|\MM{U;\mf t,\mf q}|^2}{1-U^{-1}} ~.
\end{equation}
The refined MacMahon factor entering the gauge contribution to the superconformal index can then be defined by
\begin{equation}\label{eq:Mstr_def}
    | M(\mf t,\mf q) |^2 = \lim_{U\rightarrow 1} \frac{|\MM{U;\mf t,\mf q}|^2}{1-U^{-1}} = | \MM{\mf q^{-1};\mf t,\mf q} |^2 ~,
\end{equation}
and will also accompany the residue resulting from contour pinching in the main text.

For an integer partition $\lambda$, we define
\begin{equation}
    |\lambda| \coloneq \sum_{i=1}^{\ell(\lambda)} \lambda_i \;,\quad ||\lambda||^2 \coloneq \sum_{i=1}^{\ell(\lambda)} \lambda_i^2 ~,
\end{equation}
where $\ell(\lambda)$ is the number of rows of its Young diagram. The transposed partition is denoted by $\lambda^t$, and $(i,j)\in\lambda$ labels the box in row $i$ and column $j$.

The combinatorial functions entering the instantonic contribution to the topological string partition function take the form
\begin{equation}\label{eq:ZNfun_def}
    \begin{split}
        \tilde{Z}_\nu(\mathfrak{t}, \mathfrak{q}) \coloneq& \prod_{i=1}^{\ell(\nu)} \prod_{j=1}^{\nu_i} \left(1 - \mathfrak{t}^{\nu_j^t - i + 1} \mathfrak{q}^{\nu_i - j}\right)^{-1} ~, \\
        \cN_{\lambda\mu}(U) \equiv \cN_{\lambda\mu}(U;\mf t,\mf q) \coloneq& \prod_{i,j}^\infty \frac{1-U\,\mf t^{-\lambda_j^t+i-1}\mf q^{-\mu_i+j}}{1-U\,\mf t^{i-1}\mf q^j} \\
        =& \prod_{(i,j)\in\lambda} (1-U\,\mf t^{\mu_j^t-i}\mf q^{\lambda_i-j+1}) \prod_{(i,j)\in\mu} (1-U\,\mf t^{-\lambda_j^t+i-1}\mf q^{-\mu_i+j}) ~.
    \end{split}
\end{equation}
These functions obey
\begin{equation}\label{eq:Ntrivial_identity}
    \cN_{\lambda\emptyset} \left(\frac{\mf t}{\mf q}\right) = \cN_{\emptyset\lambda} (1) = \delta_{\lambda,\emptyset} ~.
\end{equation}
To relate these expressions to the Nekrasov instanton partition function, we introduce
\begin{equation}
    \begin{split}
        \Nek_{\lambda\mu}\left(e^{-\beta u};e^{\beta\epsilon_2},e^{-\beta\epsilon_1}\right) =& \prod_{(i,j)\in\lambda} 2\;{\rm sinh} \frac\beta2\left[u+\epsilon_1(\lambda_i-j+1)+\epsilon_2(i-\mu_j^t)\right] \\
        &\;\times \prod_{(i,j)\in\mu} 2\;{\rm sinh} \frac\beta2\left[u+\epsilon_1(j-\mu_i)+\epsilon_2(\lambda_j^t-i+1)\right] ~.
    \end{split}
\end{equation}
For $U\coloneq e^{-\beta u}$, the two normalisations are related by
\begin{equation}\label{eq:NNek_rel}
    \cN_{\lambda\mu}(U;\mf t,\mf q) = \left( U\sqrt{\frac{\mf q}{\mf t}} \right)^{\frac{|\lambda|+|\mu|}{2}} \mf t^{\frac{||\mu^t||^2-||\lambda^t||^2}{4}} \mf q^{\frac{||\lambda||^2-||\mu||^2}{4}} \,\Nek_{\lambda\mu}(U;\mf t,\mf q) ~,
\end{equation}
where we again suppress the $\Omega$-deformation parameters when there is no ambiguity.

\section{Topological string partition functions}
\label{app:topstringPF}

In this appendix, we review the gluing of strip partition functions and their relation to the Nekrasov instanton partition functions. We make explicit the alternating strip parameterisation and the relation between the geometric base parameters and the instanton counting parameters.

\subsubsection*{Gluing strips}

\begin{figure}[h!]
    \centering
    \includegraphics[width=0.3\linewidth]{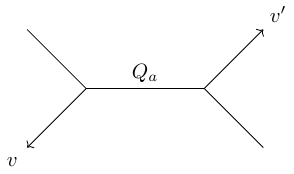}
    \caption{Local five-brane configuration defining the framing number of the internal segment with base parameter $Q_a$. The adjacent oriented charges $v$ and $v'$ give $n_{Q_a}=v'\wedge v$.}
    \label{fig:framingnumber}
\end{figure}

The refined topological string partition function of a $(p,q)$ five-brane web assembled from strips is obtained by multiplying strip contributions and summing over the partitions assigned to the glued legs. Each internal leg carries a base parameter $Q_a$, a framing contribution $f_\mu$, and a framing number $n_{Q_a}$ given by
\begin{equation}
    f_\mu(\mf t,\mf q) = (-1)^{|\mu|}\left( \frac{\mf t}{\mf q} \right)^{\frac{||\mu^t||^2}{2}}\mf q^{-\frac{\kappa_\mu}{2}} \quad,\quad n_Q = v'\,\wedge\,v = v_1'v_2 - v_2'v_1 ~,
\end{equation}
where $\kappa_\mu=||\mu||^2-||\mu^t||^2$ and the framing number is determined by the adjacent oriented five-brane charges; one such junction is illustrated in Figure \ref{fig:framingnumber}.
For two length-$N$ strips, the topological string partition function of the $N+SU(N)_\kappa+N$ gauge theory takes the form
\begin{equation}
    \begin{split}
        &\sum_{\bm\alpha} \left( \prod_{a=1}^N Q_a^{|\alpha_a|} \, f_{\alpha_a}(\mf t,\mf q)^{\,n_{Q_a}} \right) Z^{\rm strip} (\vec M, \vec N, K_1; \bm\beta,\bm\alpha) \; \left[ Z^{\rm strip} (\vec N, \vec L, K_2; \bm\alpha,\bm\gamma) \right]' ~,
    \end{split}
\end{equation}
where the prime denotes an alternate parameterisation of the strip that we describe below. As evident from the definition, the framing number depends on the precise geometry of the strips being glued and therefore encodes the effective Chern--Simons level $\kappa$ of the associated gauge group.

For the rectangular configuration engineering the $N+SU(N)_0+N$ configuration, these framing numbers vanish $n_{Q_a}=(1,1)\wedge(-1,-1)=0$, simplifying the topological string partition function to
\begin{equation}\label{eq:Ztop_SUN2N_app}
    \begin{split}
        &Z^{N+SU(N)+N} (Q_{\rm inst},\vec M, \vec L, K_1, K_2 ; \bm\beta,\bm\gamma) \\[0.1cm]
        &\hspace{3cm} = \sum_{\bm\alpha} \left(\prod_{a=1}^N Q_a^{|\alpha_a|}\right) \; Z^{\rm strip} (\vec M, \vec N, K_1; \bm\beta,\bm\alpha) \; [Z^{\rm strip} (\vec N,\vec L, K_2; \bm\alpha,\bm\gamma)]' ~.
    \end{split}
\end{equation}
The base parameters $Q_a$ can subsequently be expressed in terms of the instanton fugacity and the strip parameters, as we describe below. Due to this, we use $Q_{\rm inst}$ instead of $\vec Q$ among the arguments of the partition function, as in the main text.

\subsubsection*{Alternate parameterisation}

We now turn to explaining the prime inserted onto the second strip in \eqref{eq:Ztop_SUN2N_app}. This indicates that we use an {\it alternative} parametrisation for this strip. This changes the relation between the internal \Kah parameters of the strip and the mass parameters where we now have the following
\begin{equation}\label{eq:strip_par_alt}
    \mu_a = \frac{K}{M_aN_a}\sqrt{\frac{\mf t}{\mf q}} \quad,\quad \nu_a = \frac{M_{a+1}N_a}{K}\sqrt{\frac{\mf q}{\mf t}} ~.
\end{equation}
instead of \eqref{eq:strip_par_std}. These parameters now satisfy $\mu_a\nu_a = \frac{M_{a+1}}{M_a}$ and $\nu_a\mu_{a+1} = \frac{N_a}{N_{a+1}}$ for $a=1,\dots,N-1$. Thus, the adjacent products of the internal \Kah parameters now involve inverse ratios of the non abelian flavour fugacities relative to the standard parameterisation. The associated topological string partition function can be obtained directly from the standard expression in Eqs.\ (\ref{eq:Zstrip_pert}) and (\ref{eq:Zstrip_inst}) via the following relation
\begin{equation}\label{eq:altstripeq}
    [\cZ^{\rm strip}(M_a, N_a, K; \bm\beta,\bm\alpha)]' = \cZ^{\rm strip} \left( M_a^{-1}, N_a^{-1}, \frac{\mf q}{\mf t} K^{-1}; \bm\beta,\bm\alpha \right) ~.
\end{equation}

The consistency of this alternation prescription can be simply observed by noting that the perturbative contribution from the internal parallel legs reconstructs the correct $SU(N)$ measure:
\begin{equation}
    \begin{split}
        \left| \cZ^{\rm strip}_{\rm pert}(\vec M, \vec N, K_1; \bm\emptyset, \bm\alpha) [\cZ^{\rm strip}_{\rm pert}(\vec N, \vec L, K_2; \bm\alpha, \bm\emptyset)]' \right|^2 &\supset \prod_{1\leq a<b\leq N} \left| \cM\left( \frac{N_b}{N_a} \right)\cM\left( \frac{\mf t}{\mf q}\frac{N_b}{N_a} \right) \right|^2 \\
        &\supset \prod_{1\leq a<b\leq N} \left| \cM\left( \frac{N_a}{N_b} \right)\cM\left( \frac{N_b}{N_a} \right) \right|^2 ~,
    \end{split}
\end{equation}
where in the second line we use the reflection identity in \eqref{eq:Minv_identity}. This reproduces the contribution associated with non-zero roots of $SU(N)$ as desired; gluing two strips with the standard parameterisation would instead repeat one root orientation and thus fail to reconstruct the correct measure.

\subsubsection*{Base parameters and instanton fugacities}

Let $\mu_a^{(1)}$ and $\mu_a^{(2)}$ denote the diagonal \Kah parameters of two adjacent strips, in the notation of Figure \ref{fig:strip}. In terms of the mass parameters and the alternate parameterisation discussed above, we have
\begin{equation}
    \mu_a^{(1)} = \sqrt{\frac{\mf q}{\mf t}}\frac{M_aN_a}{K_1} \;\;,\quad \mu_a^{(2)} = \sqrt{\frac{\mf t}{\mf q}}\frac{K_2}{N_aL_a}
\end{equation}
The geometry of Figure \ref{fig:braneweb} then implies the relations
\begin{equation}
    Q_{\rm bot} = Q_1\mu_1^{(1)} \;\;,\quad Q_{\rm top} = Q_N\mu_N^{(2)} \;\;,\quad \frac{Q_{a+1}}{Q_a} = \frac{\mu_a^{(2)}}{\mu_{a+1}^{(1)}} ~.
\end{equation}
Iterating the final relation yields
\begin{equation}
    Q_{\rm bot} = Q_a \frac{\prod_{b\leq a}\mu_b^{(1)}}{\prod_{b<a}\mu_b^{(2)}} \;\;,\quad Q_{\rm top} = Q_a \frac{\prod_{b\geq a}\mu_b^{(2)}}{\prod_{b>a}\mu_b^{(1)}} ~.
\end{equation}
Geometrically, the instanton fugacity can be realised as the geometric mean of the exponentiated separation at the bottom and at the top: $Q_{\rm inst}=\sqrt{Q_{\rm top}Q_{\rm bot}}$, following \cite{Mitev:2014jza}. Consequently, we can cast the above expression into the following relation in terms of the strip parameters
\begin{equation}\label{eq:Qinst}
    Q_{\rm inst} = Q_a \times N_aM_a \prod_{b<a} M_bN_b^2L_b \times \left(\sqrt{\frac{\mf t}{\mf q}}K_1\right)^{\frac{N}{2}-a} \left(\sqrt{\frac{\mf t}{\mf q}}K_2\right)^{\frac{N}{2}-a+1} ~.
\end{equation}
This relation holds for any allowed $a=1,\dots,N$, expressing all $N$ base parameters in terms of the same instanton fugacity. Applying the construction to each adjacent pair of strips, with the appropriate parameterisation, gives the $k-1$ instanton fugacities of the full $N+SU(N)_0^{k-1}+N$ box theory.

\subsubsection*{Relation to Nekrasov}

For the isolated $N+SU(N)_0+N$ gauge theory, the external partitions are empty, and the remaining sum over $\bm\alpha$ supplies the gauge instanton expansion. Identifying the instantonic part of the topological string partition with the Nekrasov instanton partition function relies on using the relation \eqref{eq:NNek_rel} together with
\begin{equation}
    \Nek_{\lambda\emptyset}\left(\frac{\mf t}{\mf q}U^{-1}\right) = (-1)^{|\lambda|} \; \Nek_{\emptyset\lambda}\left(U\right) ~.
\end{equation}
The monomial factors depending on the strip parameters generated by this conversion combine with the base parameters through \eqref{eq:Qinst}. We use the following convention for the Nekrasov instanton partition function\footnote{With our strip gluing conventions, conversion of the refined topological string partition function produces a factor of $(-1)^{N-1}$ per instanton. Therefore the exact relation is $$Z^{N+SU(N)_0+N}_{\rm inst}(Q_{\rm inst}) = Z_{\rm inst}^{\rm Nek}\left((-1)^{N-1}Q_{\rm inst}\right) ~,$$ with all other arguments held fixed and suppressed. Throughout, $Q_{\rm inst}=\sqrt{Q_{\rm top}Q_{\rm bot}}$ retains its geometric definition; the signed argument accounts for the convention used in the Nekrasov expression.}
\begin{equation}\label{eq:NekInstPF}
    \begin{split}
        Z^{\rm Nek}_{\rm inst} (Q_{\rm inst},\vec M, \vec N,\vec L, K_1, K_2) = \sum_{\bm\alpha} \left[ \prod_{i=1}^N Q_{\rm inst}^{|\alpha_i|} \right] \frac{ \prod_{a,b=1}^N \Nek_{\emptyset\alpha_a}\left( \frac{M_bN_a}{K_1} \right) \Nek_{\emptyset\alpha_a}\left( \frac{N_aL_b}{K_2} \right) }{ \prod_{a,b=1}^N \Nek_{\alpha_a\alpha_b} \left( \frac{N_b}{N_a} \right) } ~.
    \end{split}
\end{equation}
Here $\vec M,\vec N,\vec L$ are all $SU(N)$ variables and hence satisfy the tracelessness relation
\begin{equation}
    \prod_{i=1}^N M_i = \prod_{i=1}^N N_i = \prod_{i=1}^N L_i = 1 ~.
\end{equation}
Finally, turning on a Chern--Simons level $\kappa$ amounts to introducing the following factor inside the sum over $\bm\alpha$ \cite{Tachikawa:2004ur}:
\begin{equation}
   \left(\prod_{a=1}^N N_a^{|\alpha_a|} \mf q^{\frac{-\|\alpha_a\|^{2}}{2}} \mf t^{\frac{\|\alpha_a^{t}\|^{2}}{2}}\right)^\kappa ~.
\end{equation}

\section{Three-dimensional partition functions and monopole charges}
\label{app:3dstuff}

In this appendix, we collect the conventions and tools used in the three-dimensional calculations.

\subsubsection*{$\mathbf{S}^3_b$ partition function conventions}

We work with the squashed sphere partition function whose general form was derived in \cite{Hama:2011ea} and adopt the conventions spelled out below. For a gauge theory $G$ and chiral multiplets transforming in the representations $\mathcal R$ and $\mathcal F$ of the gauge and flavour groups, respectively, the associated partition function is the following ${\rm rank}(G)$-dimensional integral over $\mathbb{R}^{{\rm rank}(G)}$:
\begin{equation}\label{eq:conventionpf}
    \cZ(\eta,\vec\nu) = \int {\rm d}\vec u \,\cZ_{\rm cl}(\eta)\, \frac{ \prod_{\vec\lambda\in\mathcal R} \prod_{\vec\omega\in\mathcal F} s_b\left( \frac{iQ}{2}(1-r)-\vec\lambda(\vec u)-\vec\omega(\vec\nu) \right) }{ \prod_{\vec\alpha\in G} s_b\left( \frac{iQ}{2}-\vec\alpha(\vec u) \right) } ~,
\end{equation}
where $\vec\alpha$ denote the roots of $G$, while $\vec\lambda$ and $\vec\omega$ run over the weights of $\mathcal R$ and $\mathcal F$; accordingly $\vec u$ and $\vec\nu$ parameterise the Cartan subalgebras of the gauge and the flavour group. The measure in \eqref{eq:conventionpf} reads
\begin{equation}
    {\rm d}\vec u = \frac{1}{|\mathcal W_G|} \prod_{a=1}^{{\rm rank}(G)} {\rm d}u_a ~,
\end{equation}
with $|\mathcal W_G|$ the order of the Weyl group of $G$. The classical prefactor encodes the FI coupling, whose parameter we denote by $\eta$
\begin{equation}
    \cZ_{\rm cl}(\eta) = {\rm exp} \left[ 2\pi i\,\eta\sum_{a=1}^{\rm rank(G)} u_a \right] ~.
\end{equation}
We write $Q=b+b^{-1}$, where $b$ is the squashing parameter of the three sphere. The double-sine function is defined by the regularised infinite product
\begin{equation}
    s_b(x) = \prod_{n_1,\,n_2\,\geq\,0} \frac{n_1 b+n_2 b^{-1}+\frac{Q}{2}-i\,x}{n_1 b+n_2 b^{-1}+\frac{Q}{2}+i\,x} ~.
\end{equation}

We describe all $\cN=4$ theories considered here in the $\cN=2^*$ language \cite{Tong:2000ky}. Concretely, we switch on a real mass $\text{Re}(m)$ for the axial symmetry $U(1)_A$, defined as the commutant inside the $\mathcal{N}=4$ R-symmetry $SU(2)\times SU(2)$ of the $U(1)_R$ subgroup that we keep manifest in $\mathcal{N}=2$ language. We parameterise mixing with the axial symmetry by
\begin{equation}
    R = R_0+r\,q_A ~.
\end{equation}
Here $R_0$ is the reference $R$ charge specified in Table \ref{chargeN2star}, $q_A$ is the axial charge, and $r$ parameterises axial mixing. As expected, the mixing realised in $\mathcal{N}=4$ theories assigns R-charge $\frac{1}{2}$ to a hypermultiplet and $1$ to an adjoint chiral.

To keep track of how $U(1)_R$ and $U(1)_A$ mix, it proves convenient to package the parameters into the holomorphic combination
\begin{equation}\label{eq:massm}
    m = \text{Re}(m)+\frac{iQ}{2}\,r ~.
\end{equation}
The charges under the two Abelian factors must then be distributed in a way compatible with the $\mathcal{N}=4$ superpotential, which for a single node takes the form
\begin{equation}
    \mathcal{W} = \text{Tr}\!\left[\,\Phi\,Q\,\widetilde{Q}\,\right] ~,
\end{equation}
where $\Phi$ is the adjoint chiral sitting inside the $\mathcal{N}=4$ vector multiplet and $Q$, $\widetilde{Q}$ are the two chiral multiplets of the hypermultiplet charged under the node at hand. Our choice of charges, both for a generic 3d $\mathcal{N}=4$ theory and for its mirror, is collected in
Table~\ref{chargeN2star}. At the level of the partition function charges under $R$ symmetry in the electric and magnetic theory are related by the following shift:
\begin{equation}
    m \;\longrightarrow\; \frac{iQ}{2}-m ~.
\end{equation}
For the  magnetic quivers throughout this work, we assign  to their hypers the charges reported for the magnetic theory in Table \ref{chargeN2star}:
\begin{table}[h]
    \centering
    \begin{tabular}{c|c|c|c}
        {Electric theory} & $Q$ & $\widetilde{Q}$ & $\Phi$ \\ \hline
        $U(1)_{A}$    & $-1$ & $-1$ & $2$ \\
        $U(1)_{R_0}$  & $1$  & $1$  & $0$
    \end{tabular}
    \qquad\qquad
    \begin{tabular}{c|c|c|c}
        {Magnetic theory} & $Q$ & $\widetilde{Q}$ & $\Phi$ \\ \hline
        $U(1)_{A}$    & $1$ & $1$ & $-2$ \\
        $U(1)_{R_0}$  & $0$ & $0$ & $2$
    \end{tabular}
    \caption{Charge assignments for the chiral fields in the $\mathcal{N}=2^*$ description. The left table refers to the conventions we adopt for a generic theory, the right one to those of its mirror. The two are related by the exchange $SU(2)_{\rm H} \leftrightarrow SU(2)_{\rm C}$.}
    \label{chargeN2star}
\end{table}

For convenience, we record the $U(N)$ measures in the $\mathcal{N}=2^*$ setup, obtained by collecting the $\mathcal{N}=2$ vector and adjoint contributions into a single object. When the $U(N)$ node gauges a Higgs branch (i.e.\ flavour) symmetry of the magnetic theory one has
\begin{equation}\label{HBmeasure}
    \Delta^{\text{HB}}_{N}(\vec{u};m) = \frac{\prod_{i,j=1}^{N} s_b\!\left(2m-\frac{iQ}{2}+u_i-u_j\right)}{\prod_{i\neq j}^{N} s_b\!\left(\frac{iQ}{2}+u_i-u_j\right)} ~,
\end{equation}
consistent with the assignments of Table \ref{chargeN2star}. 
If instead the $U(N)$ node gauges a Coulomb branch (i.e.\ topological)
symmetry, the relevant measure is
\begin{equation}\label{eq:CBmeasure}
    \Delta^{\text{CB}}_{N}(\vec{u};m) = \frac{\prod_{i,j=1}^{N} s_b\left(-2m+\frac{iQ}{2}+u_i-u_j\right)}{\prod_{i\neq j}^{N} s_b\left(\frac{iQ}{2}+u_i-u_j\right)} ~.
\end{equation}
The labels HB and CB are understood relative to the magnetic convention. For the electric theory, their roles are exchanged: Higgs branch gauging uses \eqref{eq:CBmeasure}, while Coulomb branch gauging uses \eqref{HBmeasure}.

\begin{figure}[h!]
    \centering
    \includegraphics[width=0.6\linewidth]{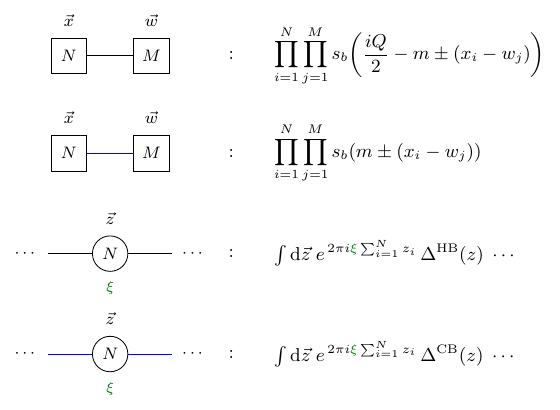}
    \caption{Hypermultiplet and gauging conventions for $S^3_b$ partition functions. Black and \textcolor{blue}{blue} edges are magnetic and electric charge assignments, respectively.}
    \label{fig:conventionsquiver}
\end{figure}

In quiver diagrams, circular nodes denote gauge groups and square nodes denote flavour symmetries. Gauging introduces the Cartan integral, the appropriate measure above and the classical FI factor. FI parameters are always shown in green. The hypermultiplets in the magnetic theory are depicted in black, while we depict those in the electric theory in \textcolor{blue}{blue}. These graphical conventions and their partition function contributions are summarised in Figure \ref{fig:conventionsquiver}.

\subsubsection*{The $T[U(N)]$ quiver tail}

\begin{figure}[h!]
    \centering
    \includegraphics[width=0.75\linewidth]{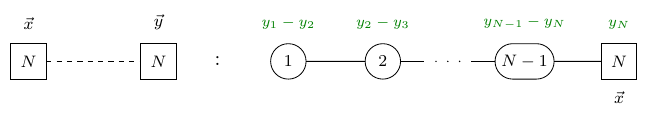}
    \caption{The $T[U(N)]$ quiver tail and the corresponding compact dashed-line notation.}
    \label{fig:tsuNappendix1}
\end{figure}

We will primarily consider theories constructed by gluing $T[U(N)]$ quiver tails. The linear quiver description of $T[U(N)]$, together with its compact dashed-line notation, is shown in Figure \ref{fig:tsuNappendix1}.
Its partition function follows from the rules introduced above and is denoted by
\begin{equation}
    \cZ_{T[U(N)]}(\vec X,\vec Y,m) ~,
\end{equation}
where $\vec X$ and $\vec Y$ parameterise the $U(N)_{\vec X}$ Higgs branch and enhanced $U(N)_{\vec Y}$ Coulomb branch symmetries, respectively. Note that with respect to the $T[SU(N)]$ theory of \cite{Gaiotto:2008ak}, the $T[U(N)]$ theory considered here has a contact term $e^{2\pi iY_N \sum_{j=1}^N X_j}$ for the flavour symmetry. Mirror symmetry exchanges the Higgs and Coulomb branches, together with their respective Cartan parameters, and acts on the axial mass as $m\to iQ/2-m$. At the level of the partition function, this amounts to
\begin{equation}
    \cZ^{T[U(N)]}(\vec{X},\vec{Y},m) = \cZ^{T[U(N)]}\left(\vec{Y},\vec{X},\frac{iQ}{2}-m\right) ~.
\end{equation}

\subsubsection*{Monopole charges}

Consider a unitary quiver with bifundamental hypermultiplets and no flavour nodes. For a bare monopole operator with magnetic charges $\mathfrak m$, we define its total charge
\begin{equation}\label{eq:monopolecharge}
    \mathcal{Q}(\mathfrak{m}) = \big(iQ-2m\big)\,\Delta(\mathfrak{m}) + \sum_a\xi_a\sum_{i=1}^{N_a}\mathfrak{m}^{(a)}_i ~,
\end{equation}
with
\begin{equation}\label{eq:GWdim}
    \Delta(\mathfrak{m}) = \frac{1}{2} \sum_{\langle a,b\rangle} \sum_{i=1}^{N_a}\sum_{j=1}^{N_b} \big|\mathfrak{m}^{(a)}_i-\mathfrak{m}^{(b)}_j\big|-\sum_a\sum_{i<j}^{N_a} \big|\mathfrak{m}^{(a)}_i-\mathfrak{m}^{(a)}_j\big| ~.
\end{equation}
Here $\Delta(\mathfrak{m})$ is the scaling dimension of the monopole operator introduced in \cite{Gaiotto:2008ak}, specialized to a quiver containing only gauge nodes. The second term in \eqref{eq:monopolecharge} encodes the charges of the monopole under the Cartan subgroup of the topological symmetry, with $\xi_a$ denoting the corresponding FI parameters.

The combination $iQ-2m$ is precisely the charge of the Coulomb branch moment map operator, where for the canonical $\cN=4$ $R$ charge assignment $m=iQ/4$ this reduces to $iQ/2$. For canonical $\mathcal{N}=4$ R-charge, \eqref{eq:monopolecharge} becomes $iQ/2$ times the monopole R-charge, supplemented by the contribution from the FI parameters. The condition for a monopole of magnetic charge $\mathfrak{m}$ to acquire a VEV is to have zero total charge $\mathcal{Q}(\mathfrak{m})=0$.

\bibliographystyle{JHEP}
\bibliography{ref}

@article{Comi:2022aqo,
    author = "Comi, Riccardo and Hwang, Chiung and Marino, Fabio and Pasquetti, Sara and Sacchi, Matteo",
    title = "{The SL(2, {\ensuremath{\mathbb{Z}}}) dualization algorithm at work}",
    eprint = "2212.10571",
    archivePrefix = "arXiv",
    primaryClass = "hep-th",
    reportNumber = "CTPU-PTC-22-28",
    doi = "10.1007/JHEP06(2023)119",
    journal = "JHEP",
    volume = "06",
    pages = "119",
    year = "2023"
}

@article{Hwang:2021ulb,
    author = "Hwang, Chiung and Pasquetti, Sara and Sacchi, Matteo",
    title = "{Rethinking mirror symmetry as a local duality on fields}",
    eprint = "2110.11362",
    archivePrefix = "arXiv",
    primaryClass = "hep-th",
    doi = "10.1103/PhysRevD.106.105014",
    journal = "Phys. Rev. D",
    volume = "106",
    number = "10",
    pages = "105014",
    year = "2022"
}

@article{Benini:2011mf,
    author = "Benini, Francesco and Closset, Cyril and Cremonesi, Stefano",
    title = "{Comments on 3d Seiberg-like dualities}",
    eprint = "1108.5373",
    archivePrefix = "arXiv",
    primaryClass = "hep-th",
    reportNumber = "PU-2391, WIS-07-11-AUG-DPPA, TAUP-2931-11",
    doi = "10.1007/JHEP10(2011)075",
    journal = "JHEP",
    volume = "10",
    pages = "075",
    year = "2011"
}

@article{vanBeestBourgetEckhardSchaferNameki2020,
  author        = {van Beest, Marieke and Bourget, Antoine and
                   Eckhard, Julius and Schafer-Nameki, Sakura},
  title         = {(5d RG-flow) Trees in the Tropical Rain Forest},
  journal       = {JHEP},
  volume        = {03},
  pages         = {241},
  year          = {2021},
  doi           = {10.1007/JHEP03(2021)241},
  eprint        = {2011.07033},
  archivePrefix = {arXiv},
  primaryClass  = {hep-th}
}

@article{BourgetHananyMiketa2020,
  author        = {Bourget, Antoine and Hanany, Amihay and Miketa, Dominik},
  title         = {Quiver origami: discrete gauging and folding},
  journal       = {JHEP},
  volume        = {01},
  pages         = {086},
  year          = {2021},
  doi           = {10.1007/JHEP01(2021)086},
  eprint        = {2005.05273},
  archivePrefix = {arXiv},
  primaryClass  = {hep-th}
}

@article{BourgetGrimmingerMartoneZafrir2021,
  author        = {Bourget, Antoine and Grimminger, Julius F. and
                   Martone, Mario and Zafrir, Gabi},
  title         = {Magnetic quivers for rank 2 theories},
  journal       = {JHEP},
  volume        = {03},
  pages         = {208},
  year          = {2022},
  doi           = {10.1007/JHEP03(2022)208},
  eprint        = {2110.11365},
  archivePrefix = {arXiv},
  primaryClass  = {hep-th}
}

@article{Aharony:1997bh,
    author = "Aharony, Ofer and Hanany, Amihay and Kol, Barak",
    title = "{Webs of (p,q) five-branes, five-dimensional field theories and grid diagrams}",
    eprint = "hep-th/9710116",
    archivePrefix = "arXiv",
    reportNumber = "IASSNS-HEP-97-113, RU-97-81, SU-ITP-97-40",
    doi = "10.1088/1126-6708/1998/01/002",
    journal = "JHEP",
    volume = "01",
    pages = "002",
    year = "1998"
}

@article{Kapustin:2011jm,
    author = "Kapustin, Anton and Willett, Brian",
    title = "{Generalized Superconformal Index for Three Dimensional Field Theories}",
    eprint = "1106.2484",
    archivePrefix = "arXiv",
    primaryClass = "hep-th",
    reportNumber = "68-2840",
    month = "6",
    year = "2011"
}

@article{Imamura:2011su,
    author = "Imamura, Yosuke and Yokoyama, Shuichi",
    title = "{Index for three dimensional superconformal field theories with general R-charge assignments}",
    eprint = "1101.0557",
    archivePrefix = "arXiv",
    primaryClass = "hep-th",
    reportNumber = "UT-11-01, TIT-HEP-607",
    doi = "10.1007/JHEP04(2011)007",
    journal = "JHEP",
    volume = "04",
    pages = "007",
    year = "2011"
}

@article{Closset:2012vp,
    author = "Closset, Cyril and Dumitrescu, Thomas T. and Festuccia, Guido and Komargodski, Zohar and Seiberg, Nathan",
    title = "{Comments on Chern-Simons Contact Terms in Three Dimensions}",
    eprint = "1206.5218",
    archivePrefix = "arXiv",
    primaryClass = "hep-th",
    reportNumber = "PUTP-2417, WIS-10-12-JUNE-DPPA",
    doi = "10.1007/JHEP09(2012)091",
    journal = "JHEP",
    volume = "09",
    pages = "091",
    year = "2012"
}

@article{Kapustin:2009kz,
    author = "Kapustin, Anton and Willett, Brian and Yaakov, Itamar",
    title = "{Exact Results for Wilson Loops in Superconformal Chern-Simons Theories with Matter}",
    eprint = "0909.4559",
    archivePrefix = "arXiv",
    primaryClass = "hep-th",
    reportNumber = "CALT-68-2750",
    doi = "10.1007/JHEP03(2010)089",
    journal = "JHEP",
    volume = "03",
    pages = "089",
    year = "2010"
}

@article{Bergman:2013aca,
    author = "Bergman, Oren and Rodr{\'\i}guez-G{\'o}mez, Diego and Zafrir, Gabi",
    title = "{5-Brane Webs, Symmetry Enhancement, and Duality in 5d Supersymmetric Gauge Theory}",
    eprint = "1311.4199",
    archivePrefix = "arXiv",
    primaryClass = "hep-th",
    doi = "10.1007/JHEP03(2014)112",
    journal = "JHEP",
    volume = "03",
    pages = "112",
    year = "2014"
}

@article{Terashima:2012ra,
    author = "Terashima, Seiji",
    title = "{Supersymmetric gauge theories on $S^4$ x $S^1$}",
    eprint = "1207.2163",
    archivePrefix = "arXiv",
    primaryClass = "hep-th",
    reportNumber = "YITP-12-55",
    doi = "10.1103/PhysRevD.89.125001",
    journal = "Phys. Rev. D",
    volume = "89",
    number = "12",
    pages = "125001",
    year = "2014"
}

@article{Awata:2008ed,
    author = "Awata, Hidetoshi and Kanno, Hiroaki",
    title = "{Refined BPS state counting from Nekrasov's formula and Macdonald functions}",
    eprint = "0805.0191",
    archivePrefix = "arXiv",
    primaryClass = "hep-th",
    doi = "10.1142/S0217751X09043006",
    journal = "Int. J. Mod. Phys. A",
    volume = "24",
    pages = "2253--2306",
    year = "2009"
}

@article{Awata:2005fa,
    author = "Awata, Hidetoshi and Kanno, Hiroaki",
    title = "{Instanton counting, Macdonald functions and the moduli space of D-branes}",
    eprint = "hep-th/0502061",
    archivePrefix = "arXiv",
    doi = "10.1088/1126-6708/2005/05/039",
    journal = "JHEP",
    volume = "05",
    pages = "039",
    year = "2005"
}

@article{Wyllard:2009hg,
    author = "Wyllard, Niclas",
    title = "{A(N-1) conformal Toda field theory correlation functions from conformal N = 2 SU(N) quiver gauge theories}",
    eprint = "0907.2189",
    archivePrefix = "arXiv",
    primaryClass = "hep-th",
    doi = "10.1088/1126-6708/2009/11/002",
    journal = "JHEP",
    volume = "11",
    pages = "002",
    year = "2009"
}

@article{Nieri:2013vba,
    author = "Nieri, Fabrizio and Pasquetti, Sara and Passerini, Filippo and Torrielli, Alessandro",
    title = "{5D partition functions, q-Virasoro systems and integrable spin-chains}",
    eprint = "1312.1294",
    archivePrefix = "arXiv",
    primaryClass = "hep-th",
    reportNumber = "DMUS-MP-13-19, PUPT-2456, LPTENS-13-23",
    doi = "10.1007/JHEP12(2014)040",
    journal = "JHEP",
    volume = "12",
    pages = "040",
    year = "2014"
}

@article{Nieri:2013yra,
    author = "Nieri, Fabrizio and Pasquetti, Sara and Passerini, Filippo",
    title = "{3d and 5d Gauge Theory Partition Functions as $q$-deformed CFT Correlators}",
    eprint = "1303.2626",
    archivePrefix = "arXiv",
    primaryClass = "hep-th",
    reportNumber = "DMUS-MP-13-07, CERN-PH-TH-2013-046, DMUS-MP-13/07, CERN-PH-TH/2013-046",
    doi = "10.1007/s11005-014-0727-9",
    journal = "Lett. Math. Phys.",
    volume = "105",
    number = "1",
    pages = "109--148",
    year = "2015"
}

@article{two,
    author = "Garavaglia, Sebastiano and Pasquetti, Sara and Singh, Palash",
    title = "From pq-webs to magnetic quivers and back II: distributional indices and {5d} theories breaking bad",
    journal = "to appear"
}

@article{Eckhard:2020jyr,
    author = {Eckhard, Julius and Sch{\"a}fer-Nameki, Sakura and Wang, Yi-Nan},
    title = "{Trifectas for T$_{N}$ in 5d}",
    eprint = "2004.15007",
    archivePrefix = "arXiv",
    primaryClass = "hep-th",
    doi = "10.1007/JHEP07(2020)199",
    journal = "JHEP",
    volume = "07",
    number = "07",
    pages = "199",
    year = "2020"
}

@article{Bourget:2019rtl,
    author = "Bourget, Antoine and Cabrera, Santiago and Grimminger, Julius F. and Hanany, Amihay and Zhong, Zhenghao",
    title = "{Brane Webs and Magnetic Quivers for SQCD}",
    eprint = "1909.00667",
    archivePrefix = "arXiv",
    primaryClass = "hep-th",
    doi = "10.1007/JHEP03(2020)176",
    journal = "JHEP",
    volume = "03",
    pages = "176",
    year = "2020"
}

@article{Aharony:2013dha,
  author        = "Aharony, Ofer and Razamat, Shlomo S. and Seiberg, Nathan and Willett, Brian",
  title         = "{3d dualities from 4d dualities}",
  journal       = "JHEP",
  volume        = "07",
  pages         = "149",
  year          = "2013",
  eprint        = "1305.3924",
  archivePrefix = "arXiv",
  primaryClass  = "hep-th",
  doi           = "10.1007/JHEP07(2013)149"
}

@article{Mitev:2014jza,
    author = "Mitev, Vladimir and Pomoni, Elli and Taki, Masato and Yagi, Futoshi",
    title = "{Fiber-Base Duality and Global Symmetry Enhancement}",
    eprint = "1411.2450",
    archivePrefix = "arXiv",
    primaryClass = "hep-th",
    reportNumber = "DESY-14-212, HU-Mathematik-2014-32, HU-EP-14-48, KIAS-P14066",
    doi = "10.1007/JHEP04(2015)052",
    journal = "JHEP",
    volume = "04",
    pages = "052",
    year = "2015"
}

@article{Katz:1996fh,
    author = "Katz, Sheldon H. and Klemm, Albrecht and Vafa, Cumrun",
    title = "{Geometric Engineering of Quantum Field Theories}",
    eprint = "hep-th/9609239",
    archivePrefix = "arXiv",
    doi = "10.1016/S0550-3213(97)00282-4",
    journal = "Nucl. Phys. B",
    volume = "497",
    pages = "173--195",
    year = "1997"
}

@article{Cremonesi:2015lsa,
    author = "Cremonesi, Stefano and Ferlito, Giulia and Hanany, Amihay and Mekareeya, Noppadol",
    title = "{Instanton Operators and the Higgs Branch at Infinite Coupling}",
    eprint = "1505.06302",
    archivePrefix = "arXiv",
    primaryClass = "hep-th",
    doi = "10.1007/JHEP04(2017)042",
    journal = "JHEP",
    volume = "04",
    pages = "042",
    year = "2017"
}

@article{Ferlito:2016grh,
    author = "Ferlito, Giulia and Hanany, Amihay",
    title = "{A tale of two cones: the Higgs Branch of Sp(n) theories with 2n flavours}",
    eprint = "1609.06724",
    archivePrefix = "arXiv",
    primaryClass = "hep-th",
    month = "9",
    year = "2016"
}

@article{Ferlito:2017xdq,
    author = "Ferlito, Giulia and Hanany, Amihay and Mekareeya, Noppadol and Zafrir, Gabi",
    title = "{3d Coulomb branch and 5d Higgs branch at infinite coupling}",
    eprint = "1712.06604",
    archivePrefix = "arXiv",
    primaryClass = "hep-th",
    doi = "10.1007/JHEP07(2018)061",
    journal = "JHEP",
    volume = "07",
    pages = "061",
    year = "2018"
}

@article{Cabrera:2018jxt,
    author = "Cabrera, Santiago and Hanany, Amihay and Yagi, Futoshi",
    title = "{Tropical Geometry and Five Dimensional Higgs Branches at Infinite Coupling}",
    eprint = "1810.01379",
    archivePrefix = "arXiv",
    primaryClass = "hep-th",
    doi = "10.1007/JHEP01(2019)068",
    journal = "JHEP",
    volume = "01",
    pages = "068",
    year = "2019"
}

@article{Hayashi:2014hfa,
    author = "Hayashi, Hirotaka and Zoccarato, Gianluca",
    title = "{Exact partition functions of Higgsed 5d $T_N$ theories}",
    eprint = "1409.0571",
    archivePrefix = "arXiv",
    primaryClass = "hep-th",
    doi = "10.1007/JHEP01(2015)093",
    journal = "JHEP",
    volume = "01",
    pages = "093",
    year = "2015"
}

@article{Alday:2009aq,
    author = "Alday, Luis F. and Gaiotto, Davide and Tachikawa, Yuji",
    title = "{Liouville Correlation Functions from Four-dimensional Gauge Theories}",
    eprint = "0906.3219",
    archivePrefix = "arXiv",
    primaryClass = "hep-th",
    doi = "10.1007/s11005-010-0369-5",
    journal = "Lett. Math. Phys.",
    volume = "91",
    pages = "167--197",
    year = "2010"
}

@article{Leung:1997tw,
    author = "Leung, Naichung Conan and Vafa, Cumrun",
    title = "{Branes and toric geometry}",
    eprint = "hep-th/9711013",
    archivePrefix = "arXiv",
    reportNumber = "HUTP-97-A090",
    doi = "10.4310/ATMP.1998.v2.n1.a4",
    journal = "Adv. Theor. Math. Phys.",
    volume = "2",
    pages = "91--118",
    year = "1998"
}

@article{Nekrasov:2003rj,
    author = "Nekrasov, Nikita and Okounkov, Andrei",
    title = "{Seiberg-Witten theory and random partitions}",
    eprint = "hep-th/0306238",
    archivePrefix = "arXiv",
    reportNumber = "ITEP-TH-36-03, PUDM-2003, IHES-P-03-43",
    doi = "10.1007/0-8176-4467-9_15",
    journal = "Prog. Math.",
    volume = "244",
    pages = "525--596",
    year = "2006"
}

@article{Aharony:1997ju,
    author = "Aharony, Ofer and Hanany, Amihay",
    title = "{Branes, superpotentials and superconformal fixed points}",
    eprint = "hep-th/9704170",
    archivePrefix = "arXiv",
    reportNumber = "RU-97-25, IASSNS-HEP-97-38",
    doi = "10.1016/S0550-3213(97)00472-0",
    journal = "Nucl. Phys. B",
    volume = "504",
    pages = "239--271",
    year = "1997"
}

@article{Iqbal:2012xm,
    author = "Iqbal, Amer and Vafa, Cumrun",
    title = "{BPS Degeneracies and Superconformal Index in Diverse Dimensions}",
    eprint = "1210.3605",
    archivePrefix = "arXiv",
    primaryClass = "hep-th",
    doi = "10.1103/PhysRevD.90.105031",
    journal = "Phys. Rev. D",
    volume = "90",
    number = "10",
    pages = "105031",
    year = "2014"
}

@article{vanBeest:2021xyt,
    author = "van Beest, Marieke and Giacomelli, Simone",
    title = "{Connecting 5d Higgs branches via Fayet-Iliopoulos deformations}",
    eprint = "2110.02872",
    archivePrefix = "arXiv",
    primaryClass = "hep-th",
    doi = "10.1007/JHEP12(2021)202",
    journal = "JHEP",
    volume = "12",
    pages = "202",
    year = "2021"
}

@article{Tachikawa:2004ur,
    author = "Tachikawa, Yuji",
    title = "{Five-dimensional Chern-Simons terms and Nekrasov's instanton counting}",
    eprint = "hep-th/0401184",
    archivePrefix = "arXiv",
    reportNumber = "UT-04-01",
    doi = "10.1088/1126-6708/2004/02/050",
    journal = "JHEP",
    volume = "02",
    pages = "050",
    year = "2004"
}

@article{Bourget:2023uhe,
    author = "Bourget, Antoine and Giacomelli, Simone and Grimminger, Julius F.",
    title = "{FI-flows of 3d $N=4$ Theories}",
    eprint = "2302.03698",
    archivePrefix = "arXiv",
    primaryClass = "hep-th",
    doi = "10.1007/JHEP04(2023)015",
    journal = "JHEP",
    volume = "04",
    pages = "015",
    year = "2023"
}

@article{Kim:2012gu,
    author = "Kim, Hee-Cheol and Kim, Sung-Soo and Lee, Kimyeong",
    title = "{5-dim Superconformal Index with Enhanced En Global Symmetry}",
    eprint = "1206.6781",
    archivePrefix = "arXiv",
    primaryClass = "hep-th",
    reportNumber = "KIAS-P12033",
    doi = "10.1007/JHEP10(2012)142",
    journal = "JHEP",
    volume = "10",
    pages = "142",
    year = "2012"
}

@article{Nekrasov:2002qd,
    author = "Nekrasov, Nikita A.",
    title = "{Seiberg-Witten prepotential from instanton counting}",
    eprint = "hep-th/0206161",
    archivePrefix = "arXiv",
    reportNumber = "ITEP-TH-22-02, IHES-P-04-22",
    doi = "10.4310/ATMP.2003.v7.n5.a4",
    journal = "Adv. Theor. Math. Phys.",
    volume = "7",
    number = "5",
    pages = "831--864",
    year = "2003"
}

@article{Aganagic:2003db,
    author = "Aganagic, Mina and Klemm, Albrecht and Marino, Marcos and Vafa, Cumrun",
    title = "{The Topological vertex}",
    eprint = "hep-th/0305132",
    archivePrefix = "arXiv",
    reportNumber = "CALT-68-2439, HUTP-03-A032, HU-EP-03-24, CERN-TH-2003-111",
    doi = "10.1007/s00220-004-1162-z",
    journal = "Commun. Math. Phys.",
    volume = "254",
    pages = "425--478",
    year = "2005"
}

@article{Iqbal:2007ii,
    author = "Iqbal, Amer and Kozcaz, Can and Vafa, Cumrun",
    title = "{The Refined topological vertex}",
    eprint = "hep-th/0701156",
    archivePrefix = "arXiv",
    doi = "10.1088/1126-6708/2009/10/069",
    journal = "JHEP",
    volume = "10",
    pages = "069",
    year = "2009"
}

@article{Iqbal:2004ne,
    author = "Iqbal, Amer and Kashani-Poor, Amir-Kian",
    title = "{The Vertex on a strip}",
    eprint = "hep-th/0410174",
    archivePrefix = "arXiv",
    reportNumber = "SMS-0402, SLAC-PUB-10804, SU-ITP-04-39",
    doi = "10.4310/ATMP.2006.v10.n3.a2",
    journal = "Adv. Theor. Math. Phys.",
    volume = "10",
    number = "3",
    pages = "317--343",
    year = "2006"
}

@article{Taki:2007dh,
    author = "Taki, Masato",
    title = "{Refined Topological Vertex and Instanton Counting}",
    eprint = "0710.1776",
    archivePrefix = "arXiv",
    primaryClass = "hep-th",
    reportNumber = "UT-07-32",
    doi = "10.1088/1126-6708/2008/03/048",
    journal = "JHEP",
    volume = "03",
    pages = "048",
    year = "2008"
}

@article{Isachenkov:2014eya,
    author = "Isachenkov, Mikhail and Mitev, Vladimir and Pomoni, Elli",
    title = "{Toda 3-Point Functions From Topological Strings II}",
    eprint = "1412.3395",
    archivePrefix = "arXiv",
    primaryClass = "hep-th",
    reportNumber = "DESY-14-236, HU-MATHEMATIK-35-2014, HU-EP-14-57",
    doi = "10.1007/JHEP08(2016)066",
    journal = "JHEP",
    volume = "08",
    pages = "066",
    year = "2016"
}

@article{Comi:2025zwu,
    author = "Comi, Riccardo and Garavaglia, Sebastiano and Giacomelli, Simone and Pasquetti, Sara and Singh, Palash",
    title = "{Breaking bad theories of class $ \mathcal{S} $}",
    eprint = "2508.21071",
    archivePrefix = "arXiv",
    primaryClass = "hep-th",
    doi = "10.1007/JHEP05(2026)075",
    journal = "JHEP",
    volume = "05",
    pages = "075",
    year = "2026"
}

@article{Mitev:2014isa,
    author = "Mitev, Vladimir and Pomoni, Elli",
    title = "{Toda 3-Point Functions From Topological Strings}",
    eprint = "1409.6313",
    archivePrefix = "arXiv",
    primaryClass = "hep-th",
    reportNumber = "DESY-14-160, HU-Mathematik-19-2014, HU-EP-14-23",
    doi = "10.1007/JHEP06(2015)049",
    journal = "JHEP",
    volume = "06",
    pages = "049",
    year = "2015"
}

@article{Gaiotto:2012xa,
    author = "Gaiotto, Davide and Rastelli, Leonardo and Razamat, Shlomo S.",
    title = "{Bootstrapping the superconformal index with surface defects}",
    eprint = "1207.3577",
    archivePrefix = "arXiv",
    primaryClass = "hep-th",
    doi = "10.1007/JHEP01(2013)022",
    journal = "JHEP",
    volume = "01",
    pages = "022",
    year = "2013"
}

@article{Cabrera:2018ann,
    author = "Cabrera, Santiago and Hanany, Amihay",
    title = "{Quiver Subtractions}",
    eprint = "1803.11205",
    archivePrefix = "arXiv",
    primaryClass = "hep-th",
    doi = "10.1007/JHEP09(2018)008",
    journal = "JHEP",
    volume = "09",
    pages = "008",
    year = "2018"
}

@article{Benini:2009gi,
    author = "Benini, Francesco and Benvenuti, Sergio and Tachikawa, Yuji",
    title = "{Webs of five-branes and N=2 superconformal field theories}",
    eprint = "0906.0359",
    archivePrefix = "arXiv",
    primaryClass = "hep-th",
    reportNumber = "PUPT-2303",
    doi = "10.1088/1126-6708/2009/09/052",
    journal = "JHEP",
    volume = "09",
    pages = "052",
    year = "2009"
}

@article{Hama:2011ea,
    author = "Hama, Naofumi and Hosomichi, Kazuo and Lee, Sungjay",
    title = "{SUSY Gauge Theories on Squashed Three-Spheres}",
    eprint = "1102.4716",
    archivePrefix = "arXiv",
    primaryClass = "hep-th",
    reportNumber = "DAMTP-2011-6, YITP-11-1",
    doi = "10.1007/JHEP05(2011)014",
    journal = "JHEP",
    volume = "05",
    pages = "014",
    year = "2011"
}

@article{Giacomelli:2023zkk,
    author = "Giacomelli, Simone and Hwang, Chiung and Marino, Fabio and Pasquetti, Sara and Sacchi, Matteo",
    title = "{Probing bad theories with the dualization algorithm. Part I}",
    eprint = "2309.05326",
    archivePrefix = "arXiv",
    primaryClass = "hep-th",
    reportNumber = "USTC-ICTS/PCFT-23-24, DMUS-MP-23/12",
    doi = "10.1007/JHEP04(2024)008",
    journal = "JHEP",
    volume = "04",
    pages = "008",
    year = "2024"
}

@article{Giacomelli:2024laq,
    author = "Giacomelli, Simone and Hwang, Chiung and Marino, Fabio and Pasquetti, Sara and Sacchi, Matteo",
    title = "{Probing bad theories with the dualization algorithm. Part II.}",
    eprint = "2401.14456",
    archivePrefix = "arXiv",
    primaryClass = "hep-th",
    doi = "10.1007/JHEP07(2024)165",
    journal = "JHEP",
    volume = "07",
    pages = "165",
    year = "2024"
}

@article{Gaiotto:2008ak,
    author = "Gaiotto, Davide and Witten, Edward",
    title = "{S-Duality of Boundary Conditions In N=4 Super Yang-Mills Theory}",
    eprint = "0807.3720",
    archivePrefix = "arXiv",
    primaryClass = "hep-th",
    doi = "10.4310/ATMP.2009.v13.n3.a5",
    journal = "Adv. Theor. Math. Phys.",
    volume = "13",
    number = "3",
    pages = "721--896",
    year = "2009"
}

@article{Bottini:2021vms,
    author = "Bottini, Lea E. and Hwang, Chiung and Pasquetti, Sara and Sacchi, Matteo",
    title = "{4d S-duality wall and SL(2, \ensuremath{\mathbb{Z}}) relations}",
    eprint = "2110.08001",
    archivePrefix = "arXiv",
    primaryClass = "hep-th",
    doi = "10.1007/JHEP03(2022)035",
    journal = "JHEP",
    volume = "03",
    pages = "035",
    year = "2022"
}

@article{Tong:2000ky,
    author = "Tong, David",
    title = "{Dynamics of N=2 supersymmetric Chern-Simons theories}",
    eprint = "hep-th/0005186",
    archivePrefix = "arXiv",
    reportNumber = "MIT-CTP-2985, KCL-TH-00-28",
    doi = "10.1088/1126-6708/2000/07/019",
    journal = "JHEP",
    volume = "07",
    pages = "019",
    year = "2000"
}

\end{document}